\documentclass[aps,prd,preprintnumbers,nofootinbib,superscriptaddress,
fleqn,floatfix,tightenlines,twocolumn,10pt]{revtex4} 
\usepackage{amsmath,amsfonts,amssymb,amscd,amsxtra,amsthm}
\usepackage{graphicx}  
\usepackage{epstopdf}
\usepackage{dcolumn}  
\usepackage{bm}          
\usepackage{slashed}
\usepackage{cancel}
\usepackage{float}
\usepackage{mathtools}
\usepackage{amsbsy}
\usepackage{amstext}
\usepackage{tabularx}
\usepackage{enumitem}  
\usepackage{array} 
\usepackage{tikz}
\usepackage{multirow}
\usepackage{physics}
\usepackage{makecell}
\usepackage{subcaption}
\usepackage{dsfont}
\usepackage{hyperref}

\usepackage[utf8]{inputenc} 
\usepackage{booktabs} 
\usepackage[normalem]{ulem} 

\renewcommand{\sout}{\bgroup \color{red} \ULdepth=-.5ex \ULset}

\makeatletter

\begin{document}
\title{Relativistic energy-momentum tensor distributions in a polarized nucleon} 

\author{Ho-Yeon Won}
\email[E-mail: ]{hoyeon.won@polytechnique.edu}
\affiliation{CPHT, CNRS, \'Ecole Polytechnique, Institut Polytechnique de Paris,
91120 Palaiseau, France}

\author{C\'edric Lorc\'e}
\email[E-mail: ]{cedric.lorce@polytechnique.edu}
\affiliation{CPHT, CNRS, \'Ecole Polytechnique, Institut Polytechnique de Paris,
91120 Palaiseau, France}

\date {\today}
\begin{abstract}
We study in detail the relativistic distributions of energy, 
longitudinal momentum, longitudinal energy flux, 
and axial momentum flux inside nucleons based on the quantum phase-space formalism. Similar to recent studies on the electromagnetic current, we include the effects of the nucleon polarization and show that the latter are essential for understanding how the Breit frame distributions transform under a longitudinal Lorentz boost. We also explicitly demonstrate that, in the infinite-momentum frame, these distributions allow one to recover not only the ``good'' but also the ``bad'' components of the light-front energy-momentum tensor distributions. 
\end{abstract}
\maketitle
\section{Introduction}
A critical challenge in hadronic physics is to unravel 
the internal structure of the nucleons.
It is also one of the key questions 
that the future Electron-Ion Collider (EIC) 
experiments~\cite{Accardi:2012qut,AbdulKhalek:2021gbh}
in the USA aims to address, namely
\textit{``How are partons inside the nucleon distributed 
in both momentum and position space?''}.
The energy-momentum tensor (EMT) emerged as 
a promising candidate for addressing these questions
in the framework of Quantum chromodynamics (QCD). 
It plays a crucial role in encoding information 
about the internal structure, including the mass, spin, 
and mechanical properties of the nucleon, see~\cite{Polyakov:2018zvc,Burkert:2023wzr} for recent reviews on both theoretical and experimental aspects.

Similar to the electromagnetic current, the matrix elements of the EMT are described in terms of form factors (FFs), which are Lorentz-invariant functions encoding the spatial extension and structure of the nucleon. These EMT FFs can be accessed indirectly through generalized parton distributions~\cite{Ji:1996ek,Diehl:2003ny}. A chiral decomposition of the EMT motivated the introduction of the P-odd partner of the EMT, which also allows the study of the spin-orbit correlations~\cite{Lorce:2014mxa,Kim:2024cbq,Bhattacharya:2024sno,Lorce:2025ayr}. In addition, the relation 
between the EMT and transverse-momentum dependent 
parton distributions has recently been explored in~\cite{Lorce:2023zzg}.

Spatial distributions are defined via a Fourier transform of the matrix elements evaluated in some particular frame. Three-dimensional EMT distributions in the Breit frame (BF) have been introduced in~\cite{Polyakov:2002yz}. BF distributions are known to be plagued by relativistic recoil corrections which spoil their interpretation as genuine densities. Alternatively, the light-front (LF) formalism allows one to define genuine two-dimensional relativistic densities in the Drell-Yan frame (DYF)~\cite{Burkardt:2000za,Burkardt:2002hr,Miller:2007uy,Miller:2010nz}. The LF EMT distributions for an unpolarized nucleon with vanishing transverse momentum have been introduced in~\cite{Lorce:2018egm} and shown to coincide with the EMT distributions in the infinite-momentum frame (IMF). These LF distributions have then been generalized to arbitrary values of the nucleon transverse momentum~\cite{Freese:2021czn}. As a result of the Galilean symmetry of the LF formalism in the transverse plane, the intrinsic LF distributions simply coincide with those defined at vanishing transverse momentum. An annoying feature of this formalism is that transverse nucleon polarization induces distortions in e.g.~the LF charge~\cite{Carlson:2007xd} and LF momentum~\cite{Abidin:2008sb} distributions. 

An interpolation between the BF and IMF EMT distributions can be defined within a phase-space approach, where the relativistic distributions are regarded only as quasi-densities~\cite{Lorce:2018egm}. However, only unpolarized distributions were considered and a complete understanding of how BF EMT distributions evolve into IMF EMT distribution could not be reached. Recent studies on the frame dependence of the electromagnetic charge and current distributions~\cite{Lorce:2020onh,Lorce:2022jyi,Chen:2022smg} clarified the situation and revealed that the Wigner spin rotation plays a crucial role in this matter.

In this work, we extend the analysis of~\cite{Lorce:2018egm} by including the nucleon polarization. 
The paper is organized as follows.
In Sec.~\ref{sec.2}, 
we remind the parametrization of the EMT matrix elements and some of its key properties.
Then, 
in Sec.~\ref{sec.3}, 
we review the definition of the relativistic spatial distributions within the quantum phase-space approach, and we analyze the distributions of energy, longitudinal momentum, 
longitudinal energy flux, and axial momentum flux 
for both unpolarized and transversely polarized nucleons. Switching to the LF components, we show in Sec.~\ref{sec.4} that these distributions coincide in IMF limit with the LF ones.
After a summary of our results, we added three appendices where we (A) discuss some details regarding the value of the dipole mass used for quark spin FF, (B) present an alternative parametrization of the EMT matrix elements directly in terms of the BF multipole FFs, and (C) show additional plots for the longitudinal momentum and axial momentum flux distributions.

\section{Matrix elements of the energy-momentum tensor \label{sec.2}}
According to General relativity (GR), the EMT is a symmetric rank-2 tensor.
As a result, many early studies on the EMT in particle physics adopted the symmetric form, also known as the Belinfante-improved EMT~\cite{BELINFANTE1939887,Pauli:1940dq}.
It should, however, be kept in mind that GR is a classical field theory, whereas the concept of spin, which is directly related to the antisymmetric part of the EMT, is of a quantum nature.
Hence, restricting the analysis to the symmetric EMT alone is not well justified in the context of particle physics, see Ref.~\cite{Leader:2013jra} for further details and discussions. In the present work, we consider the (more general) asymmetric form of the EMT, which allow us to decompose the total angular momentum (TAM) into spin and orbital angular momentum (OAM) contributions.

As a result of the Poincar\'e symmetry,
the matrix elements of the local gauge-invariant and 
asymmetric EMT current~\cite{Leader:2013jra}
can be parametrized for a spin-1/2 target with mass $M$~\cite{Ji:1996ek,Bakker:2004ib}
in terms of five EMT FFs
as follows:
\begin{widetext}
\begin{align}
    \mel{p^{\prime},s^{\prime}}{\hat{T}_{a}^{\mu\nu}\left(0\right)}{p,s} 
& = \bar{u}\left(p^{\prime},s^{\prime}\right)
    \Bigg[
    \frac{P^{\{\mu}\gamma^{\nu\}}}{2}  
    A_{a}(Q^2)
  + \frac{\Delta^{\mu}\Delta^{\nu}-g^{\mu\nu}\Delta^{2}}{4M}  
    D_{a}(Q^2)
  + M g^{\mu\nu} 
    \bar{C}_{a}(Q^2)\cr
&   \hspace{3cm}
  + \frac{iP^{\{\mu}\sigma^{\nu\}\rho}\Delta_{\rho}}{4M} B_{a}(Q^2)
  - \frac{iP^{[\mu}\sigma^{\nu]\rho}\Delta_{\rho}}{2M}  
    S_{a}(Q^2)
    \Bigg]
    u\left(p,s\right), 
\label{Parametrization}
\end{align}
\end{widetext}
where the subscript $a$ stands for a specific type of constituent,
namely quark or gluon.
For simplicity, we use the notations 
$a^{\{\mu}b^{\nu\}}=a^\mu b^\nu+a^\nu b^\mu$ and $a^{[\mu}b^{\nu]}=a^\mu b^\nu-a^\nu b^\mu$.
The four-momentum eigenstates are normalized as 
$\braket{p^{\prime},s^{\prime}}{p,s}=
2P^{0}\left(2\pi\right)^{3}\delta^{(3)}
\left(\bm{p}^{\prime}-\bm{p}\right)\delta_{s^{\prime}s}$
with the canonical spin polarizations $s^{\prime},s$.
$P=(p'+p)/2$ and $\Delta=p'-p$ denote the average momentum and momentum transfer, respectively. Since $\Delta$ is a spacelike four-vector, we use for convenience the Lorentz-invariant variable $Q^2=-\Delta^2$. 

The spin FF for quarks~$S_{q}$ is related to 
the axial-vector FF via the QCD equation of motion~\cite{Bakker:2004ib,Leader:2013jra,Lorce:2017wkb},
while it vanishes for gluons
\begin{align}
  S_{q}(Q^2)
= \frac{1}{2}
  G_{A}^{q}(Q^2),
  \qquad
  S_{G}(Q^2)
= 0.
\end{align}
The latter follows from the impossibility of writing down a local and gauge-invariant gluon spin operator~\cite{Leader:2013jra}. 
Consequently, the gluon TAM appears in a pure orbital form, which can be regarded as an ``effective'' one. 
The total EMT FFs are obtained by summing over the quark and gluon contributions 
\begin{align}
    \mathcal F(Q^{2})
& = \mathcal F_{q}(Q^{2})
  +\mathcal F_{G}(Q^{2}), \cr
  \mathcal  F_{q}(Q^{2})
& = \sum_{f=u,d,\cdots}
  \mathcal  F_{f}(Q^{2}),
\end{align}
where $\mathcal F=A,B,D,\bar{C},S$.
Since the symmetrized total EMT is not renormalized~\cite{Collins:1976yq}, the total FFs $A$, $B$, $D$, 
and $\bar{C}$ are scheme and scale independent.
The conservation of total linear and angular momenta imply in addition the following constraints~\cite{Leader:2013jra,Teryaev:1999su,Lowdon:2017idv,Cotogno:2019xcl} 
\begin{align}
    A
    (0)
  = 1,  
    \hspace{.8cm}
    B
    (0)
  = 0,
    \hspace{.8cm}
    \bar{C}
    (Q^{2})
  = 0.
\label{symmetry}
\end{align}
In contrast, $D(0)$ is not constrained by Poincar\'e symmetry but is conjectured to be negative in hadrons based on stability arguments~\cite{Perevalova:2016dln,Lorce:2025oot}.
The vanishing of the total $\bar{C}$ FF
follows directly from the EMT conservation $\partial_{\mu}\hat{T}^{\mu\nu}=0$. This does not mean, however, that the individual quark and gluon contributions $\bar C_a(Q^2)$ are also zero.

In the present work, we will adopt a simple multipole Ansatz for the EMT FFs to illustrate our results 
\begin{align}
   \mathcal F_{a}(Q^{2})
& = \frac{\mathcal F_{a}(0)}{\left(1+Q^{2}/\Lambda_{\mathcal F_{a}}^{2}\right)^{n_{F}}}.
\end{align}
Such a form is supported, e.g., by calculations in the chiral quark-soliton model~\cite{Goeke:2007fp,Won:2023cyd,Won:2022cyy,Won:2023ial,Won:2023zmf}.
The parameters of the multipole Ansatz for $A$, $B$, $\bar{C}$, 
and $D$ FFs are given in Ref.~\cite{Lorce:2018egm}.
Regarding the quark spin FF, 
we will use the dipole mass 
$\Lambda_{S_{q}}=1.00$ GeV and the normalization
$S_{q}(0)=\frac{1}{2}g_{A}^{0}=0.17$.
Changing the value of this dipole mass strongly affects the sign of the OAM FF, $L_{a}(Q^{2})=\frac{1}{2}[A_{a}(Q^{2})+B_a(Q^2)]-S_{a}(Q^{2})$, the nucleon spin polarization, 
and the dipole structure of the longitudinal OAM distribution in the BF, see Appendix~\ref{app:1} for a detailed discussion.

\section{Distributions in elastic frame \label{sec.3}}
A class of Lorentz frames, called the elastic frame (EF), has been introduced in Ref.~\cite{Lorce:2017wkb} to study the relativistic spatial distributions of TAM, OAM, and spin inside the nucleon. 
The formalism developed into a relativistic phase-space picture~\cite{Lorce:2018zpf} and was then used to study the spatial distributions of the EMT~\cite{Lorce:2018egm}, the electromagnetic current~\cite{Lorce:2020onh,Kim:2021kum,Lorce:2022jyi,Chen:2022smg,Kim:2022wkc,Chen:2023dxp,Hong:2023tkv}, and the axial-vector current~\cite{Chen:2024oxx,Chen:2024ksq}, along with their dependence on the hadron momentum. 

Whereas the former EMT study~\cite{Lorce:2018egm} was limited to an unpolarized nucleon target, we include in the present work the contributions from the nucleon polarization. As demonstrated in the cases of the electromagnetic~\cite{Chen:2022smg} and axial-vector~\cite{Chen:2024oxx} currents, the latter is essential to a complete understanding of the dependence of the spatial distribution on the nucleon momentum.  
\newline

\subsection{General expressions}

Following the phase-space formalism, 
we can express the 2D spatial distributions of the EMT in the EF as
\begin{align}
&   \hspace{-0.5cm}
    T_{a}^{\mu\nu}
    \left(\bm{b}_{\perp},P_{z};s^{\prime},s\right)
  = \int dr_{z}\, 
    \expval{\hat{T}_{a}^{\mu\nu}\left(\bm{r}\right)}_{\bm{R},\bm{P}=P_{z}\hat{\bm{e}}_{z}}^{s^{\prime},s}\cr
&   \hspace{-0.5cm}
  = \int \frac{d^{2}\Delta_{\perp}}{\left(2\pi\right)^{2}}\,
    e^{-i\bm{\Delta}_{\perp}\cdot\bm{b}_{\perp}}
    \left.
    \frac{\mel{p^{\prime},s^{\prime}}{\hat{T}_{a}^{\mu\nu}\left(0\right)}{p,s}}{2P^{0}}
    \right|_{\mathrm{EF}},
\label{EMTdistribution_EF}
\end{align}
where the spatial axes are chosen so that the average momentum takes the form 
$\bm{P}=\left(\bm{0}_{\perp},P_{z}\right)$.
The impact parameter coordinates,
$\bm{b}_{\perp}=\left(\bm{r}_{\perp}-\bm{R}_{\perp},0\right)$,
represent the transverse position relative to the center of the system. 
The integral over the longitudinal component in position space enforces $\Delta_{z}=0$ and hence
the elastic condition, i.e.~zero energy transfer 
$\Delta^{0}=\bm{\Delta}\cdot\bm{P}/P^{0}=0$.
As a result, the 2D distributions in the EF  
remain time-independent. Given that $P^{2}=M^{2}\left(1+\tau\right)$ with $\tau=Q^{2}/4M^{2}$,
the average energy in the EF is $P^{0}=p^{\prime0}=p^{0}=
\sqrt{P_{z}^{2}+M^{2}\left(1+\tau\right)}$.
In the limit $P_{z}\to0$, the EF coincides with the transverse BF.
Conversely, as $P_{z}\to\infty$, the EF approaches the IMF. 

If we denote by $\Lambda$ the boost from the BF momentum $p_\mathrm{BF}$ to some momentum $p$, the EMT matrix elements transform according to~\cite{Jacob:1959at,Durand:1962zza} 
\begin{align}
&   \hspace{-0.5cm} 
    \mel{p^{\prime},s^{\prime}}{\hat{T}_{a}^{\mu\nu}\left(0\right)}{p,s}  \cr
& = \sum_{s_{\mathrm{BF}}^{\prime},s_{\mathrm{BF}}}
    D_{s_{\mathrm{BF}}s}^{(j)}
    \left(p_{\mathrm{BF}},\Lambda\right)
    D_{s_{\mathrm{BF}}^{\prime}s^{\prime}}^{*(j)}    
    \left(p_{\mathrm{BF}}^{\prime},\Lambda\right) \cr
&   \hspace{0.5cm}  \times    
    \Lambda_{\phantom{\mu}\alpha}^{\mu}
    \Lambda_{\phantom{\nu}\beta}^{\nu}
    \mel{p_{\mathrm{BF}}^{\prime},s_{\mathrm{BF}}^{\prime}}
    {\hat{T}_{a}^{\alpha\beta}\left(0\right)}
    {p_{\mathrm{BF}},s_{\mathrm{BF}}},
\label{LT}
\end{align}
where $D^{(j)}$ is the Wigner rotation matrix for spin-$j$ particles.
For spin-1/2 particles, the latter can be written as follows:
\begin{align}
    D_{s^{\prime}s}^{(1/2)}
    \left(p,\Lambda\right)
  = \cos{\frac{\theta}{2}} \delta_{s^{\prime}s}
  + i \sin{\frac{\theta}{2}}
    \frac{\left(\bm{p}\times\bm{\sigma}_{s^{\prime}s}\right)_{z}}{\left|\bm{p}_{\perp}\right|}.
\end{align}
By comparing the expressions on the left-hand side of Eq.~\eqref{LT},
obtained directly in the EF using 
Dirac bilinears~\cite{Lorce:2017isp}, with the generic Lorentz 
transformation on the right-hand side from the transverse BF 
($P_{z}=0$ and $\Delta_{z}=0$),  
we determine the Wigner rotation angle 
\begin{align}
    \cos{\theta}
& = \frac{P^{0}+M\left(1+\tau\right)}{\left(P^{0}+M\right)\sqrt{1+\tau}}, \cr
    \sin{\theta}
& = - \frac{\sqrt{\tau}P_{z}}{\left(P^{0}+M\right)\sqrt{1+\tau}},
\label{WR}
\end{align}
and the Lorentz boost parameters
\begin{align}
    \gamma
= \frac{P^{0}}{\sqrt{P^{2}}},
    \qquad
    \beta
  = \frac{P_{z}}{P^{0}}.
\label{WR2}
\end{align}
These coincide precisely with the results obtained 
in Ref.~\cite{Chen:2022smg} in the case of the electromagnetic current. Note that both the Wigner rotation angle and the Lorentz boost parameters depend on the momentum transfer.


Since the total four-momentum operator is defined as $\hat P^\mu=\int d^3r\,\hat{T}^{0\mu}\left(\bm{r}\right)$, we find that
\begin{align}
    \int d^{2}b_{\perp}\,
    T^{\mu\nu}
    \left(\bm{b}_{\perp},P_{z};s^{\prime},s\right)=M \frac{u^{\mu}u^{\nu}}{u^{0}}
    \delta_{s^{\prime}s}
\label{integralEMT}
\end{align}
with $u^\mu=\gamma_{P}\left(1,\bm{0}_{\perp},\beta_{P}\right)$ the four-velocity of the system.
The Lorentz boost parameters are given by 
\begin{align}
    \gamma_{P}
 = \frac{E_{P}}{M}, \qquad
    \beta_{P}
  = \frac{P_{z}}{E_{P}}
\end{align}
with $E_{P}= \sqrt{P_{z}^{2}+M^{2}}$, and coincide with the forward limit $\Delta\to0$ of the expressions in Eq.~\eqref{WR2}. It is clear from Eq.~\eqref{integralEMT} that the spatial distributions $T^{00}$, $T^{03}$, $T^{30}$, and $T^{33}$ diverge in the IMF. Since our goal is to discuss the distortions induced by a Lorentz boost along the $z$-direction, we factor out the trivial Lorentz factors and define EMT distributions with $P_z$-independent normalizations. Namely, we consider similar to Ref.~\cite{Lorce:2018egm}
\begin{align}
    t_{a}^{\mu\nu}
&:= \gamma_{P}
    T_{a}^{\mu\nu}
  = \begingroup
    \renewcommand{\arraystretch}{2} 
    \setlength{\tabcolsep}{15pt} 
    \begin{pmatrix}
    \gamma_{P}^{2}\rho_{a}            & \gamma_{P}\mathcal{P}_{a}^{i} & \gamma_{P}^{2}\mathcal{P}_{a}^{z}\\
    \gamma_{P}\mathcal{I}_{a}^{i}     & \Pi_{a}^{ij}          & \gamma_{P}\Pi_{a}^{iz} \\
    \gamma_{P}^{2}\mathcal{I}_{a}^{z} & \gamma_{P}\Pi_{a}^{zi}        & \gamma_{P}^{2}\Pi_{a}^{zz} 
    \end{pmatrix}
    \endgroup,
\label{defEMTdis}
\end{align}
where the indices $i,j=x,y$ denote transverse components.

In this paper, we focus on the ``longitudinal'' components $T^{00}$, $T^{03}$, $T^{30}$, and $T^{33}$. The other components of the EMT do not mix with these under a longitudinal Lorentz boost and will be discussed in a future work~\cite{Won:2024ll}. We find that the EF amplitudes are given by
\begin{widetext}
\begin{subequations}\label{EFampl}
\begin{align}
    \left.\frac{\mel{p^{\prime},s^{\prime}}{\hat{T}_{a}^{00}\left(0\right)}{p,s}}{2P^{0}}\right|_{\mathrm{EF}}
& = \gamma
    \left[
    \cos{\theta}
    \delta_{s^{\prime}s}
  + \sin{\theta}    
    \frac{\left(\bm{\sigma}_{s^{\prime}s}\times i\bm{\Delta}_{\perp}\right)_{z}}{2M\sqrt{\tau}}
    \right]
    M
    \left[
    E_{a}(Q^2)
  + \beta^{2}
    F_{a}(Q^2)\right]  \cr
& + \gamma
    \beta
    \left[
    \cos{\theta} 
    \frac{\left(\bm{\sigma}_{s^{\prime}s}\times i\bm{\Delta}_{\perp}\right)_{z}}{2M\sqrt{\tau}}
  - \sin{\theta} \delta_{s^{\prime}s}
    \right]
    2M\sqrt{\tau}
    J_{a}(Q^2),
\label{T00}
\end{align}
\begin{align}
    \left.\frac{\mel{p^{\prime},s^{\prime}}{\hat{T}_{a}^{03}\left(0\right)}{p,s}}{2P^{0}}\right|_{\mathrm{EF}}
& = \gamma
    \beta
    \left[
    \cos{\theta}
    \delta_{s^{\prime}s}
  + \sin{\theta}    
    \frac{\left(\bm{\sigma}_{s^{\prime}s}\times i\bm{\Delta}_{\perp}\right)_{z}}{2M\sqrt{\tau}}
    \right]
    M
    \left[E_{a}(Q^2)
  + F_{a}(Q^2)\right]  \cr
& + \gamma
    \left[
    \cos{\theta} 
    \frac{\left(\bm{\sigma}_{s^{\prime}s}\times i\bm{\Delta}_{\perp}\right)_{z}}{2M\sqrt{\tau}}
  - \sin{\theta} \delta_{s^{\prime}s}
    \right] 
    M\sqrt{\tau}
    \left[
    \left(1+\beta^{2}\right)
    J_{a}(Q^2)
  - \left(1-\beta^{2}\right)
    S_{a}(Q^2)
    \right],
\label{T03}
\end{align}
\begin{align}
    \left.\frac{\mel{p^{\prime},s^{\prime}}{\hat{T}_{a}^{33}\left(0\right)}{p,s}}{2P^{0}}\right|_{\mathrm{EF}}
& = \gamma
    \left[
    \cos{\theta}
    \delta_{s^{\prime}s}
  + \sin{\theta}    
    \frac{\left(\bm{\sigma}_{s^{\prime}s}\times i\bm{\Delta}_{\perp}\right)_{z}}{2M\sqrt{\tau}}
    \right]
    M
    \left[\beta^{2}
    E_{a}(Q^2) 
  + F_{a}(Q^2)\right]\cr
& + \gamma
    \beta
    \left[
    \cos{\theta} 
    \frac{\left(\bm{\sigma}_{s^{\prime}s}\times i\bm{\Delta}_{\perp}\right)_{z}}{2M\sqrt{\tau}}
  - \sin{\theta} 
    \delta_{s^{\prime}s}
    \right]
    2M\sqrt{\tau}
    J_{a}(Q^2),
\label{T33}
\end{align}
\label{EFmatrixelement}
\end{subequations}
\end{widetext}
where we defined energy, TAM and force FFs as 
\begin{align}
    E_{a}
& = A_{a}
  - \tau
    B_{a}
  + \bar{C}_{a}
  + \tau 
    D_{a},  \cr
    \hspace{-1.0cm}
    J_{a}
& = \frac{A_{a}
    + B_{a}}{2},  \cr    
    \hspace{-1.0cm}
    F_{a}
& = - \bar{C}_{a}
    - \tau 
      D_{a}.
\label{EJF}
\end{align}
These are the natural combinations that appear in the BF. 
The matrix elements of $\hat{T}_{a}^{30}$ are obtained through the substitution $S_{a}\mapsto-S_{a}$ 
in the matrix elements of $\hat{T}_{a}^{03}$. Since $\beta\to 1$ when $P_z\to \infty$, all the amplitudes in Eq.~\eqref{EFampl} become equal in the IMF.

The normalized distributions of energy $\rho_{a}$, longitudinal momentum $\mathcal{P}_{a}^{z}$, longitudinal energy flux $\mathcal{I}_{a}^{z}$, and axial momentum flux (i.e.~longitudinal flux of longitudinal momentum) $\Pi^{zz}_a$ can therefore be expressed as 
\begin{widetext}
\begin{align}
    g_{a}
    (\bm{b}_{\perp},P_{z};s^{\prime},s)  
  = \int \frac{d^{2}\Delta_{\perp}}{\left(2\pi\right)^{2}}\,
    e^{-i\bm{\Delta}_{\perp}\cdot\bm{b}_{\perp}} 
    \left[
    \delta_{s^{\prime}s}
    \,\widetilde{g}_{a}^{U}
    (Q^{2},P_{z})
  + \frac{\left(\bm{\sigma}_{s^{\prime}s}\times i\bm{\Delta}_{\perp}\right)_{z}}{2M}
    \,\widetilde{g}_{a}^{T}
    (Q^{2},P_{z})\right]
\label{spatial_distribution}
\end{align}  
\end{widetext}
with $g=\rho,\mathcal{P}^{z},\mathcal{I}^{z},\Pi^{zz}$. The superscripts $U$ and $T$ denote the unpolarized and transversely polarized contributions, respectively. Note that these EMT distributions are not affected by the nucleon longitudinal polarization.

The unpolarized EF distributions are defined as~\cite{Lorce:2018egm}
\begin{align}
    g_{a}
    (b,P_{z})
&:= \frac{1}{2} 
    \sum_{s^{\prime},s}
    g_{a}
    (\bm{b}_{\perp},P_{z};s^{\prime},s)
    \delta_{s^{\prime}s}\cr
& = \int \frac{d^{2}\Delta_{\perp}}{\left(2\pi\right)^{2}}\,
    e^{-i\bm{\Delta}_{\perp}\cdot\bm{b}_{\perp}} 
    \widetilde{g}_{a}^{U}
    (Q^{2},P_{z})
\end{align}
with $b=\left|\bm{b}_{\perp}\right|$. The axial symmetry is broken by the nucleon transverse polarization. In the following, we will consider the case of a nucleon polarized in the $x$-direction. Its spin state is then given by $\ket{s_{x}=\pm \frac{1}{2}}=\left(\ket{s=+\frac{1}{2}}\pm\ket{s=-\frac{1}{2}}\right)/\sqrt{2}$ and 
the corresponding EF distributions are defined as 
\begin{align}
&   \hspace{-0.7cm}
    g_{a}^{P}
    (\bm{b}_{\perp},P_{z})
 := g_{a}
    (\bm{b}_{\perp},P_{z};s'_{x}=+\tfrac{1}{2},s_{x}=+\tfrac{1}{2}),  \cr
  &\quad\qquad= \int \frac{d^{2}\Delta_{\perp}}{\left(2\pi\right)^{2}}\,
    e^{-i\bm{\Delta}_{\perp}\cdot\bm{b}_{\perp}} \cr
    &\qquad\quad
    \times\left[
    \widetilde{g}_{a}^{U}
    (Q^{2},P_{z})
  + \frac{i\Delta_{y}}{2M}\,
    \widetilde{g}_{a}^{T}
    (Q^{2},P_{z})\right].
\end{align}

\subsection{Energy distribution  \label{subsec:1}}

The spin-independent and spin-dependent amplitudes defining the EF energy distribution are given by
\begin{widetext}
\begin{subequations}\label{energy_amplitudes}
\begin{align}
    \widetilde{\rho}_{a}^{U}
    (Q^{2},P_{z})
& = \frac{1}{\gamma_P}\frac{P^{0}\left[P^{0}+M\left(1+\tau\right)\right]}{\left(P^{0}+M\right)\left(1+\tau\right)}  
    \left\{
    E_{a}(Q^2)
  + \frac{2\tau P_{z}^{2}}{P^{0}\left[P^{0}+M\left(1+\tau\right)\right]}
    J_{a}(Q^2)
  + \left(\frac{P_{z}}{P^{0}}\right)^{2}
    F_{a}(Q^2)
    \right\},
\label{rho_U}
\end{align}  
\begin{align}
    \widetilde{\rho}_{a}^{T}
    (Q^{2},P_{z})
& =\frac{1}{\gamma_P} \frac{P^{0}P_{z}}{\left(P^{0}+M\right)\left(1+\tau\right)}
    \left\{
  - E_{a}(Q^2)
  + \frac{2\left[P^{0}+M\left(1+\tau\right)\right]}{P^{0}}
    J_{a}(Q^2)
  - \left(\frac{P_{z}}{P^{0}}\right)^{2}
    F_{a}(Q^2)
    \right\},
\label{rho_T}
\end{align}
\label{rho}
\end{subequations}
\end{widetext}
and the total EF energy distribution is normalized as
\begin{align}
    \int d^{2} b_{\perp}\,\rho
    (\bm{b}_{\perp},P_{z};s^{\prime},s)
& =\widetilde g^U(0,P_{z})\delta_{s^{\prime}s}\cr
&= M\delta_{s^{\prime}s}
\end{align}
since $E(0)=1$ and $F(0)=0$.

\begin{figure*}[htbp!]
  \centering
  \textbf{EF energy distributions for an unpolarized nucleon}

  \vspace{0.2cm} 
  
  \begin{minipage}{0.42\textwidth}
    \centering
    Quark
  \end{minipage}
  \hspace{-0.05\textwidth}
  \begin{minipage}{0.42\textwidth}
    \centering
    Gluon
  \end{minipage}

  \begin{minipage}{0.40\textwidth}
    \centering
    \includegraphics[width=\textwidth]{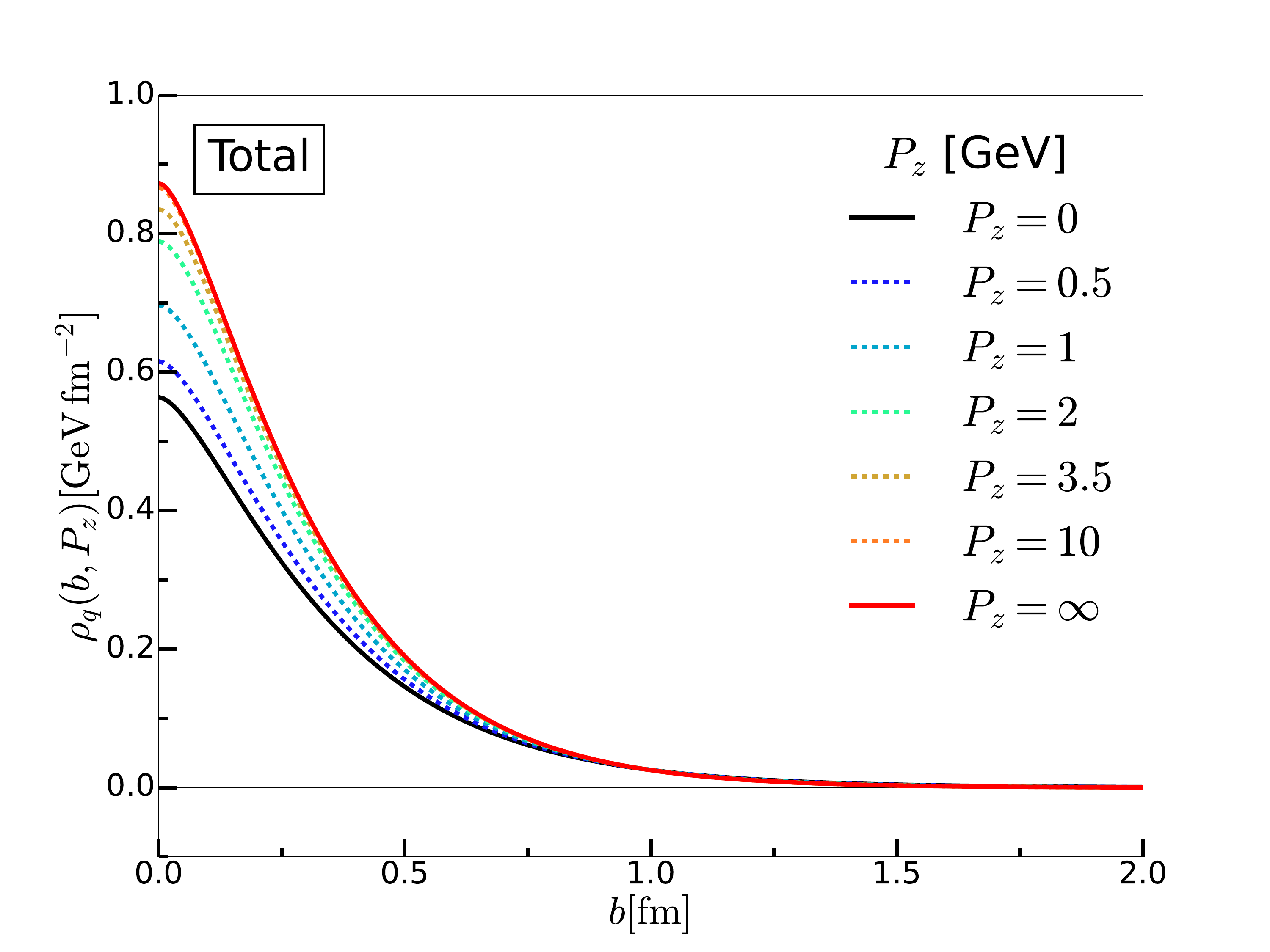}
  \end{minipage}
  \hspace{-0.6cm}
  \begin{minipage}{0.40\textwidth}
    \centering
    \includegraphics[width=\textwidth]{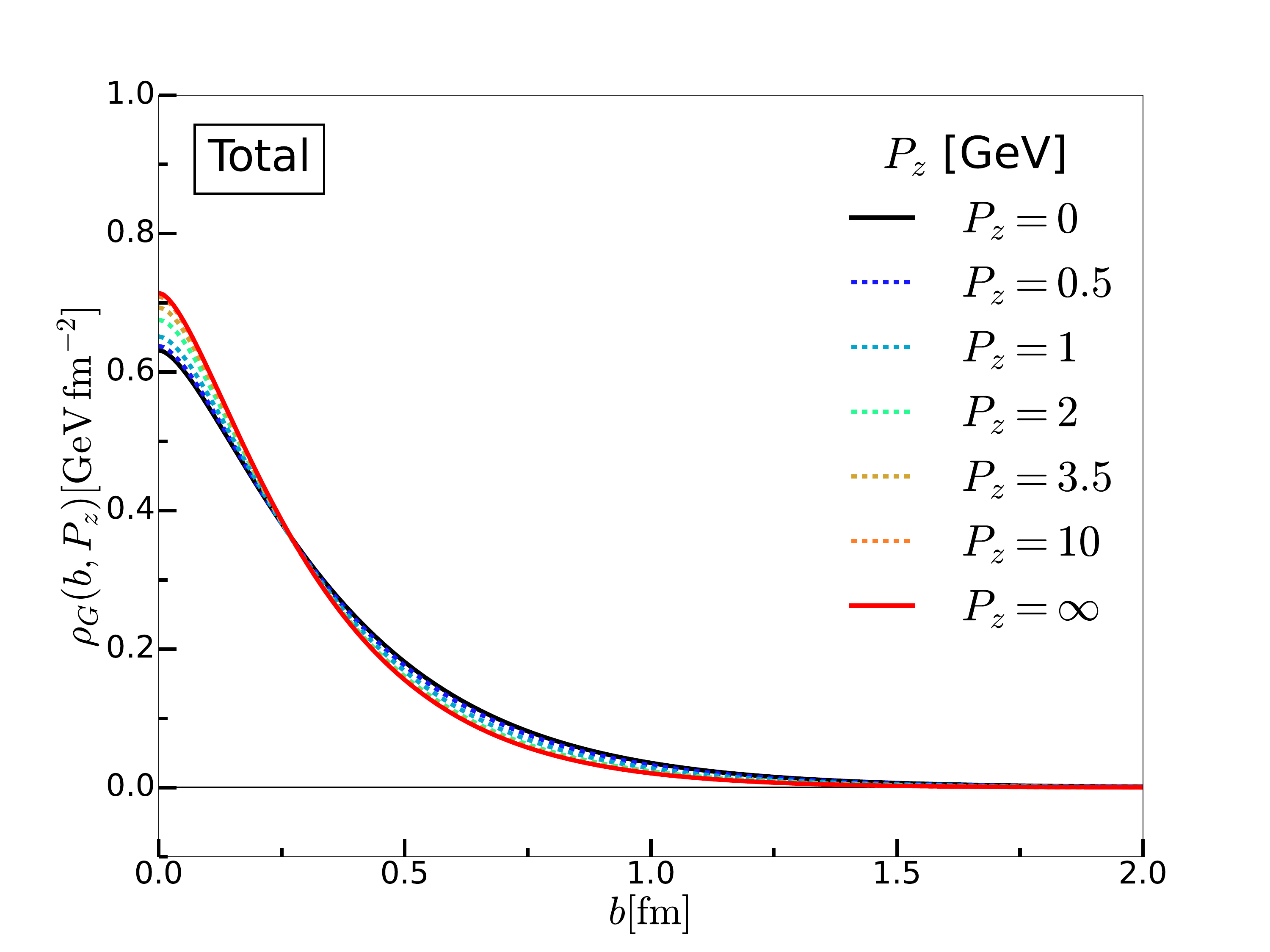}
  \end{minipage}
  
  \vspace{-0.4cm} 
  
  \begin{minipage}{0.40\textwidth}
    \centering
    \includegraphics[width=\textwidth]{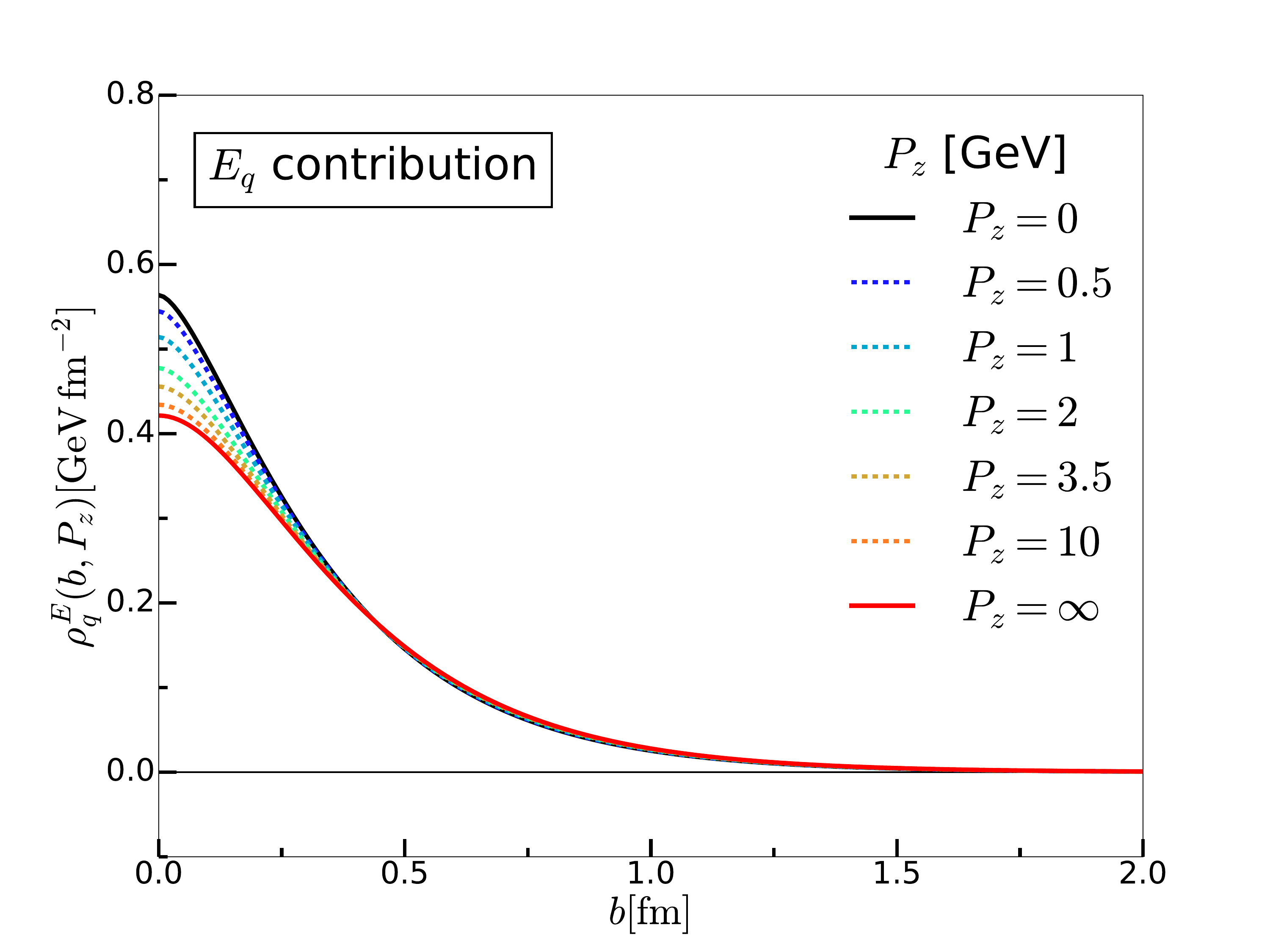}
  \end{minipage}
  \hspace{-0.6cm}
  \begin{minipage}{0.40\textwidth}
    \centering
    \includegraphics[width=\textwidth]{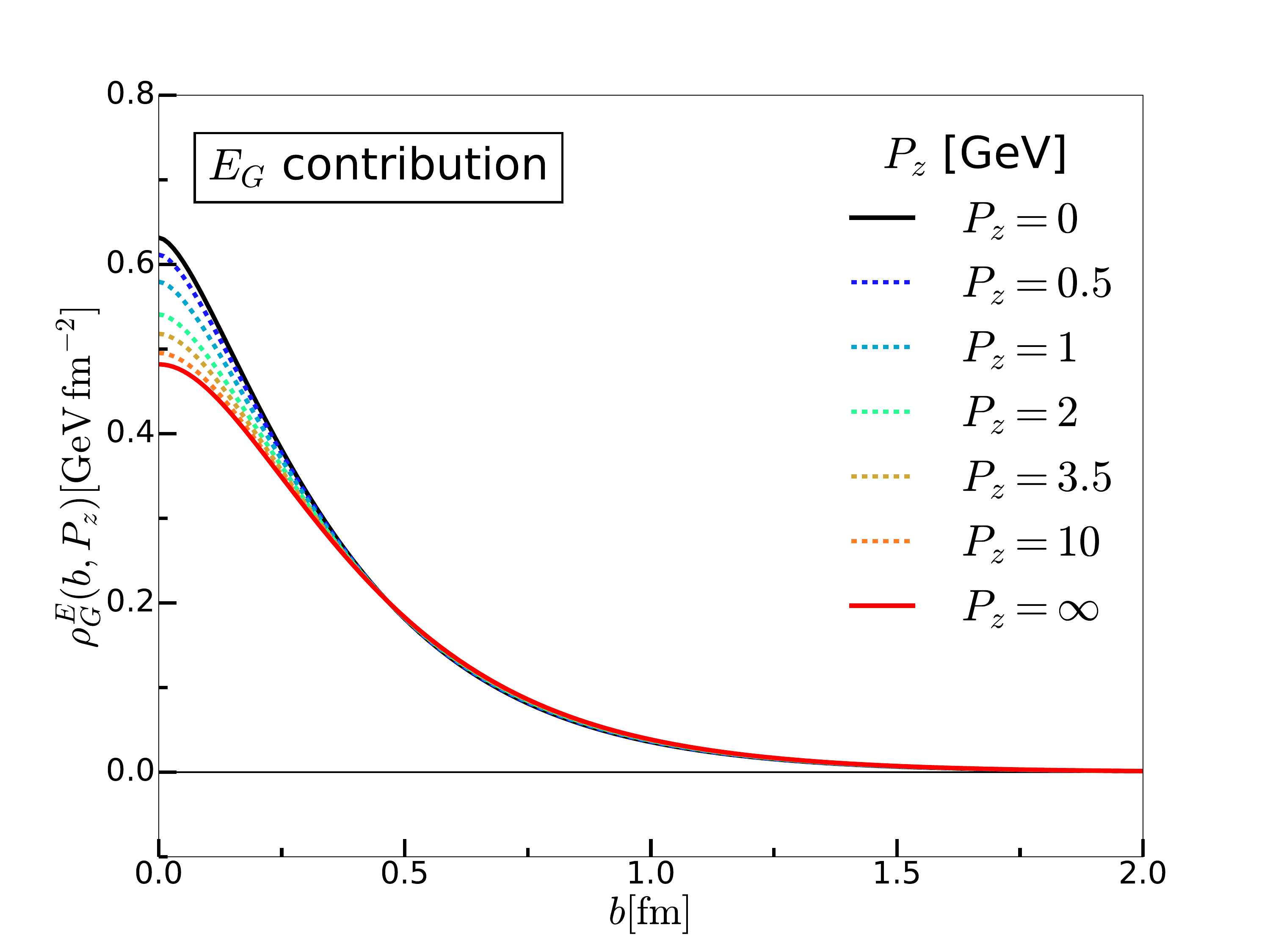}
  \end{minipage}
  
  \vspace{-0.4cm} 
  
  \begin{minipage}{0.40\textwidth}
    \centering
    \includegraphics[width=\textwidth]{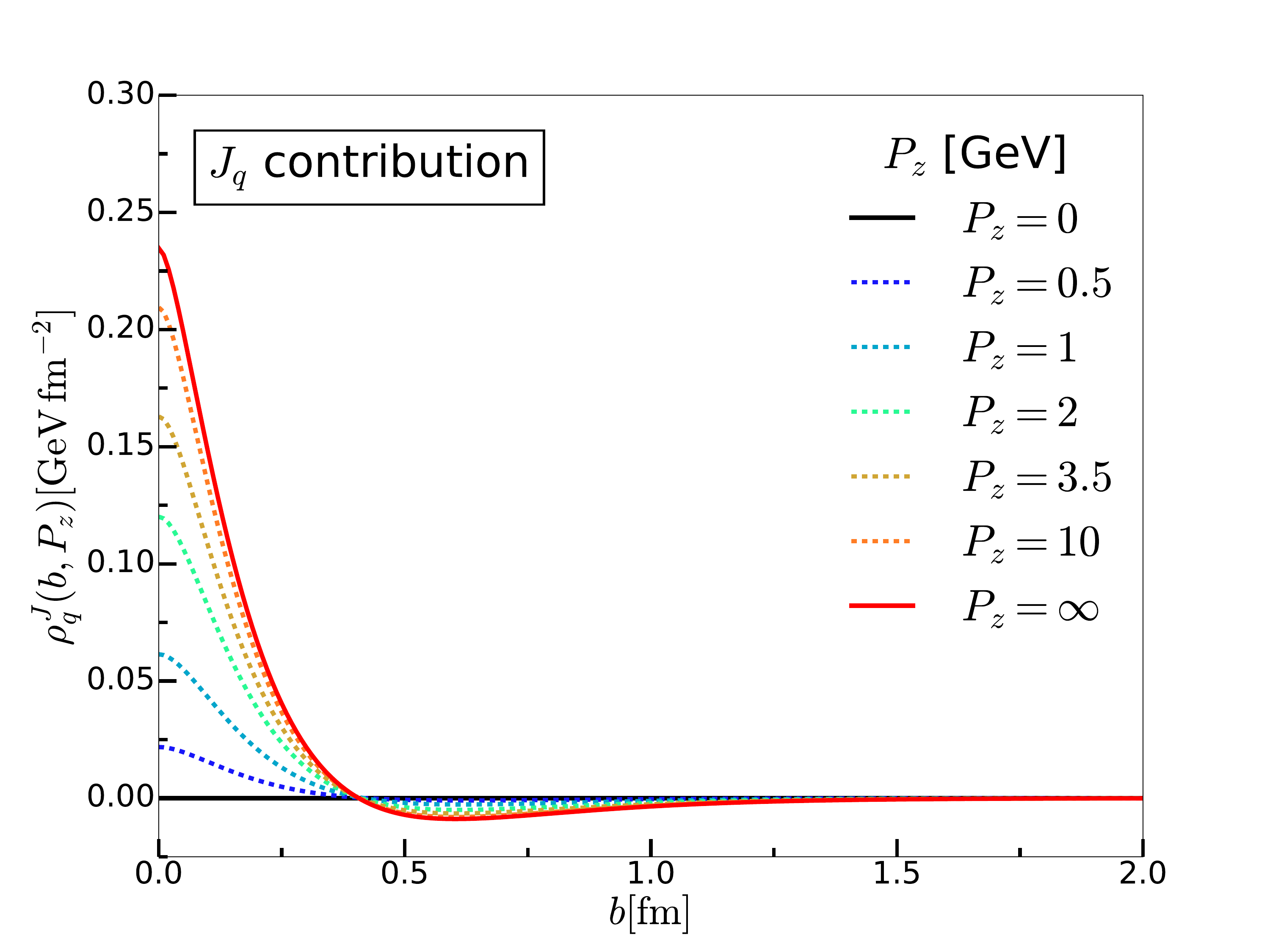}
  \end{minipage}
  \hspace{-0.6cm}
  \begin{minipage}{0.40\textwidth}
    \centering
    \includegraphics[width=\textwidth]{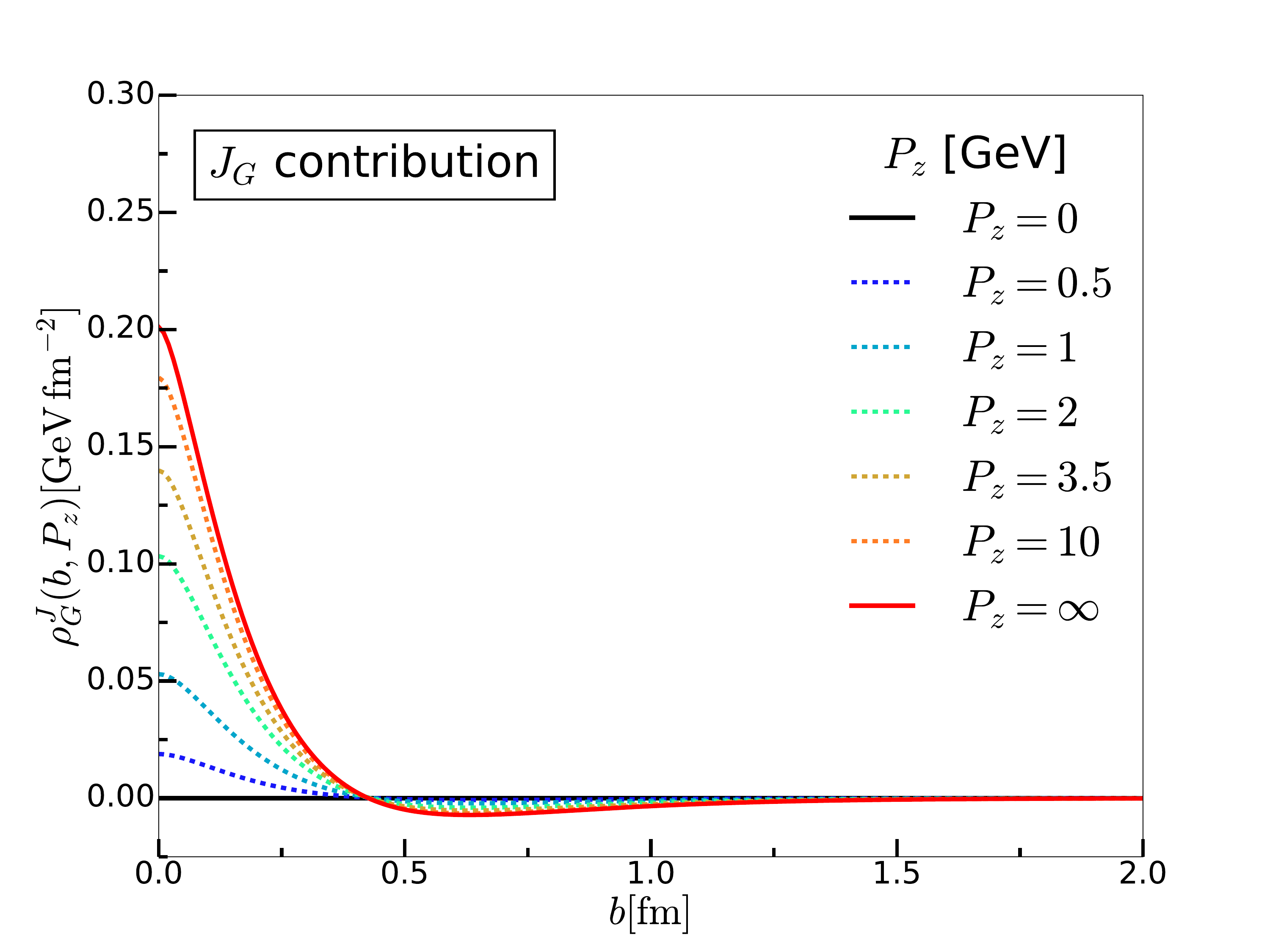}
  \end{minipage}
  
  \vspace{-0.4cm} 
  
  \begin{minipage}{0.40\textwidth}
    \centering
    \includegraphics[width=\textwidth]{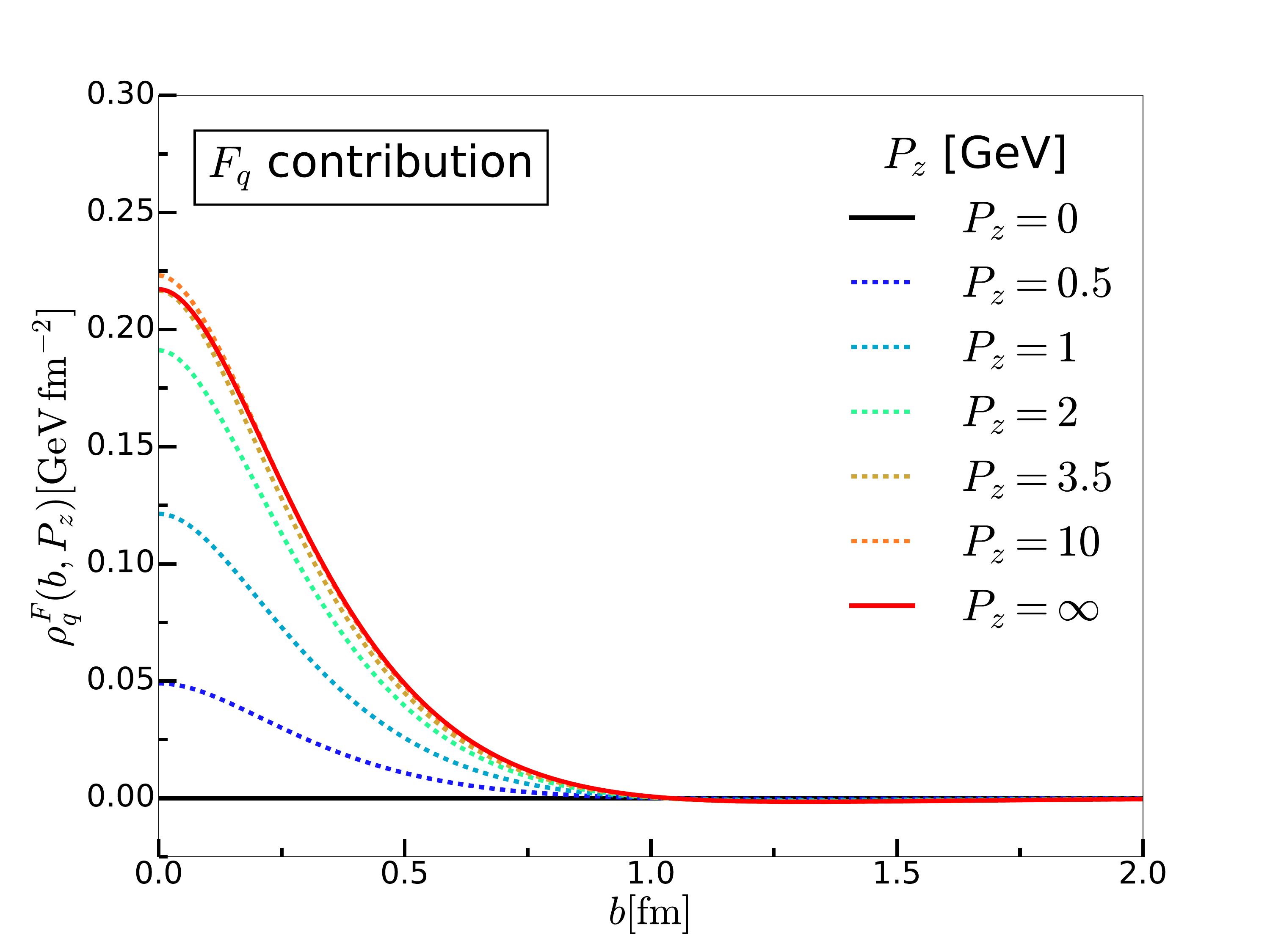}
  \end{minipage}
  \hspace{-0.6cm}
  \begin{minipage}{0.40\textwidth}
    \centering
    \includegraphics[width=\textwidth]{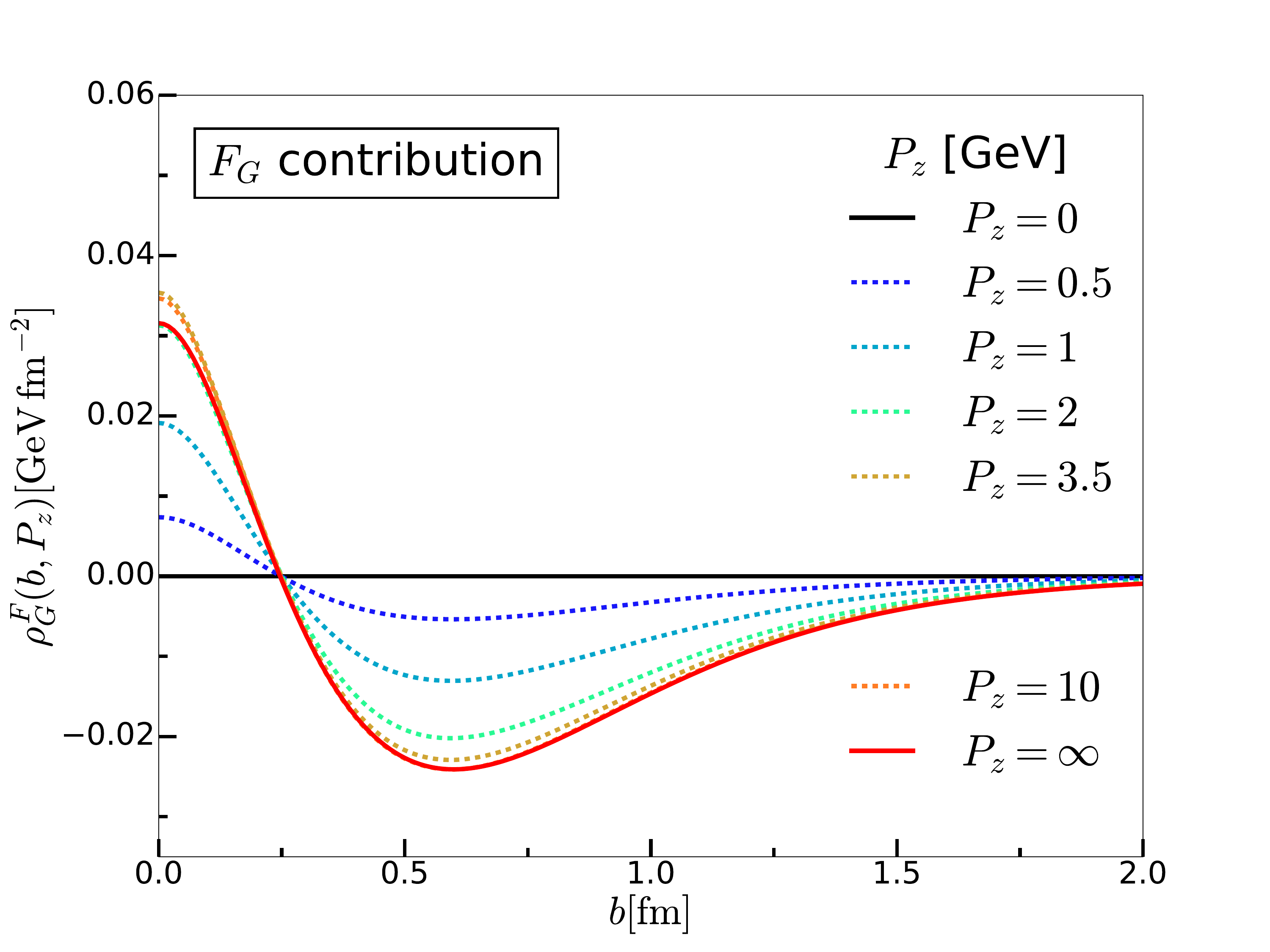}
  \end{minipage}
  
  \caption{Radial EF energy distributions in an unpolarized nucleon for different values of the nucleon momentum. The total energy distribution is normalized to $M$. The left (right) column corresponds to the quark (gluon) contribution. The first row depicts the sum of the three contributions shown in the other rows. Based on the simple multipole model of Ref.~\cite{Lorce:2018egm} for the EMT FFs.}
  \label{fig:1}
\end{figure*}
\begin{figure*}[htbp!]
  \centering
  \textbf{EF energy distributions for a transversely polarized nucleon}

  \vspace{0.2cm} 
  
  \begin{minipage}{0.42\textwidth}
    \centering
    Quark
  \end{minipage}
  \hspace{-0.05\textwidth}
  \begin{minipage}{0.42\textwidth}
    \centering
    Gluon
  \end{minipage}

  \begin{minipage}{0.40\textwidth}
    \centering
    \includegraphics[width=\textwidth]{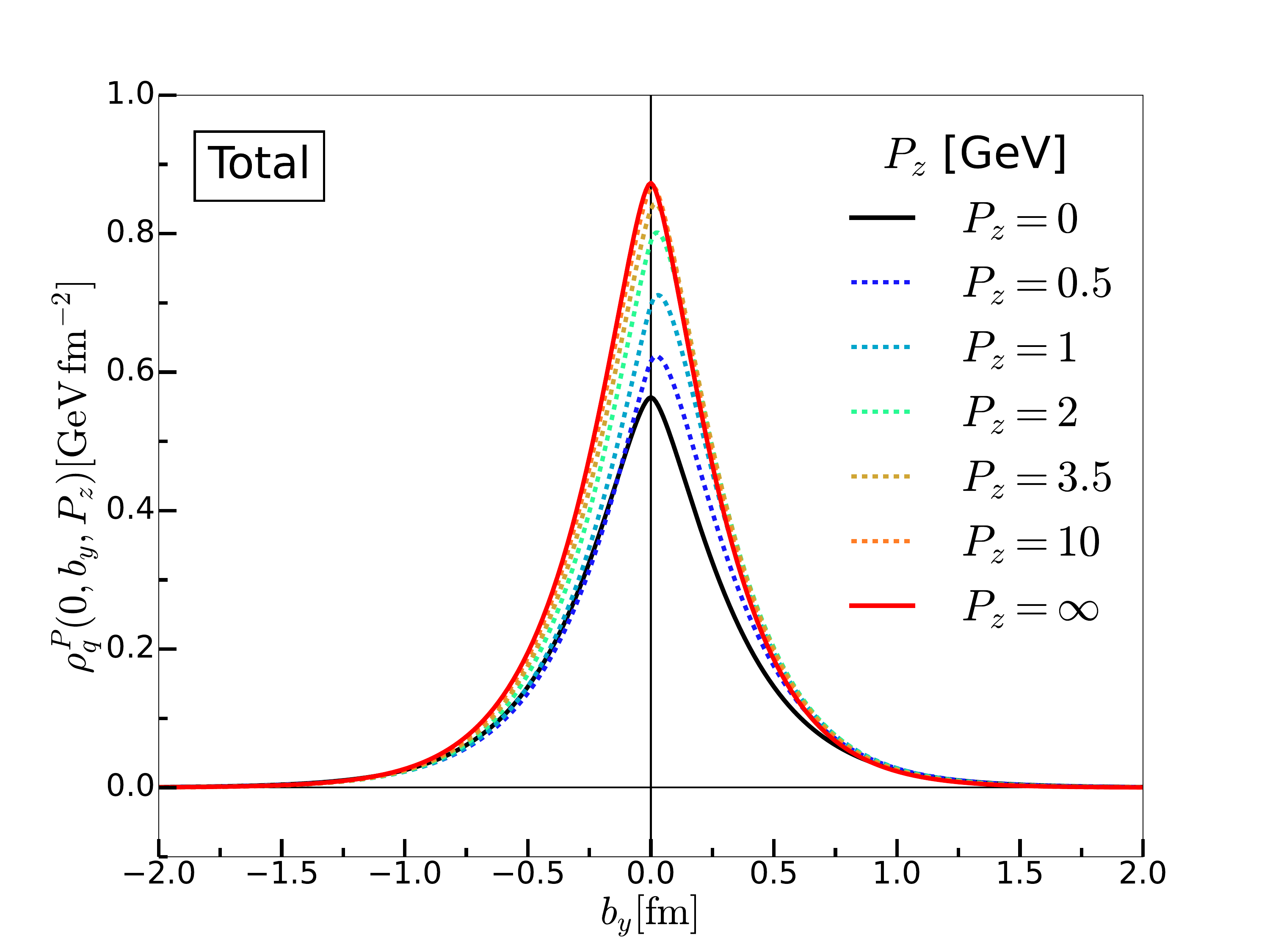}
  \end{minipage}
  \hspace{-0.6cm}
  \begin{minipage}{0.40\textwidth}
    \centering
    \includegraphics[width=\textwidth]{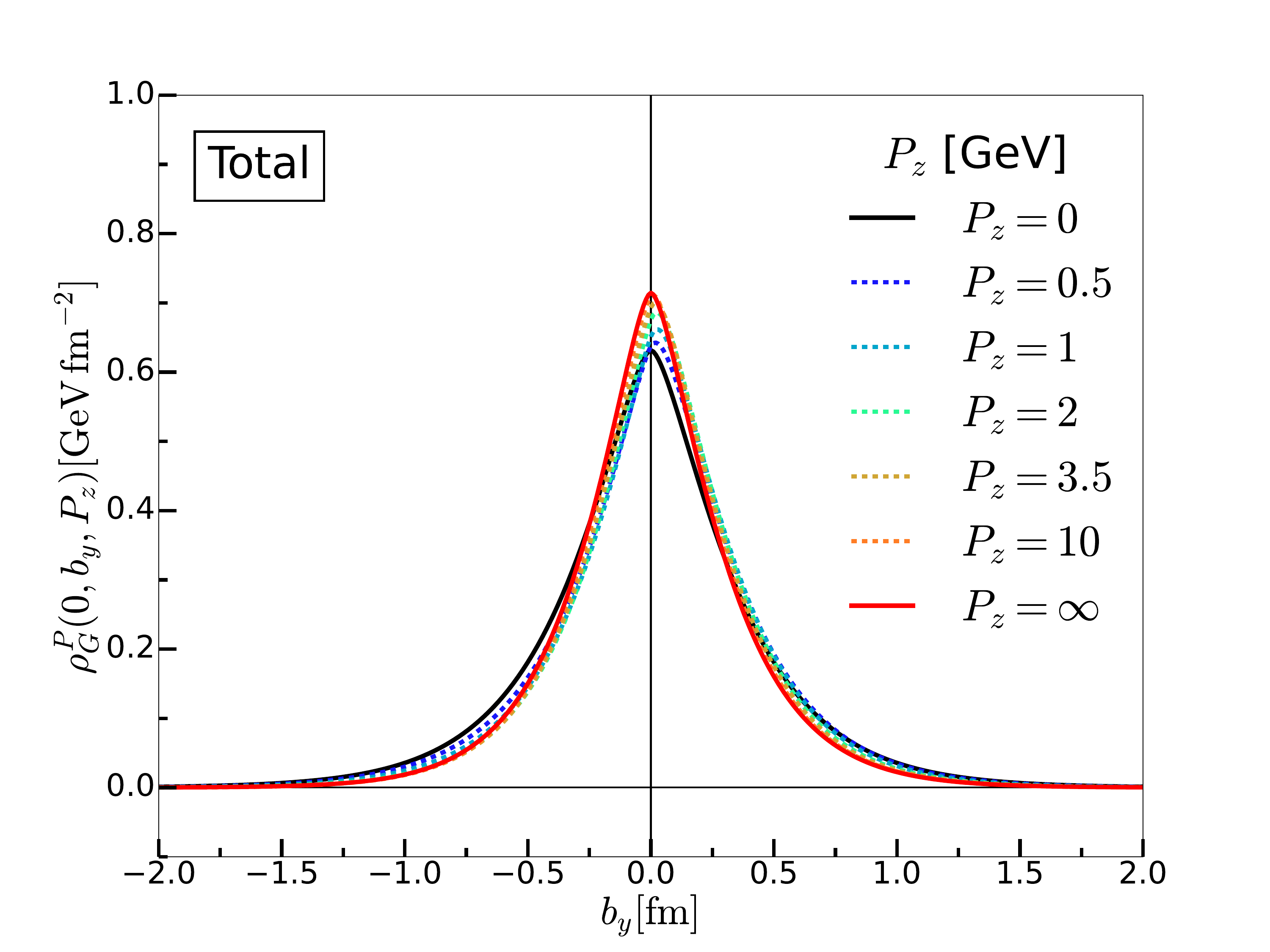}
  \end{minipage}
  
  \vspace{-0.4cm} 
  
  \begin{minipage}{0.40\textwidth}
    \centering
    \includegraphics[width=\textwidth]{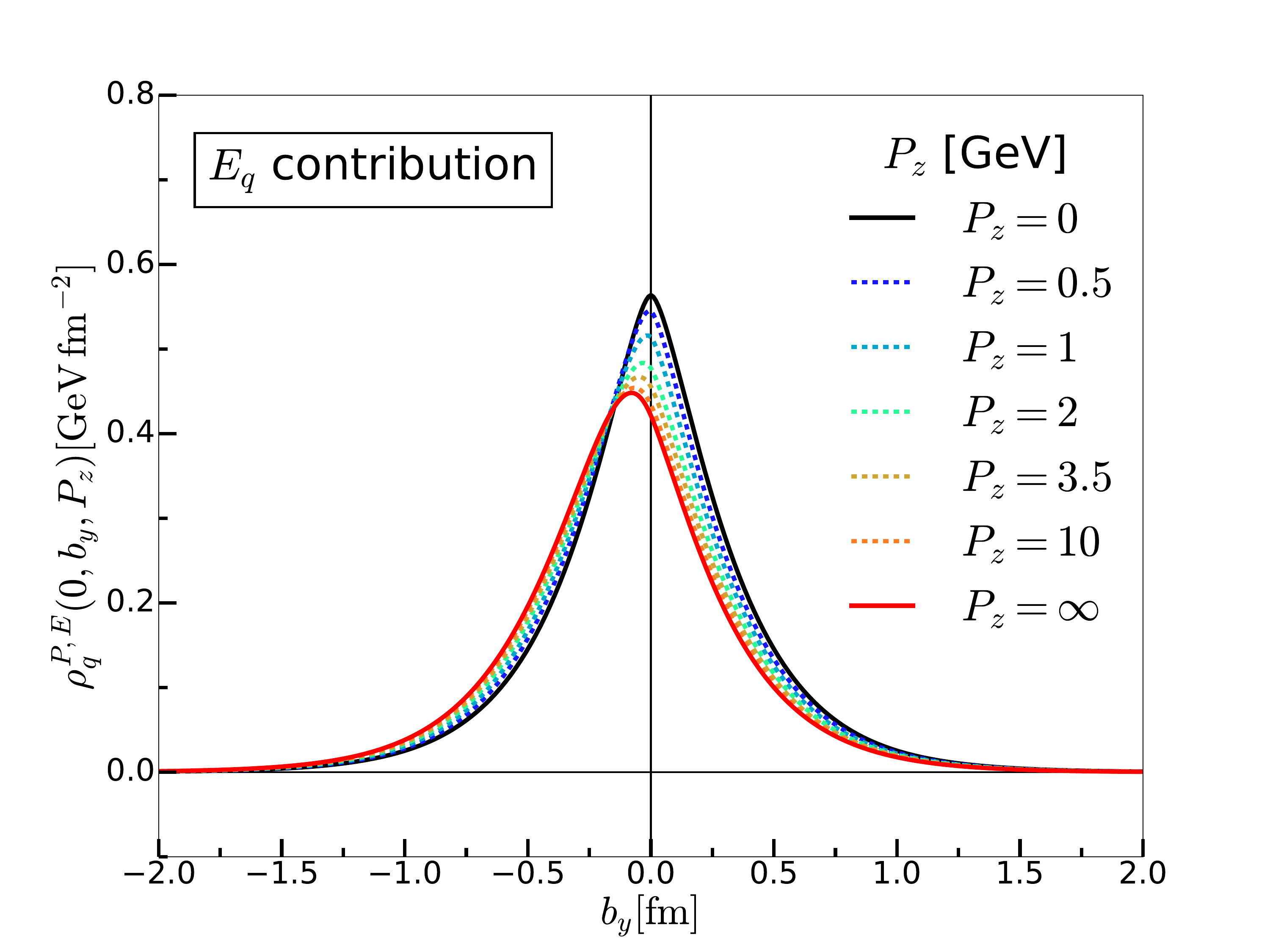}
  \end{minipage}
  \hspace{-0.6cm}
  \begin{minipage}{0.40\textwidth}
    \centering
    \includegraphics[width=\textwidth]{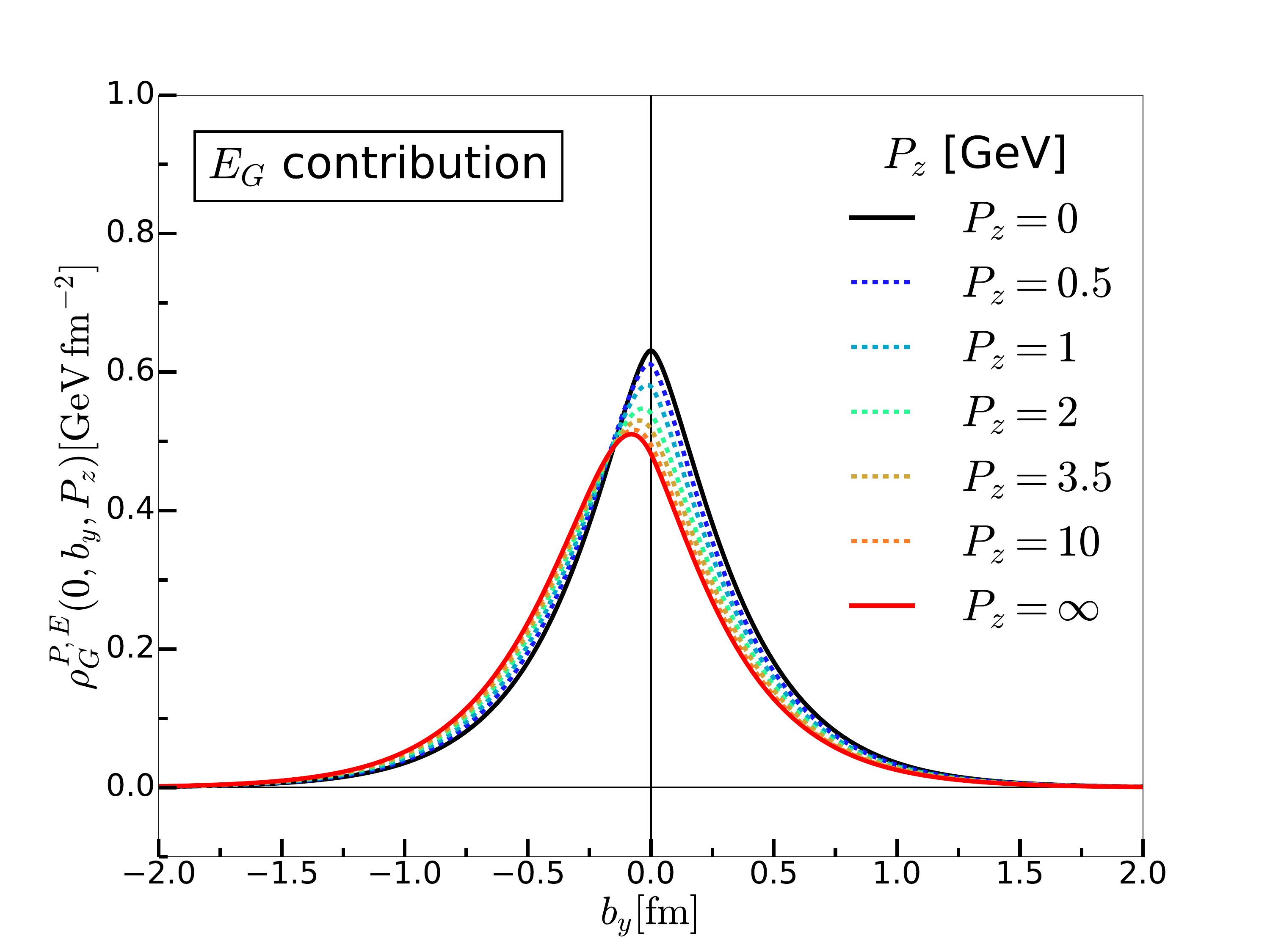}
  \end{minipage}
  
  \vspace{-0.4cm} 
  
  \begin{minipage}{0.40\textwidth}
    \centering
    \includegraphics[width=\textwidth]{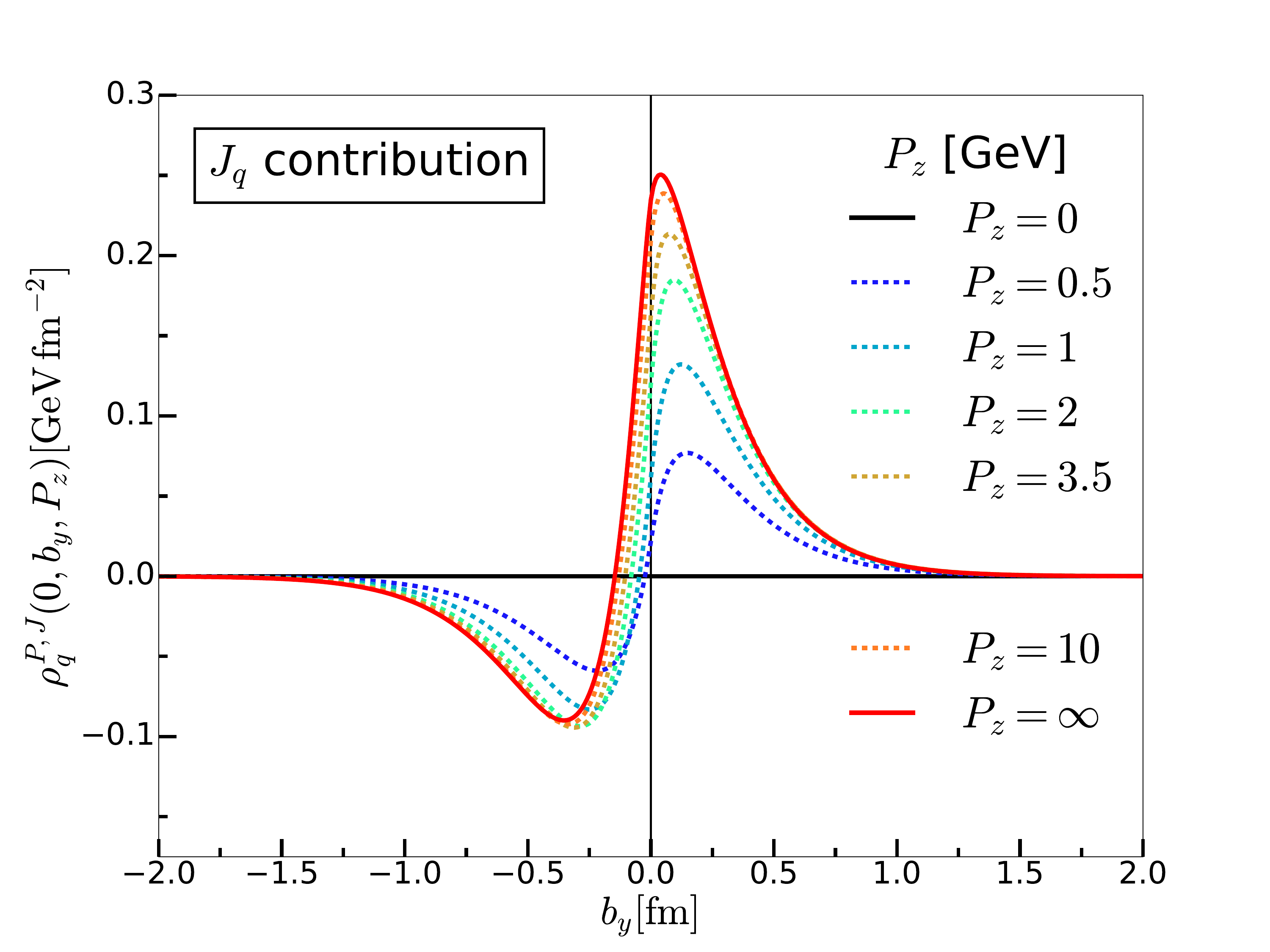}
  \end{minipage}
  \hspace{-0.6cm}
  \begin{minipage}{0.40\textwidth}
    \centering
    \includegraphics[width=\textwidth]{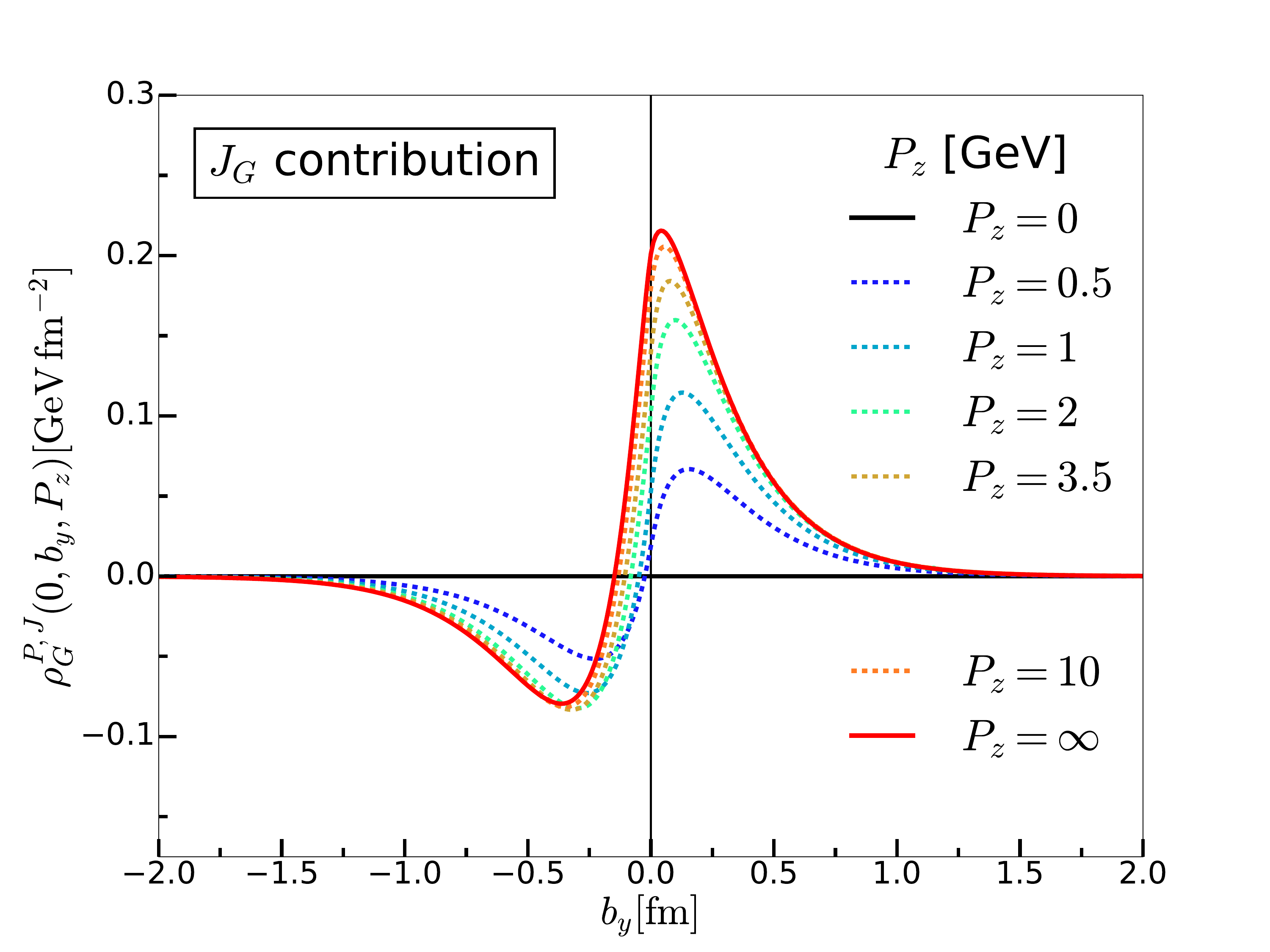}
  \end{minipage}
  
  \vspace{-0.4cm} 
  
  \begin{minipage}{0.40\textwidth}
    \centering
    \includegraphics[width=\textwidth]{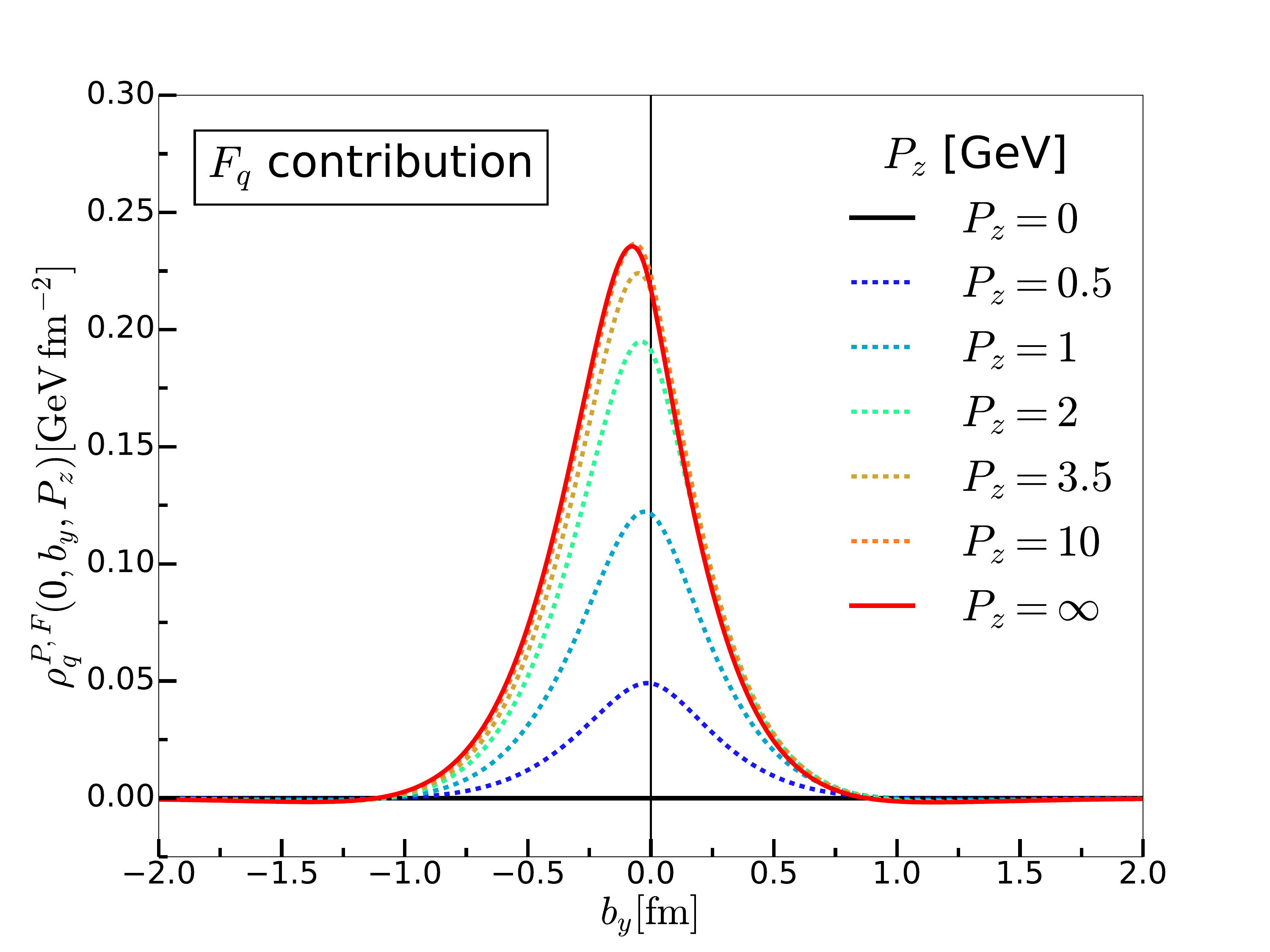}
  \end{minipage}
  \hspace{-0.6cm}
  \begin{minipage}{0.40\textwidth}
    \centering
    \includegraphics[width=\textwidth]{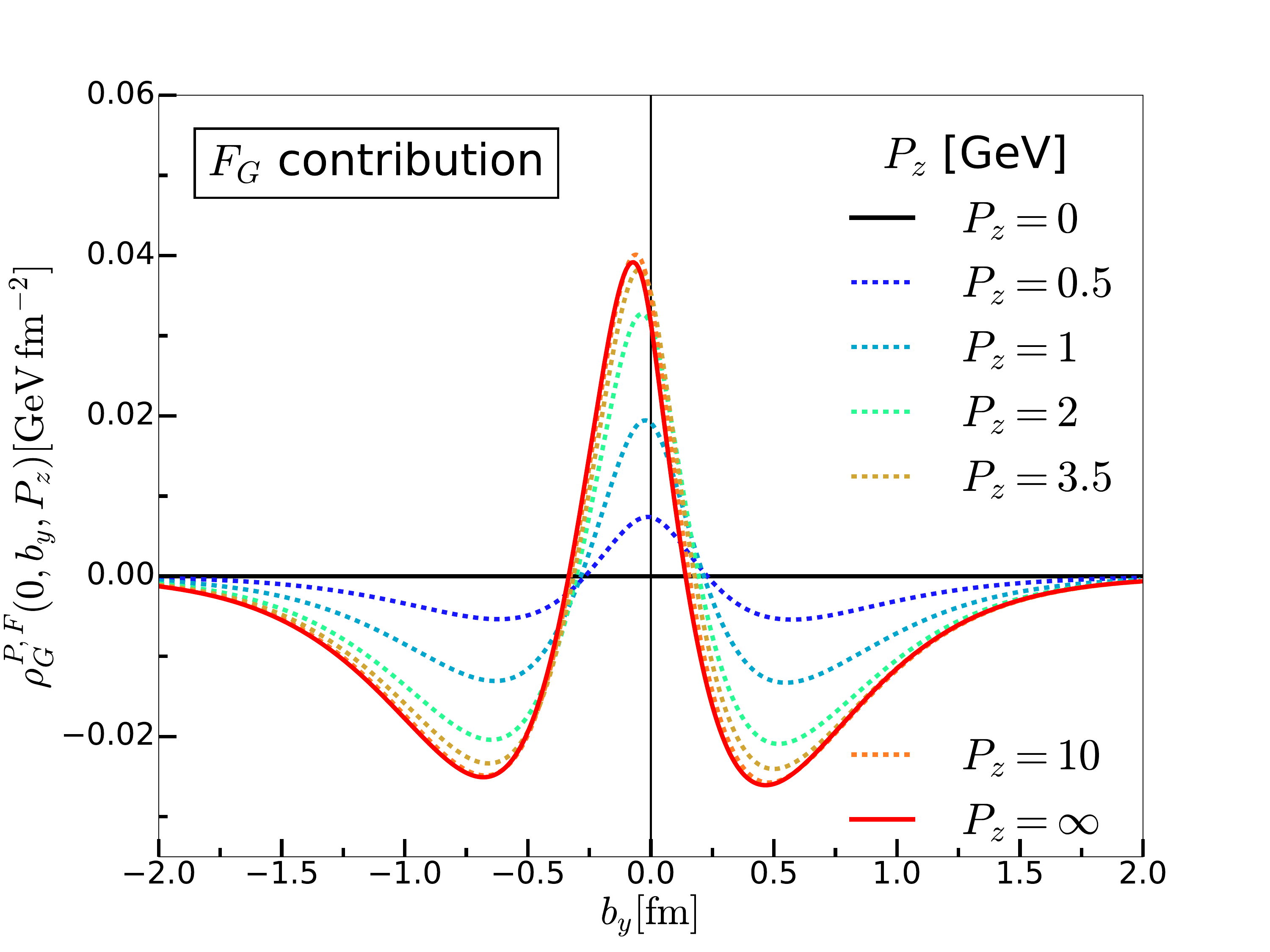}
  \end{minipage}
  \caption{EF energy distributions at $b_x=0$ in a nucleon polarized along the $x$-axis for different values of the nucleon momentum. The total energy distribution is normalized to $M$. The left (right) column corresponds to the quark (gluon) contribution. The first row depicts the sum of the three contributions shown in the other rows. Based on the simple multipole model of Ref.~\cite{Lorce:2018egm} for the EMT FFs.}
  \label{fig:2}
\end{figure*}

In the transverse BF, 
the energy amplitudes reduce to
\begin{align}
    \widetilde{\rho}_{a}^{U}
    (Q^{2},0)
& = M
    E_{a}(Q^2), \cr
    \widetilde{\rho}_{a}^{T}
    (Q^{2},0)
& = 0,
\end{align}
in agreement with the results in Refs.~\cite{Polyakov:2018zvc,Lorce:2018egm}. All the terms associated with explicit factors of $P_z$ in Eq.~\eqref{energy_amplitudes} can then be understood as contributions induced by the Lorentz boost from the BF to the EF. In the IMF, we find that 
\begin{align}
    \widetilde{\rho}_{a}^{U}
    (Q^{2},\infty)
& = M
    A_{a}
    (Q^2)  , \cr
    \widetilde{\rho}_{a}^{T}
    (Q^{2},\infty)
& = M
    B_{a}
    (Q^2).
\label{IMFenergy}
\end{align}

In Fig.~\ref{fig:1}, we show the quark and gluon contributions to the EF energy distribution 
in an unpolarized nucleon at various values of $P_{z}$.
Since these distributions are axially symmetric, only the radial profile is displayed. 
As $P_{z}$ increases, the contribution associated with $E_a$  decreases near the center while the induced contributions associated with $J_a$ and $F_a$ increase. 
We observe that the contribution associated with $F_q$ is much larger than that associated with $F_G$. This difference finds its origin in the fact that $\bar C_q(Q^2)=-\bar C_G(Q^2)$.

\begin{figure*}[htbp]
  \centering
  \textbf{Total EF energy distribution for a transversely polarized nucleon}

  \vspace{0.5cm} 

  \includegraphics[width=1.0\textwidth]{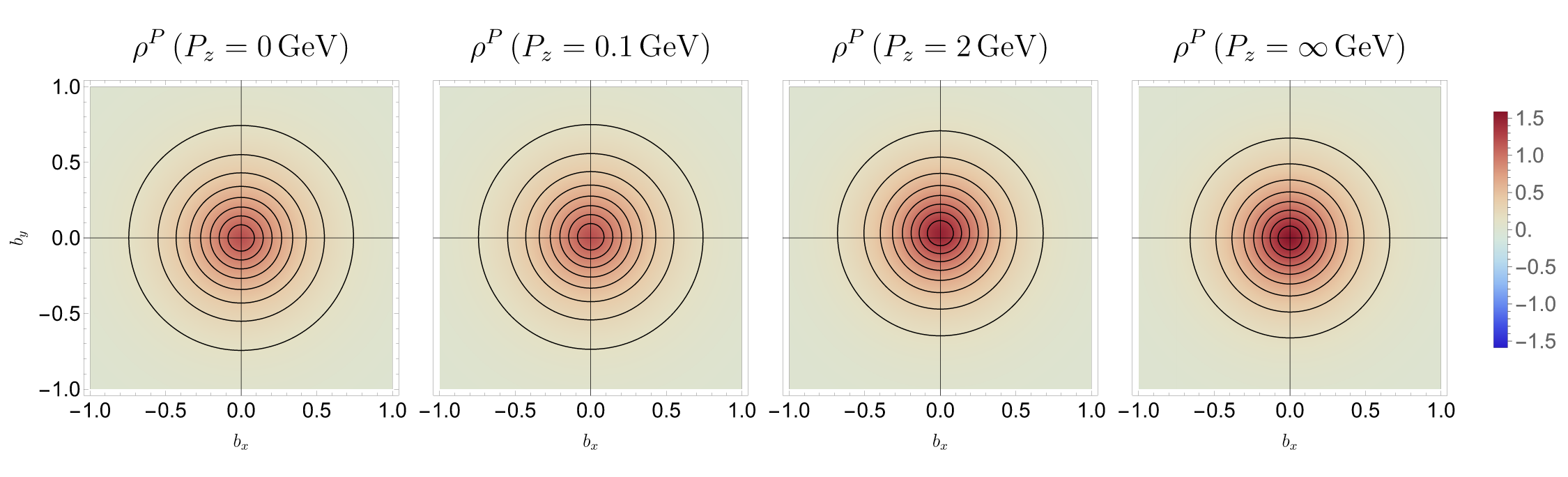}
  \hspace{0cm}

  \caption{Total EF energy distributions in the transverse plane of a nucleon polarized along the $x$-axis for different values of the nucleon momentum. The total energy distribution is normalized to $M$. Based on the simple multipole model of Ref.~\cite{Lorce:2018egm} for the EMT FFs.}

  \label{fig:3}
\end{figure*}

In Fig.~\ref{fig:2},
we illustrate the same distributions for the nucleon polarized along the $x$-direction. Since the axial symmetry is broken in this case, we show the profile at $b_x=0$. We find that the dipole shift induced by the contribution associated with $J_a$ goes 
in the opposite direction to those associated with $E_a$ and $F_a$. In the absence of Wigner rotation, only the former would play a role.
As shown in Fig.~\ref{fig:3}, 
a dipolar distortion along the $y$-axis appears in the total EF energy distribution for non-zero $P_z$. The fact that this dipolar distortion disappears in the IMF is an artifact of the naive multipole model of Ref.~\cite{Lorce:2018egm}, which assumed $B_q(Q^2)=-B_G(Q^2)$ due to the lack of information on the gluon GFFs. In a more realistic description, we expect a priori $B_q(Q^2)\neq-B_G(Q^2)$ and hence a dipolar distortion in the IMF similar to what is observed for the charge distribution~\cite{Kim:2021kum,Chen:2022smg}. 

The total EF energy dipole moment
\begin{align}
&\hspace{-.75cm}\int d^2b_\perp\,\bm b_\perp\,\rho(\bm b_\perp,P_z;s',s)=\frac{(\bm e_z\times\bm \sigma_{s's})_z}{2M}\widetilde\rho^T(0,P_z)\cr
& =(\bm e_z\times\bm \sigma_{s's})_z\,\frac{P_z}{E_P}\left[J(0)-\frac{E_P}{E_P+M}\frac{E(0)}{2}\right]
\end{align} 
has the same structure as the EF electric dipole moment~\cite{Chen:2023dxp}. The first contribution corresponds to the longitudinal boost of a rest-frame transverse TAM and has the expected form $\bm\beta_P\times\bm J$. The second contribution comes from a sideways shift of the center of
spin, defining the origin of our coordinate system, with respect to the relativistic center of
mass in a moving frame~\cite{Lorce:2018zpf,Lorce:2021gxs}. In the IMF, this dipole moment vanishes as a consequence of $B(0)=0$.

\subsection{Longitudinal momentum and longitudinal energy flux distributions  \label{subsec:2}}
The spin-independent and spin-dependent amplitudes defining the longitudinal momentum distribution in the EF are given by
\begin{widetext}
\begin{subequations}\label{momentum_amplitudes}
\begin{align}
    \widetilde{\mathcal{P}}_{a}^{z,U}
    (Q^{2},P_{z})
& = \frac{1}{\gamma_P} 
    \frac{P_{z}\left[P^{0}+M\left(1+\tau\right)\right]}{\left(P^{0}+M\right)\left(1+\tau\right)}  \cr
&   \quad  \times
    \left\{
    E_{a}(Q^2)
  + \frac{\tau P^{2}}{P^{0}\left[P^{0}+M\left(1+\tau\right)\right]}
    \left[\left(\frac{2P_{z}^{2}}{P^{2}}+1\right)
    J_{a}(Q^2)
  - S_{a}(Q^2)\right]   
  + F_{a}(Q^2)
    \right\},
\label{PzU}
\end{align}
\begin{align}
    \widetilde{\mathcal{P}}_{a}^{z,T}
    (Q^{2},P_{z})
& = \frac{1}{\gamma_P}
    \frac{P^2}{\left(P^{0}+M\right)\left(1+\tau\right)}\cr
&   \quad  \times
    \left\{
  - \frac{P_{z}^{2}}{P^2}
    E_{a}(Q^2) 
  + \frac{P^{0}+M\left(1+\tau\right)}{P^{0}}     
    \left[
    \left(\frac{2P_{z}^{2}}{P^{2}}+1\right)
    J_{a}(Q^2)
  - S_{a}(Q^2)
    \right]
  - \frac{P_{z}^{2}}{P^2}
    F_{a}(Q^2)
    \right\}.
\label{PzT}
\end{align}
\label{Pz}
\end{subequations}
\end{widetext}
For the longitudinal energy flux distribution in the EF, the spin-independent and 
spin-dependent amplitudes $\widetilde{\mathcal{I}}_{a}^{z,U}$ and $\widetilde{\mathcal{I}}_{a}^{z,T}$ are obtained 
by reversing the sign of the spin FF.
The total EF distribution of longitudinal momentum 
is normalized as
\begin{align}
    \int d^{2} b_{\perp}\;
        \mathcal{P}^{z}
    (\bm{b}_{\perp},P_{z};s^{\prime},s)
& = M \beta_{P}
    \delta_{s^{\prime}s},
\end{align}
and similarly for the total EF distribution of longitudinal energy flux $\mathcal{I}^z$.

\begin{figure*}[htb!]
  \centering
  \textbf{Total EF distributions of longitudinal momentum and longitudinal energy flux for a transversely polarized nucleon}
  \vspace{0.5cm} 

  \includegraphics[width=1.0\textwidth]{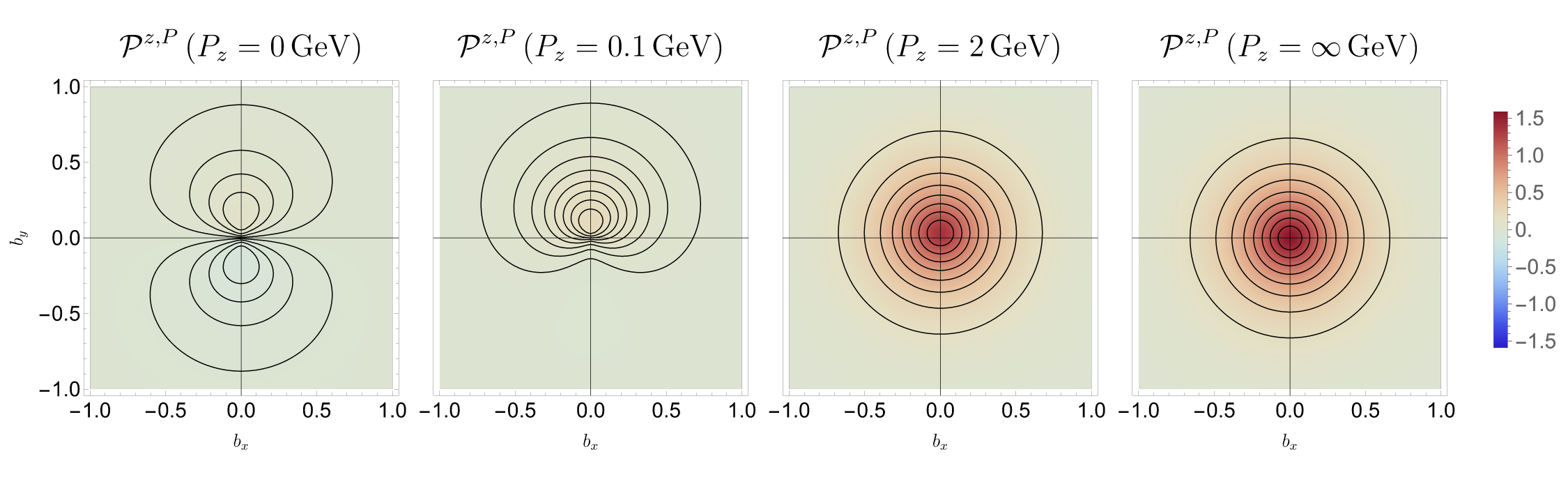}
  \hspace{0cm}

  \vspace{-0.5cm} 

  \includegraphics[width=1.0\textwidth]{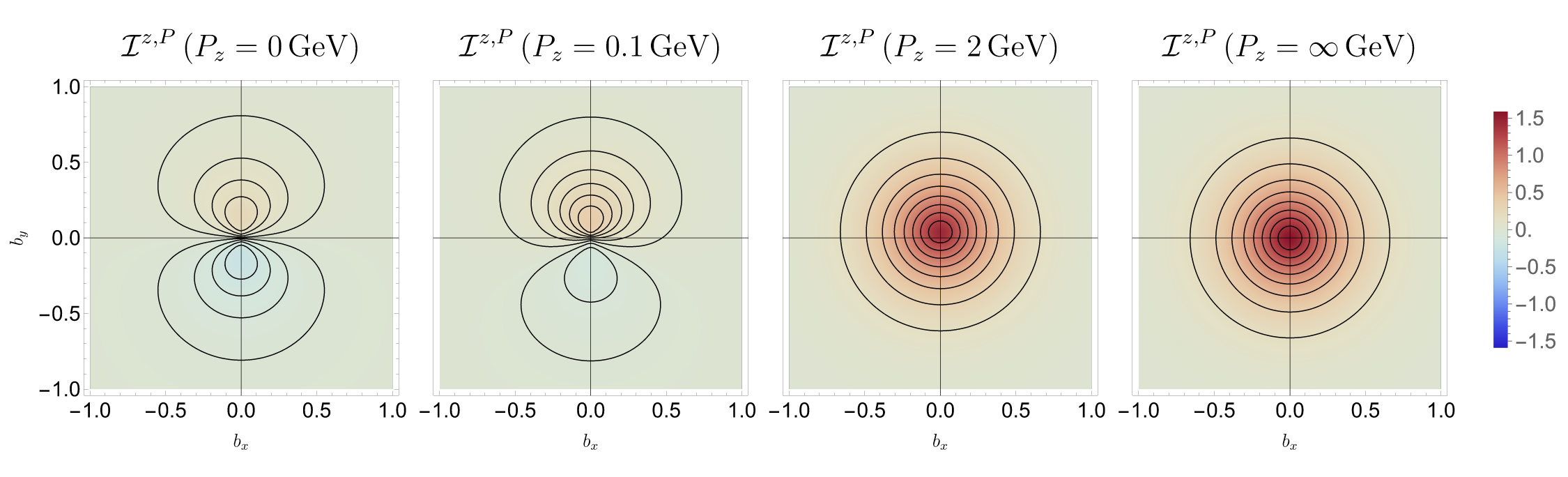}
  \hspace{0cm}

  \caption{Total EF frame distributions of longitudinal momentum and longitudinal energy flux in the transverse plane of a nucleon polarized along the $x$-axis for different values of the nucleon momentum. The total distributions are normalized to $M\beta_P$. Based on the simple multipole model of Ref.~\cite{Lorce:2018egm} for the EMT FFs.}

  \label{fig:6}
\end{figure*}

In the transverse BF, the longitudinal 
momentum and longitudinal energy flux amplitudes reduce to
\begin{align}
    \widetilde{\mathcal{P}}_{a}^{z,U}
    (Q^{2},0)
& = \widetilde{\mathcal{I}}_{a}^{z,U}
    (Q^{2},0)
  = 0, \cr
    \widetilde{\mathcal{P}}_{a}^{z,T}
    (Q^{2},0)
& = M
    \left[
    J_{a}
    (Q^2)
  - S_{a}
    (Q^2)
    \right],  \cr
    \widetilde{\mathcal{I}}_{a}^{z,T}
    (Q^{2},0)
& = M
    \left[
    J_{a}
    (Q^2)
  + S_{a}
    (Q^2)
    \right],
\label{Limit_P^z}
\end{align}
in agreement with the results in Refs.~\cite{Polyakov:2018zvc,Lorce:2018egm}. All the terms associated with explicit factors of $P_z$ in Eq.~\eqref{momentum_amplitudes} can then be understood as contributions induced by the Lorentz boost from the BF to the EF. In the IMF, we find that 
\begin{align}
    \widetilde{\mathcal{P}}^{z,U}_a
    (Q^{2},\infty)
& = \widetilde{\mathcal{I}}_{a}^{z,U}
    (Q^{2},\infty)
  = M
    A_{a}
    (Q^2)  , \cr
    \widetilde{\mathcal{P}}_{a}^{z,T}
    (Q^{2},\infty)
& = \widetilde{\mathcal{I}}_{a}^{z,T}
    (Q^{2},\infty)
  = M
    B_{a}
    (Q^2).
\label{IMFlongitudinal_momentum}
\end{align}

In Fig.~\ref{fig:6}, we compare the total EF distributions of longitudinal momentum and longitudinal energy flux in the transverse plane for non-zero $P_z$. The differences result directly from the fact that the contribution associated with $S_a$ enters with an opposite sign in these two distributions. In the IMF, however, $\mathcal P^z$ and $\mathcal I^z$ become equal because the spin contribution is kinematically suppressed at large $P_z$.

Contrary to the energy dipole moment, the dipole moment of the longitudinal momentum distribution does not vanish in the BF, but is given by
\begin{align}
\hspace{-0.6cm}\int d^2b_\perp\,\bm b_\perp\,\mathcal P^z(\bm b_\perp,0;s',s)&=(\bm e_z\times\bm \sigma_{s's})_z\,\frac{1}{2}L(0),
\end{align} 
where $L_{a}(Q^{2})=J_{a}(Q^{2})-S_{a}(Q^{2})$.
Since the 3D BF momentum distributions
\begin{align}
& 
    \mathcal P^\mu
    \left(\bm{r};s^{\prime},s\right)
  = \expval{\hat{T}^{0\mu}\left(\bm{r}\right)}_{\bm{0},\bm{0};s^{\prime},s}\cr
  & \quad 
  = \int \frac{d^{3}\Delta}{\left(2\pi\right)^{3}}\,
    e^{-i\bm{\Delta}\cdot\bm{r}}
    \left.
    \frac{\mel{p^{\prime},s^{\prime}}{\hat{T}^{0\mu}\left(0\right)}{p,s}}{2P^{0}}
    \right|_{\mathrm{BF}}
\end{align}
are axially symmetric about the polarization axis, we indeed expect for a nucleon polarized along the $x$-direction that
\begin{equation}
    \int d^3r\, y\mathcal P^z(\bm r)=\frac{1}{2}\int d^3r \left[y\mathcal P^z(\bm r)-z\mathcal P^y(\bm r)\right].
\end{equation}
In other words, the dipole moment in the BF momentum distribution is a direct manifestation of the internal OAM (recall that, in this work, the gluon spin contribution is effectively represented in an orbital form).

Similarly, the dipole moment of the longitudinal energy flux distribution is given in the BF by 
\begin{align}
&\hspace{-0.6cm}\int d^2b_\perp\,\bm b_\perp\,\mathcal I^z(\bm b_\perp,0;s',s)\cr
&=(\bm e_z\times\bm \sigma_{s's})_z\,\frac{1}{2}\left[L(0)+2S(0)\right],
\end{align} 
In particular, even in the absence of internal OAM, there is still a circulation of energy due to the intrinsic spin.

\subsection{Axial momentum flux distribution \label{subsec:3}}
\begin{figure*}[htb!]
  \centering
  \textbf{Total EF distribution of axial momentum flux for a transversely polarized nucleon}

  \vspace{0.5cm} 

  \includegraphics[width=1.0\textwidth]{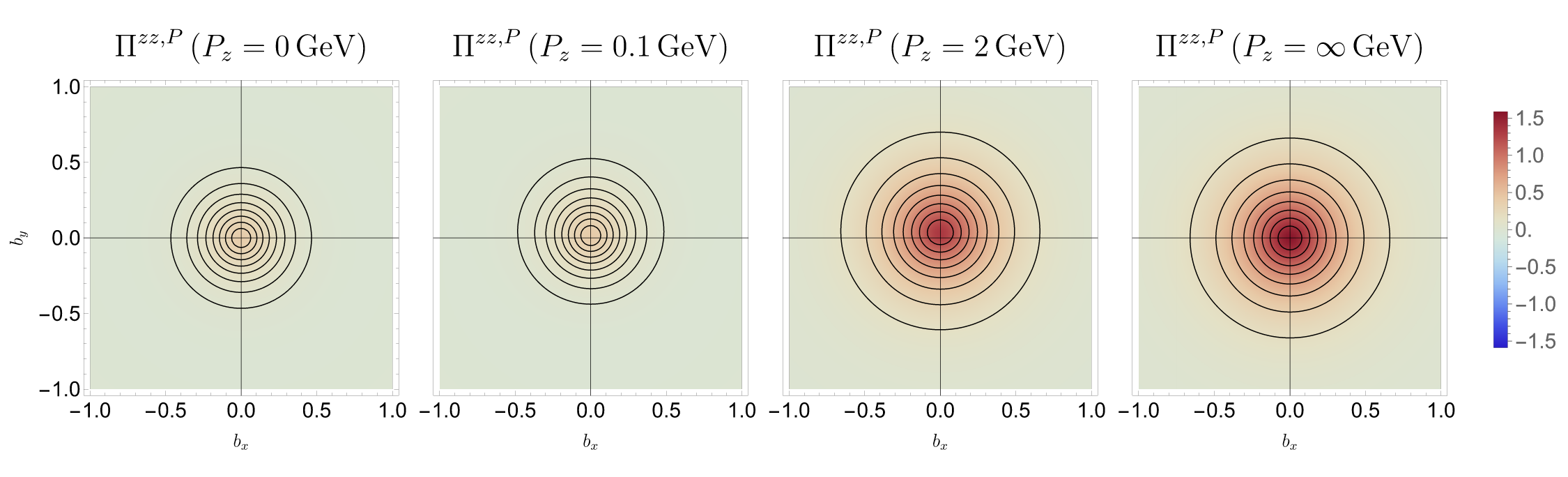}
  \hspace{0cm}

  \caption{Total EF frame distributions of axial momentum flux in the transverse plane of a nucleon polarized along the $x$-axis for different values of the nucleon momentum. The total distributions are normalized to $M\beta^2_P$. Based on the simple multipole model of Ref.~\cite{Lorce:2018egm} for the EMT FFs.}

  \label{fig:9}
\end{figure*} 

The spin-independent and spin-dependent amplitudes defining the axial momentum flux distribution in the EF are given by
\begin{widetext}
\begin{subequations}\label{thrust_amplitudes}
\begin{align}
    \widetilde{\Pi}_{a}^{zz,U}
    (Q^{2},P_{z})
& = \frac{1}{\gamma_P}
    \frac{P^{0}\left[P^{0}+M\left(1+\tau\right)\right]}{\left(P^{0}+M\right)\left(1+\tau\right)}
    \left\{
    \left(\frac{P_{z}}{P^{0}}\right)^{2}
    E_{a}(Q^2)
  + \frac{2\tau P_{z}^{2}}{P^{0}\left[P^{0}+M\left(1+\tau\right)\right]}
    J_{a}(Q^2)
  + F_{a}(Q^2)
    \right\},
\label{sigma_U}\end{align}
\begin{align}
    \widetilde{\Pi}_{a}^{zz,T}
    (Q^{2},P_{z})
& = \frac{1}{\gamma_P}
    \frac{P^{0}P_{z}}{\left(P^{0}+M\right)\left(1+\tau\right)}
    \left\{
  - \left(\frac{P_{z}}{P^{0}}\right)^{2}
    E_{a}(Q^2)
  + \frac{2\left[P^{0}+M\left(1+\tau\right)\right]}{P^{0}}
    J_{a}(Q^2)
  - F_{a}(Q^2)
    \right\},
\label{sigma_T}
\end{align}
\label{sigmaz}
\end{subequations}
\end{widetext}
and the total EF distribution of axial momentum flux is normalized as 
\begin{align}
    \int d^{2}b_{\perp}\,
    \Pi^{zz}
    (\bm{b}_{\perp},P_{z};s^{\prime},s)
  = M \beta_{P}^{2}
    \delta_{s^{\prime}s}.
\end{align}
In the transverse BF, 
the axial momentum flux amplitudes reduce to
\begin{align}\label{BFthrustampl}
    \widetilde{\Pi}_{a}^{zz,U}
    (Q^{2},0)
& = MF_{a}
    (Q^2), \cr
    \widetilde{\Pi}_{a}^{zz,T}
    (Q^{2},0)
& = 0.
\end{align}
in agreement with the results in Refs.~\cite{Polyakov:2018zvc,Lorce:2018egm}. All the terms associated with explicit factors of $P_z$ in Eq.~\eqref{thrust_amplitudes} can then be understood as contributions induced by the Lorentz boost from the BF to the EF. In the IMF, we find that 
\begin{align}
    \widetilde{\Pi}_{a}^{zz,U}
    (Q^{2},\infty)
& = M
    A_{a}
    (Q^2), \cr
    \widetilde{\Pi}_{a}^{zz,T}
    (Q^{2},\infty)
& = M
    B_{a}
    (Q^2),
\label{IMFlongitudinal_normal_force}
\end{align}
just like in the cases of the EF energy, longitudinal
momentum, and longitudinal energy flux distributions, see Eqs.~\eqref{IMFenergy}
and~\eqref{IMFlongitudinal_momentum}.

In Fig.~\ref{fig:9}, we show the total EF distributions of axial momentum flux for non-zero $P_z$. The differences with the total EF energy distributions displayed in Fig.~\ref{fig:3} are simply due to the different kinematical weighting of the contributions associated with $E_a$ and $F_a$, see Eqs.~\eqref{energy_amplitudes} and~\eqref{thrust_amplitudes}. This explains also why the dipole moment of the total axial momentum flux distribution in the EF
\begin{align}
&\hspace{-.75cm}\int d^2b_\perp\,\bm b_\perp\,\Pi^{zz}(\bm b_\perp,P_z;s',s)=(\bm e_z\times\bm \sigma_{s's})_z\cr
& \quad\times\frac{P_z}{E_P}\left[J(0)-\frac{E_P}{E_P+M}\,\left(\frac{P_z}{E_P}\right)^2\frac{E(0)}{2}\right]
\end{align} 
differs from the total EF energy dipole moment, except in the limit $P_z\to\infty$.

\section{Distributions on the light front \label{sec.4}} 

A key feature of the light-front (LF) formalism~\cite{Dirac:1949cp,Brodsky:1997de} is that it exhibits Galilean symmetry in the transverse plane. As a result, spatial densities with probabilistic interpretation and free of relativistic recoil corrections can, in some cases, be defined within this framework~\cite{Burkardt:2000za,Burkardt:2002hr}. This applies, in particular, to the EMT component $T^{++}$~\cite{Burkardt:2002hr,Abidin:2008sb}, which is interpreted as the LF density of longitudinal momentum. 

By adapting the phase-space formalism to the LF picture, one can define 2D LF distributions for all the EMT components as follows~\cite{Lorce:2018egm}
\begin{align}
&   \hspace{-0.5cm}
    T_{a}^{\mu\nu}
    \left(\bm{b}_{\perp},P^+;\lambda^{\prime},\lambda\right)= \int \frac{d^{2}\Delta_{\perp}}{\left(2\pi\right)^{2}}\,
    e^{-i\bm{\Delta}_{\perp}\cdot\bm{b}_{\perp}}\cr
& \hspace{1.8cm} \times
    \left.
    \frac{_\text{LF}\!\mel{p^{\prime},\lambda^{\prime}}{\hat{T}_{a}^{\mu\nu}\left(0\right)}{p,\lambda}_\text{LF}}{2P^+}
    \right|_{\mathrm{DYF}},
\label{EMTdistribution_LF}
\end{align}
where $|p,\lambda\rangle_\text{LF}$ is a LF four-momentum eigenstate with LF helicity $\lambda$. LF components are defined as $x^\mu=[x^+,x^-,\bm x_\perp]$ with $x^\pm=(x^0+x^3)/\sqrt{2}$. The symmetric Drell-Yan frame (DYF), characterized by $\Delta^+=0$ and $\bm P_\perp=\bm 0_\perp$, ensures that the LF energy transfer vanishes $\Delta^-=(\bm \Delta_\perp\cdot\bm P_\perp-\Delta^+ P^-)/P^+=0$. It follows that the 2D LF distributions remain $x^+$-independent~\cite{Lorce:2017wkb}.

Focusing of the ``longitudinal'' LF components $T^{++}$, $T^{+-}$, $T^{-+}$, and $T^{--}$, we find that the LF amplitudes are given by
\begin{widetext}
\begin{subequations}
\begin{align}
   _\text{LF}\!\mel{p^{\prime},\lambda^{\prime}}{\hat{T}_{a}^{++}\left(0\right)}{p,\lambda}_\text{LF}\Big|_\text{DYF}
& = 2\left(P^{+}\right)^{2}
    \Bigg\{
    \delta_{\lambda^{\prime}\lambda}
    A_{a}(Q^2)
    + \frac{\left(\bm{\sigma}_{\lambda^{\prime}\lambda}\times i\bm{\Delta}_{\perp}\right)_{z}}{2M}
   B_{a}(Q^2)
    \Bigg\},
\label{T++}
\end{align}
\begin{align}
    _\text{LF}\!\mel{p^{\prime},\lambda^{\prime}}{\hat{T}_{a}^{+-}\left(0\right)}{p,\lambda}_\text{LF}\Big|_\text{DYF}
& = \frac{2P^+P^-}{1+\tau}
    \Bigg\{
    \delta_{\lambda^{\prime}\lambda}
    \left[
    E_{a}(Q^2)-F_{a}(Q^2)+2\tau S_{a}(Q^2)
    \right]\cr
&   \hspace{2cm}
  - \frac{\left(\bm{\sigma}_{\lambda^{\prime}\lambda}\times i\bm{\Delta}_{\perp}\right)_{z}}{2M}
    \left[E_{a}(Q^2)-F_{a}(Q^2)-2
    S_{a}(Q^2)
    \right]
    \Bigg\},
\label{T+-}
\end{align}
\begin{align}
    _\text{LF}\!\mel{p^{\prime},\lambda^{\prime}}{\hat{T}_{a}^{--}\left(0\right)}{p,\lambda}_\text{LF}\Big|_\text{DYF}
& = \frac{2\left(P^{-}\right)^{2}}{1+\tau}
    \Bigg\{
    \delta_{\lambda^{\prime}\lambda}
    \left[
    (1-\tau)A_{a}(Q^2)-2\tau
    B_{a}(Q^2)
    \right]\cr
&   \hspace{2cm}
  - \frac{\left(\bm{\sigma}_{\lambda^{\prime}\lambda}\times i\bm{\Delta}_{\perp}\right)_{z}}{2M}
    \left[
    2A_{a}(Q^2)+(1-\tau)B_{a}(Q^2)
    \right]
    \Bigg\}.
\label{T--}
\end{align}
\label{LFamplitudes}
\end{subequations}
\end{widetext}
The matrix element of $\hat{T}^{-+}_a$ is obtained by reversing the sign of $S_{a}$ in Eq.~\eqref{T+-}. Note that we can write $P^-=M^2(1+\tau)/(2P^+)$.

Similar to the analysis presented in Ref.~\cite{Chen:2022smg} for the electromagnetic current, we can relate these LF amplitudes to the corresponding IMF amplitudes. Indeed, combining the results in Eq.~\eqref{EFmatrixelement}, we find that 
\begin{widetext}
\begin{subequations}
\begin{align}
    \mel{p^{\prime},s^{\prime}}{\hat{T}_{a}^{++}\left(0\right)}{p,s}\Big|_\text{EF}
& = \frac{2\left(P^{+}\right)^{2}}{\sqrt{1+\tau}}
    \Bigg\{
    \delta_{s^{\prime}s}
    \left[
    \cos{\theta}
    \left(E_{a}(Q^2)+F_{a}(Q^2)\right)
  - 2\sqrt{\tau}
    \sin{\theta}
    J_{a}(Q^2)
    \right]\cr
&   \hspace{2cm}
  + \frac{\left(\bm{\sigma}_{s^{\prime}s}\times i\bm{\Delta}_{\perp}\right)_{z}}{2M\sqrt{\tau}}
    \left[
    \sin{\theta}
    \left(E_{a}(Q^2)+F_{a}(Q^2)\right)
  + 2\sqrt{\tau}
    \cos{\theta}
    J_{a}(Q^2)
    \right]
    \Bigg\},
\label{T++IMF}
\end{align}
\begin{align}
    \mel{p^{\prime},s^{\prime}}{\hat{T}_{a}^{+-}\left(0\right)}{p,s}\Big|_\text{EF}
& = \frac{2P^{+}P^{-}}{\sqrt{1+\tau}}
    \Bigg\{
    \delta_{s^{\prime}s}
    \left[
    \cos{\theta}
    \left(E_{a}(Q^2)-F_{a}(Q^2)\right)
  - 2\sqrt{\tau}
    \sin{\theta}
    S_{a}(Q^2)
    \right]\cr
&   \hspace{2cm}
  + \frac{\left(\bm{\sigma}_{s^{\prime}s}\times i\bm{\Delta}_{\perp}\right)_{z}}{2M\sqrt{\tau}}
    \left[
    \sin{\theta}
    \left(E_{a}(Q^2)-F_{a}(Q^2)\right)
  + 2\sqrt{\tau}
    \cos{\theta}
    S_{a}(Q^2)
    \right]
    \Bigg\},
\label{T+-IMF}
\end{align}
\begin{align}
    \mel{p^{\prime},s^{\prime}}{\hat{T}_{a}^{--}\left(0\right)}{p,s}\Big|_\text{EF}
& = \frac{2\left(P^{-}\right)^{2}}{\sqrt{1+\tau}}
    \Bigg\{
    \delta_{s^{\prime}s}
    \left[
    \cos{\theta}
    \left(E_{a}(Q^2)+F_{a}(Q^2)\right)
  + 2\sqrt{\tau}
    \sin{\theta}
    J_{a}(Q^2)
    \right]\cr
&   \hspace{2cm}
  + \frac{\left(\bm{\sigma}_{s^{\prime}s}\times i\bm{\Delta}_{\perp}\right)_{z}}{2M\sqrt{\tau}}
    \left[
    \sin{\theta}
    \left(E_{a}(Q^2)+F_{a}(Q^2)\right)
  - 2\sqrt{\tau}
    \cos{\theta}
    J_{a}(Q^2)
    \right]
    \Bigg\}.
\label{T--IMF}
\end{align}
\label{IMFamplitudes}
\end{subequations}
\end{widetext}
In general, canonical polarization $s$ differs from LF helicity $\lambda$. However, the difference disappears in the IMF, where the Wigner rotation~\eqref{WR} reduces to
\begin{align}
    \hspace{-0.6cm}
    \lim_{P_{z}\to\infty}
    \cos{\theta}
& = \frac{1}{\sqrt{1+\tau}}, 
    \hspace{0.4cm}
    \lim_{P_{z}\to\infty}
    \sin{\theta}
  = - \frac{\sqrt{\tau}}{\sqrt{1+\tau}}.
\end{align}
The spin-independent and spin-dependent contributions 
to the IMF matrix element of $\hat{T}^{++}$ are then driven by the following linear combinations of EMT FFs 
\begin{subequations}
\begin{align}
    \lim_{P_{z}\to\infty}
    \frac{\cos{\theta}
    \left(E_{a}+F_{a}\right)
  - 2\sqrt{\tau}
    \sin{\theta}
    J_{a}}{\sqrt{1+\tau}}
  = A_{a},
\end{align}
\begin{align}
    \lim_{P_{z}\to\infty}
    \frac{\sin{\theta}
    \left(E_{a}+F_{a}\right)
  + 2\sqrt{\tau}
    \cos{\theta}
    J_{a}}{\sqrt{\tau}\sqrt{1+\tau}}
  = B_{a},
\end{align}
\end{subequations}
in agreement with the LF amplitude~\eqref{T++}. Similarly, we find for the unpolarized and polarized contributions to the IMF matrix element of $\hat{T}^{+-}$
\begin{subequations}
\begin{align}
&   \hspace{-0.5cm}
    \lim_{P_{z}\to\infty}
    \frac{\cos{\theta}
    \left(E_{a}-F_{a}\right)
  - 2\sqrt{\tau}
    \sin{\theta}
    S_{a}}{\sqrt{1+\tau}}\cr
&   \hspace{0.4cm}
  = \frac{E_{a}
  - F_{a}
  + 2 \tau S_{a}}{1+\tau},
\end{align}
\begin{align}
&   \hspace{-0.5cm}
    \lim_{P_{z}\to\infty}
    \frac{\sin{\theta}
    \left(E_{a}-F_{a}\right)
  + 2\sqrt{\tau}
    \cos{\theta}
    S_{a}}{\sqrt{\tau}\sqrt{1+\tau}} \cr
&   \hspace{0.4cm}
  = \frac{- E_{a}
  + F_{a}
  + 2 S_{a}}{1+\tau},
\end{align}
\end{subequations}
and of $\hat{T}^{--}$ 
\begin{subequations}
\begin{align}
&   \hspace{-0.5cm}
    \lim_{P_{z}\to\infty}
    \frac{\cos{\theta}
    \left(E_{a}+F_{a}\right)
  + 2\sqrt{\tau}
    \sin{\theta}
    J_{a}}{\sqrt{1+\tau}}\cr
&   \hspace{0.4cm}
  = \frac{\left(1-\tau\right)
    A_{a}
  - 2 \tau B_{a}}{1+\tau}, 
\end{align}
\begin{align}
&   \hspace{-0.5cm}
    \lim_{P_{z}\to\infty}
    \frac{\sin{\theta}
    \left(E_{a}+F_{a}\right)
  - 2\sqrt{\tau}
    \cos{\theta}
    J_{a}}{\sqrt{\tau}\sqrt{1+\tau}}\cr
&   \hspace{0.4cm}
  = \frac{- 2 A_{a}
  - \left(1-\tau\right)
    B_{a}}{1+\tau},
\end{align}
\end{subequations}
which also agree with the LF expressions in Eqs.~\eqref{T+-} and~\eqref{T--}.

\section{Summary \label{sec.5}}
In this paper, 
we have studied the relativistic distributions 
of the local gauge-invariant and asymmetric energy-momentum tensor (EMT) current 
within the framework of the quantum phase-space formalism 
for the nucleons,
taking into account for the first time the nucleon transverse polarization.
We have focused on the ``longitudinal'' components of the EMT and defined the corresponding scaled distributions of energy, 
longitudinal momentum, longitudinal energy flux, and 
axial momentum flux, including their decomposition into quark and gluon contributions. The frame dependence of these distributions has then been studied in detail and illustrated using a simple multipole model for the EMT form factors.

More specifically, we have demonstrated that the spin structure of the EMT matrix elements is the simplest in the Breit frame, which can be thought of as the average rest frame of the system. The latter is therefore the natural frame for defining the intrinsic EMT distributions. We have also explicitly shown that the structure of the relativistic distributions for arbitrary values of the nucleon longitudinal momentum results from a combination of both the familiar mixing of the rank-two tensor components under a longitudinal boost and the Wigner spin rotation. In the infinite-momentum frame, all the ``longitudinal'' distributions become equal. The mixing of components under a longitudinal boost can be avoided by switching to the light-front components. In this form, we have shown that not only the ``good'' $T^{++}$ but also the ``bad'' $T^{+-},T^{-+},T^{--}$ distributions coincide in the infinite-momentum frame with those defined within the light-front formalism.

\section{Acknowledgments \label{sec.6}}
This work is supported by the France Excellence scholarship 
through Campus France funded by the French government 
(Ministère de l’Europe et des Aﬀaires Étrangères), 
Grant No. 141295X (H.-Y.W.).

\appendix

\section{Quark axial singlet dipole mass \label{app:1}}

\begin{figure*}[htbp!]
  \centering

  \includegraphics[width=0.45\textwidth]{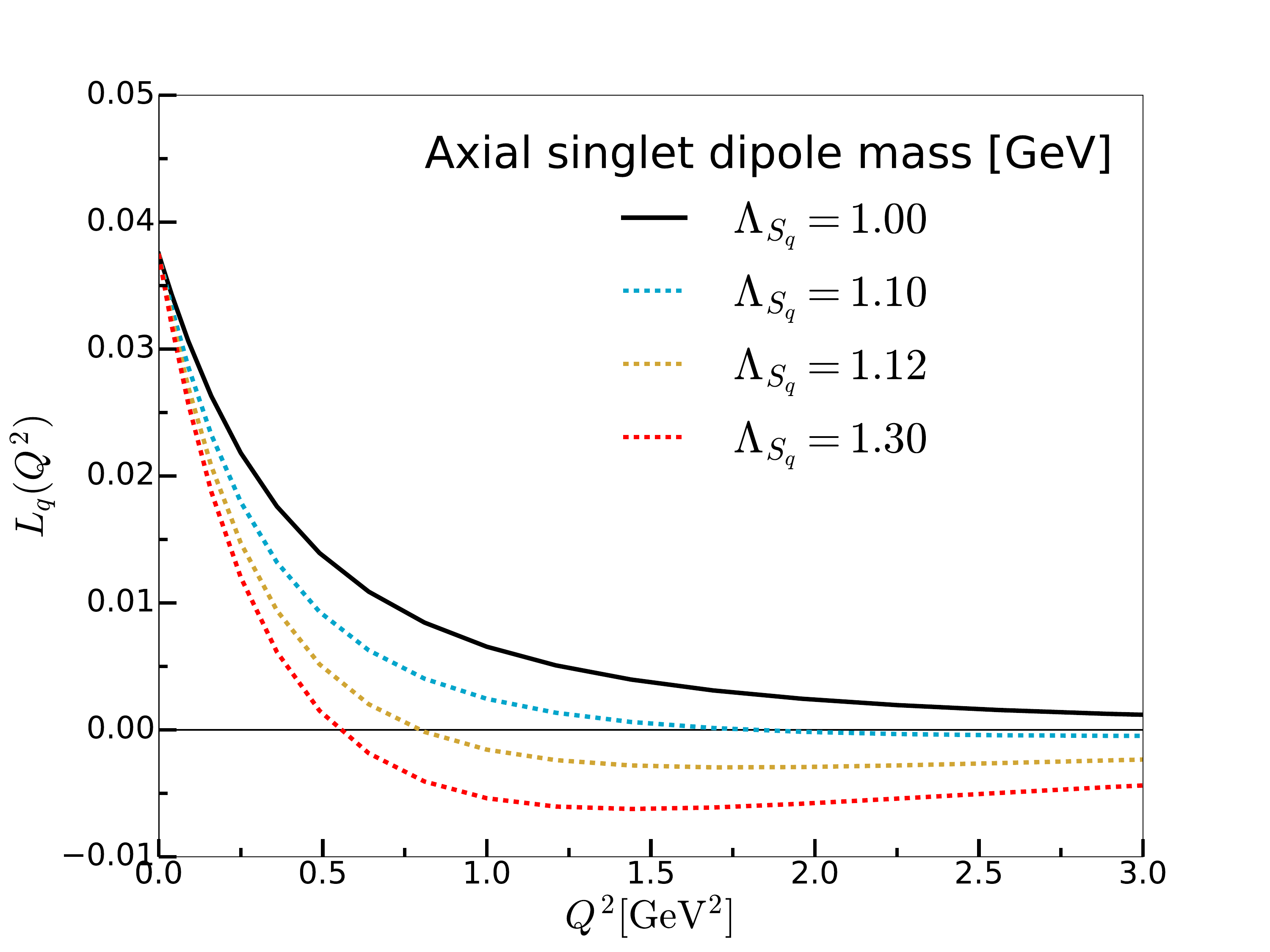}

  \caption{Dependence of the 
  quark OAM FF on the quark axial singlet dipole mass 
  $\Lambda_{S_{q}}$.}
  \label{fig:10}
\end{figure*}
\begin{figure*}[htbp!]
  \centering

  \includegraphics[width=1.0\textwidth]{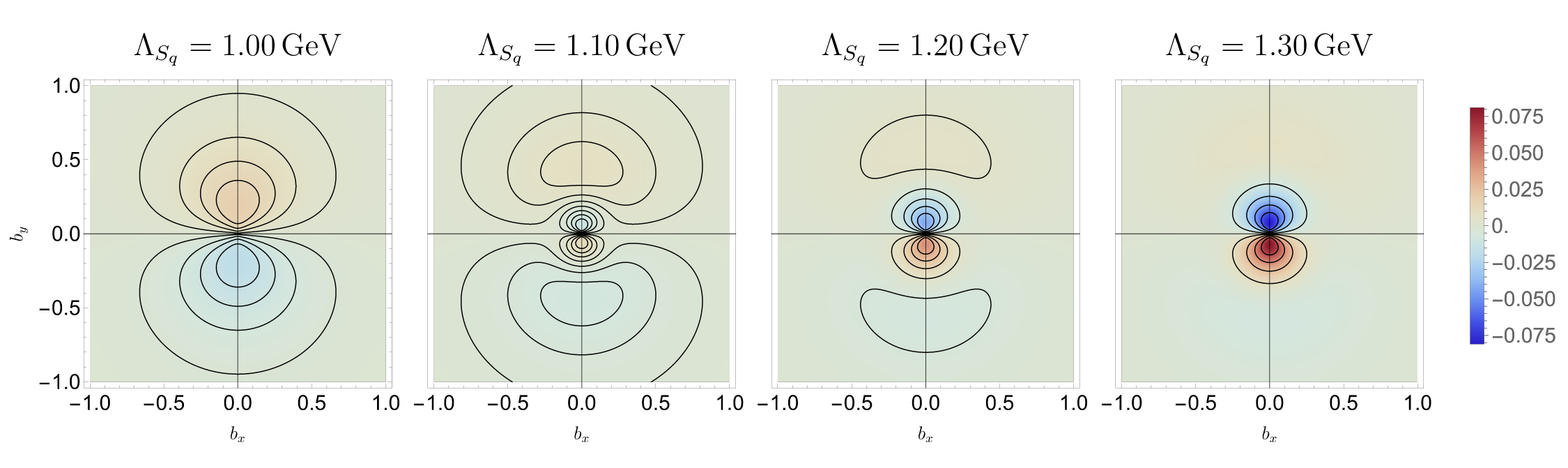}
  \hspace{0cm}

  \caption{Evolution of the BF distribution of longitudinal momentum in the transverse plane for different values of the quark axial singlet dipole mass $\Lambda_{S_{q}}$.}
  \label{fig:11}
\end{figure*}

In this Appendix,
we discuss in more detail the parametrization of the quark spin FF used
to illustrate the EMT distributions in the EF. 
Many studies have been devoted to the axial 
iso-vector FF, based either on experiments 
involving neutrino-nucleus scattering~\cite{NOMAD:2009qmu,MiniBooNE:2010xqw, MINERvA:2013bcy,
MINERvA:2013kdn,MINOS:2014axb,T2K:2014hih,T2K:2016jor} 
or on a variety of theoretical frameworks~\cite{Kim:2022rfw,Butkevich:2018hll,Amaro:2015lga,A1:1999kwj,
Gonzalez-Jimenez:2012snp,Nieves:2011pp,Alexandrou:2020okk,
Sufian:2018qtw,Gupta:2017dwj,Alexandrou:2017hac,Alexandrou:2010hf,
Alexandrou:2007eyf}, see also the review~\cite{Bernard:2001rs}. 
In contrast, the axial flavor-singlet FF is less well known.

In order to determine an appropriate value for the axial singlet dipole mass, we focus on the BF longitudinal momentum distribution in the transverse plane for a nucleon polarized along the $x$-axis
\begin{align}
    \hspace{-0.4cm}
    \mathcal{P}_{a}^{z,P}
    \left(\bm{b}_{\perp},0\right)
& = M
    \int \frac{d^{2}\Delta_{\perp}}{\left(2\pi\right)^{2}}\,
    e^{-i\bm{\Delta}_{\perp}\cdot\bm{b}_{\perp}}
    \frac{i\Delta_{y}}{2M}
    L_{a}(Q^2).
\end{align}
Since the gluon spin FF is zero, 
we consider only the quark part.

In Fig.~\ref{fig:10}, we show how the quark OAM FF changes 
as the axial singlet dipole mass $\Lambda_{S_q}$ varies.
The value of $Q^2$ at which the quark OAM FF becomes negative decreases with increasing values of $\Lambda_{S_q}$. This sign change leads to a node in the BF longitudinal momentum distribution along the radial direction in the transverse plane, depicted in Fig.~\ref{fig:11}. While the value of the total quark OAM $L_q(0)$ does not depend on the axial singlet dipole mass, its spatial distribution does. For $\Lambda_{S_{q}}=1.00$ GeV, the polarity of the dipole pattern agrees with our expectation of a positive quark OAM along the $x$-axis. For larger values of the dipole mass, the dipole pattern with the expected polarity spreads out in the transverse plane, diminishes in magnitude, and is gradually overshadowed by a new dipole pattern with opposite polarity. For the sake of simplicity, in this work we have adopted the value $\Lambda_{S_{q}}=1.00$ GeV to ensure that $L_q(Q^2)>0$ and thus avoid the appearance of a dipole pattern with the unexpected polarity. Note, however, that the polarity flip is in principle allowed and has been observed in the spatial distribution of $d$-quark average transverse momentum inside a longitudinally polarized nucleon, calculated within some light-front quark models~\cite{Lorce:2011ni}.

\section{Breit frame multipole form factors}

Since the physics is more transparent when expressed in terms of the BF amplitudes, we find as with the electromagnetic current~\cite{Lorce:2020onh} that the EMT matrix elements can alternatively be parametrized in terms of the BF multipole FFs 
\begin{widetext}
\begin{align}
    \mel{p^{\prime},s^{\prime}}{\hat{T}_{a}^{\mu\nu}\left(0\right)}{p,s} 
& = \bar{u}\left(p^{\prime},s^{\prime}\right)
    \Bigg[
    M\,\frac{P^{\mu}P^{\nu}}{P^2}E_a(Q^2)  
  + \frac{\Delta^{\mu}\Delta^{\nu}-\frac{1}{3}g^{\mu\nu}_P\Delta^{2}}{4M}  
    F_{2,a}(Q^2)
  - M g^{\mu\nu}_P 
    F_{0,a}(Q^2)\cr
&   \hspace{2.2cm}
  + \frac{iP^{\{\mu}\epsilon^{\nu\}\alpha\beta\rho}\Delta_{\alpha}P_\beta\gamma_\rho\gamma_5}{2P^2} J_{a}(Q^2)
  - \frac{i}{2}\epsilon^{\mu\nu\alpha\beta}\Delta_\alpha\gamma_\beta\gamma_5\,
    S_{a}(Q^2)
    \Bigg]
    u\left(p,s\right), 
\label{Parametrization2}
\end{align}
\end{widetext}
where $g^{\mu\nu}_P=g^{\mu\nu}-P^\mu P^\nu/P^2$ is the projector onto the subspace orthogonal to $P$. The force FF driving the axial momentum flux in the transverse BF~\eqref{BFthrustampl} is given by a linear combination of a monopole and a quadrupole FF, viz. $F_a=F_{0,a}-\frac{\tau}{3}F_{2,a}$ with $F_{0,a}=-\bar{C}_{a}-\frac{2}{3}\tau D_{a}$ and $F_{2,a}=D_{a}$.

\section{Spatial distributions of longitudinal momentum and axial momentum flux}

\begin{figure*}[htbp]
  \centering
  \textbf{EF longitudinal momentum distributions for an unpolarized nucleon}

  \vspace{0.2cm} 

  \begin{minipage}{0.37\textwidth}
    \centering
    Quark
  \end{minipage}
  \hspace{-0.05\textwidth}
  \begin{minipage}{0.37\textwidth}
    \centering
    Gluon
  \end{minipage}

  \begin{minipage}{0.35\textwidth}
    \centering
    \includegraphics[width=\textwidth]{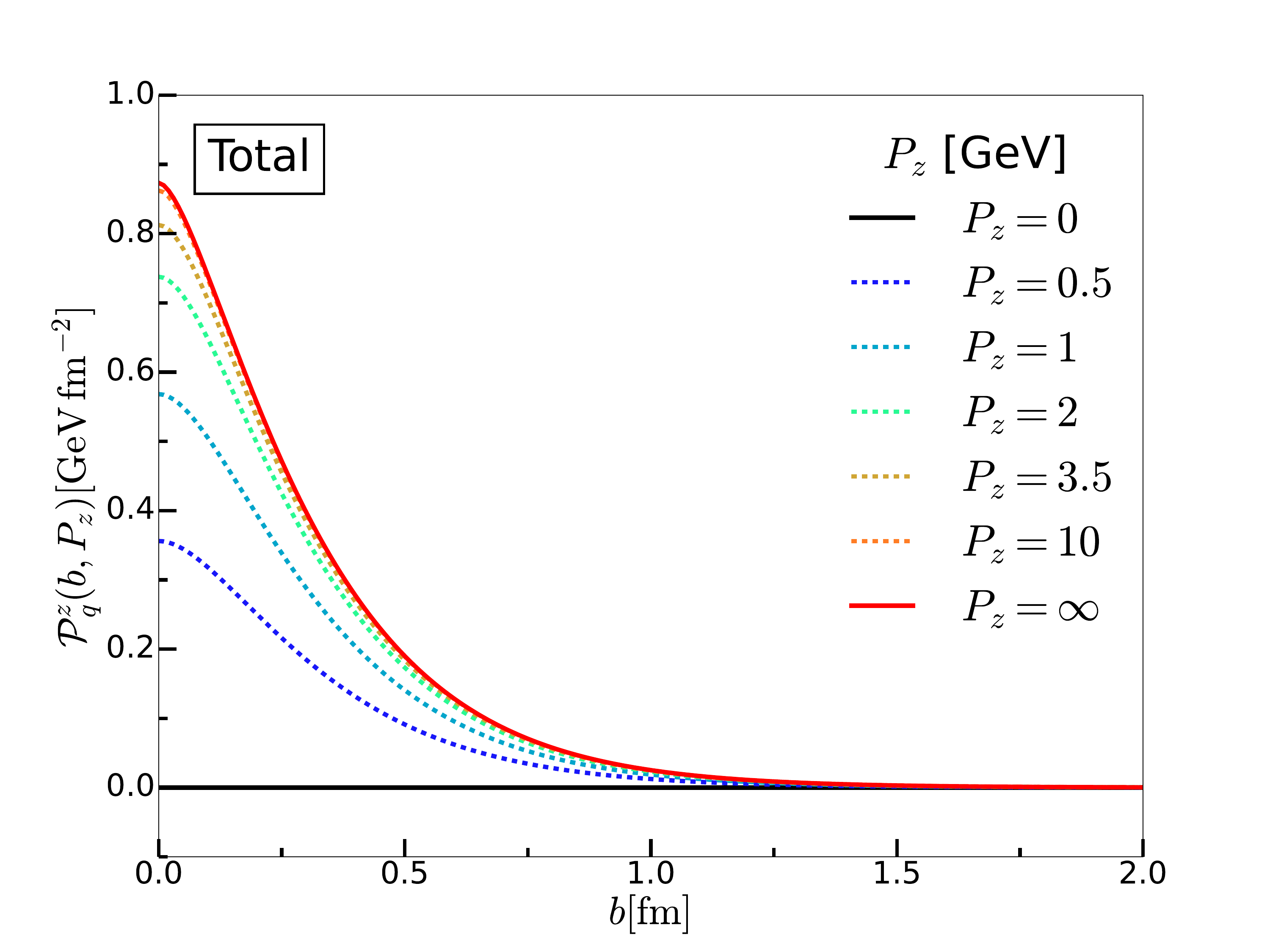}
  \end{minipage}
  \hspace{-0.6cm}
  \begin{minipage}{0.35\textwidth}
    \centering
    \includegraphics[width=\textwidth]{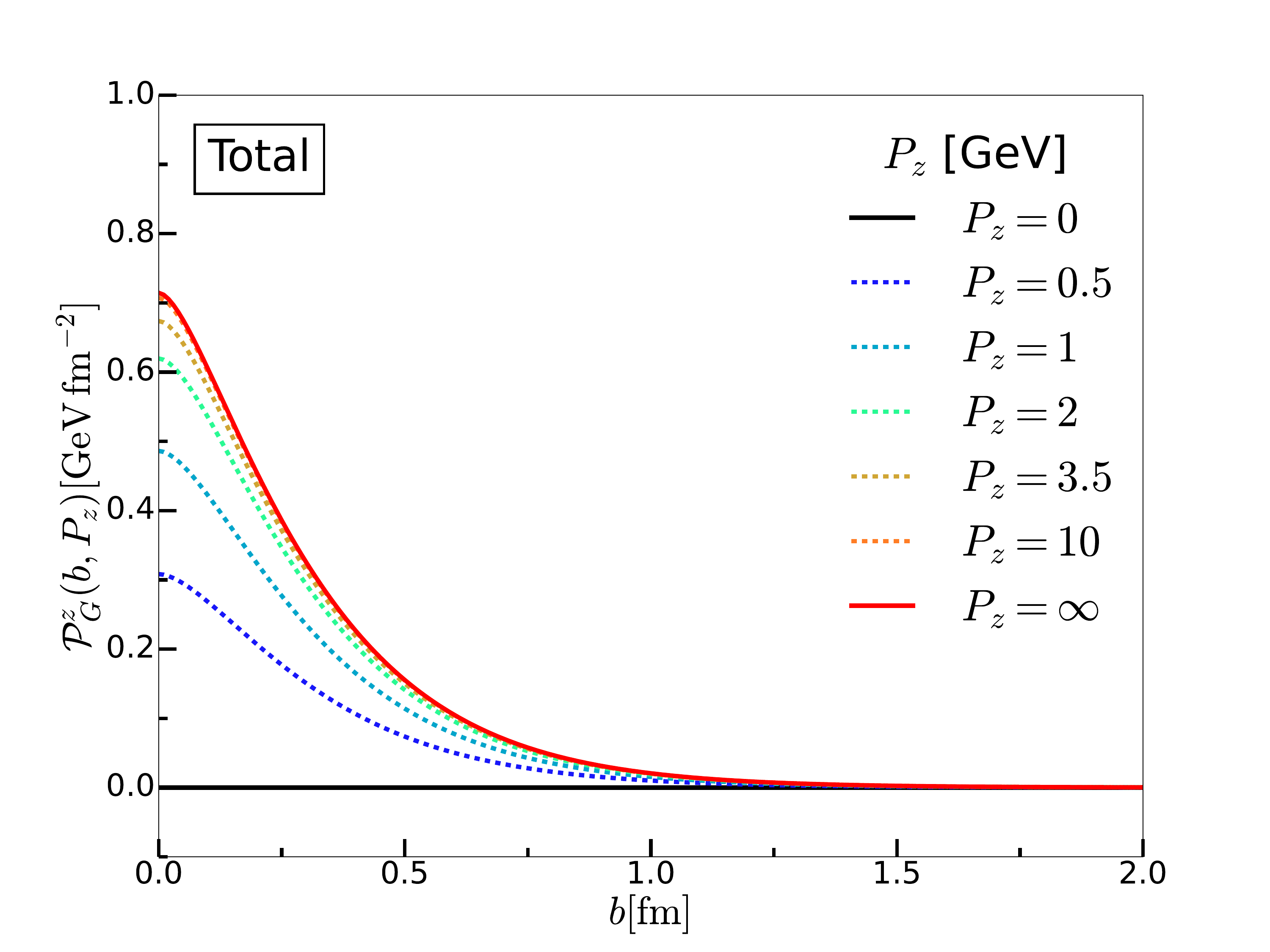}
  \end{minipage}
  
  \vspace{-0.35cm} 
  
  \begin{minipage}{0.35\textwidth}
    \centering
    \includegraphics[width=\textwidth]{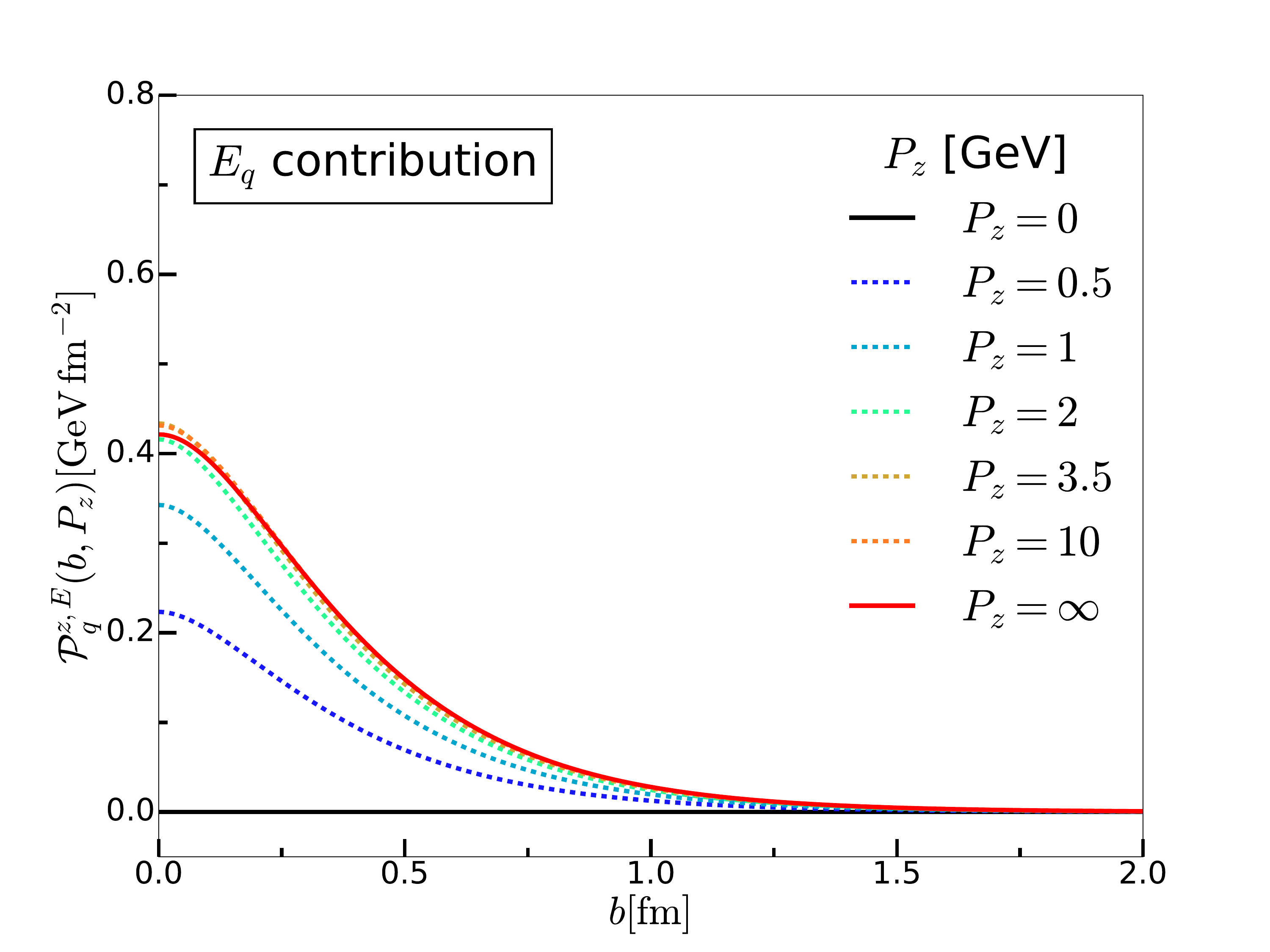}
  \end{minipage}
  \hspace{-0.6cm}
  \begin{minipage}{0.35\textwidth}
    \centering
    \includegraphics[width=\textwidth]{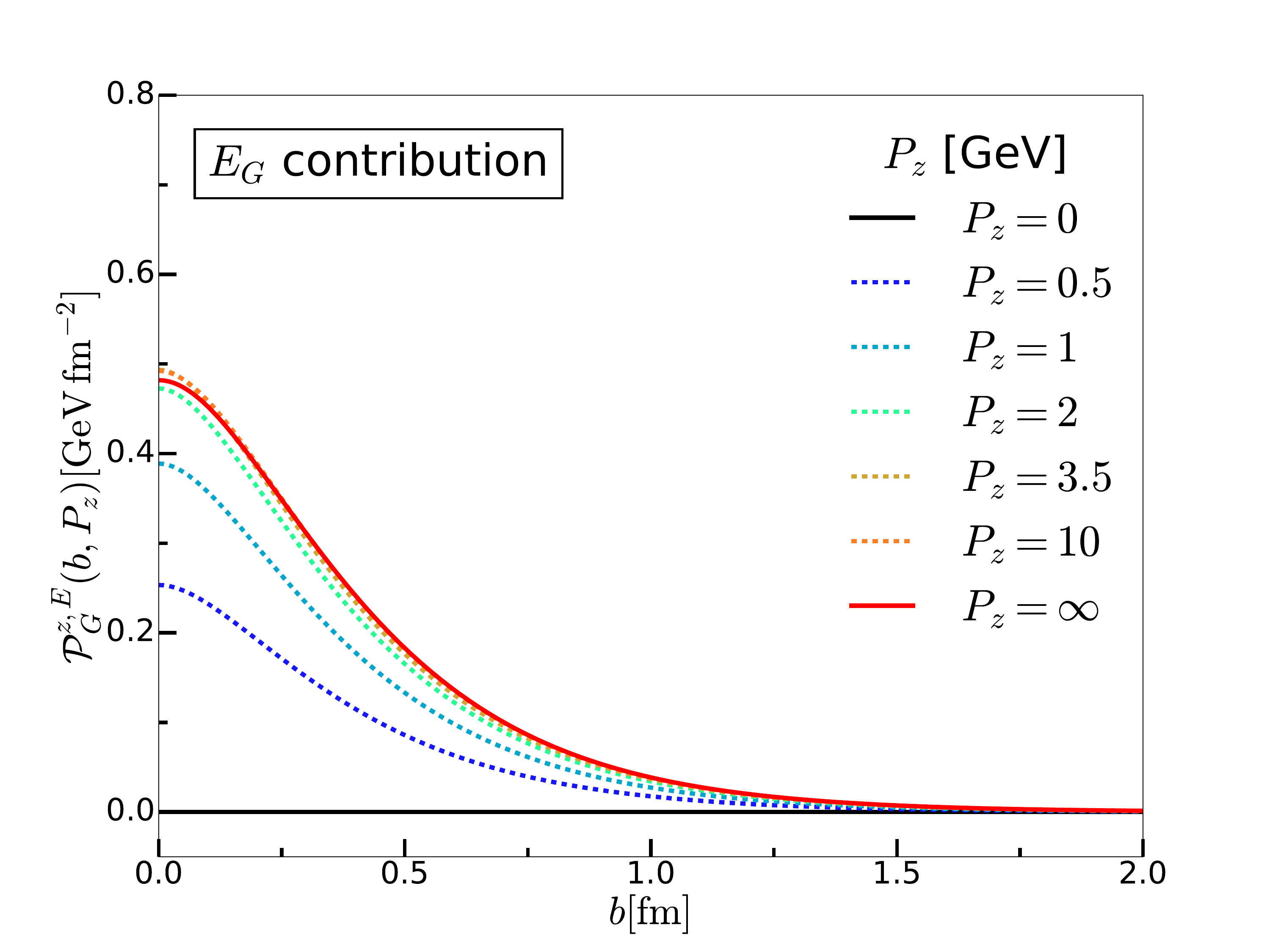}
  \end{minipage}
  
  \vspace{-0.35cm} 
  
  \begin{minipage}{0.35\textwidth}
    \centering
    \includegraphics[width=\textwidth]{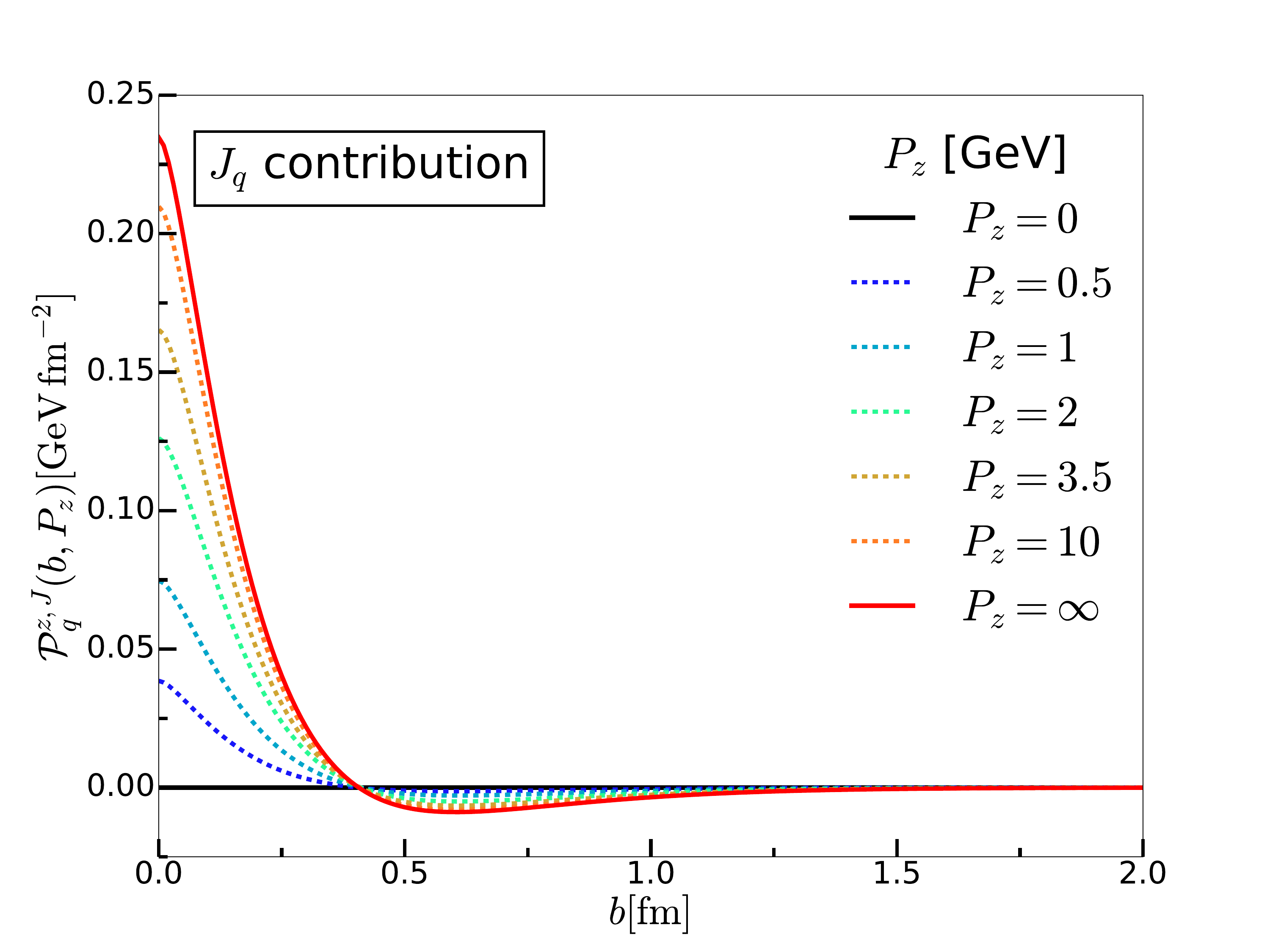}
  \end{minipage}
  \hspace{-0.6cm}
  \begin{minipage}{0.35\textwidth}
    \centering
    \includegraphics[width=\textwidth]{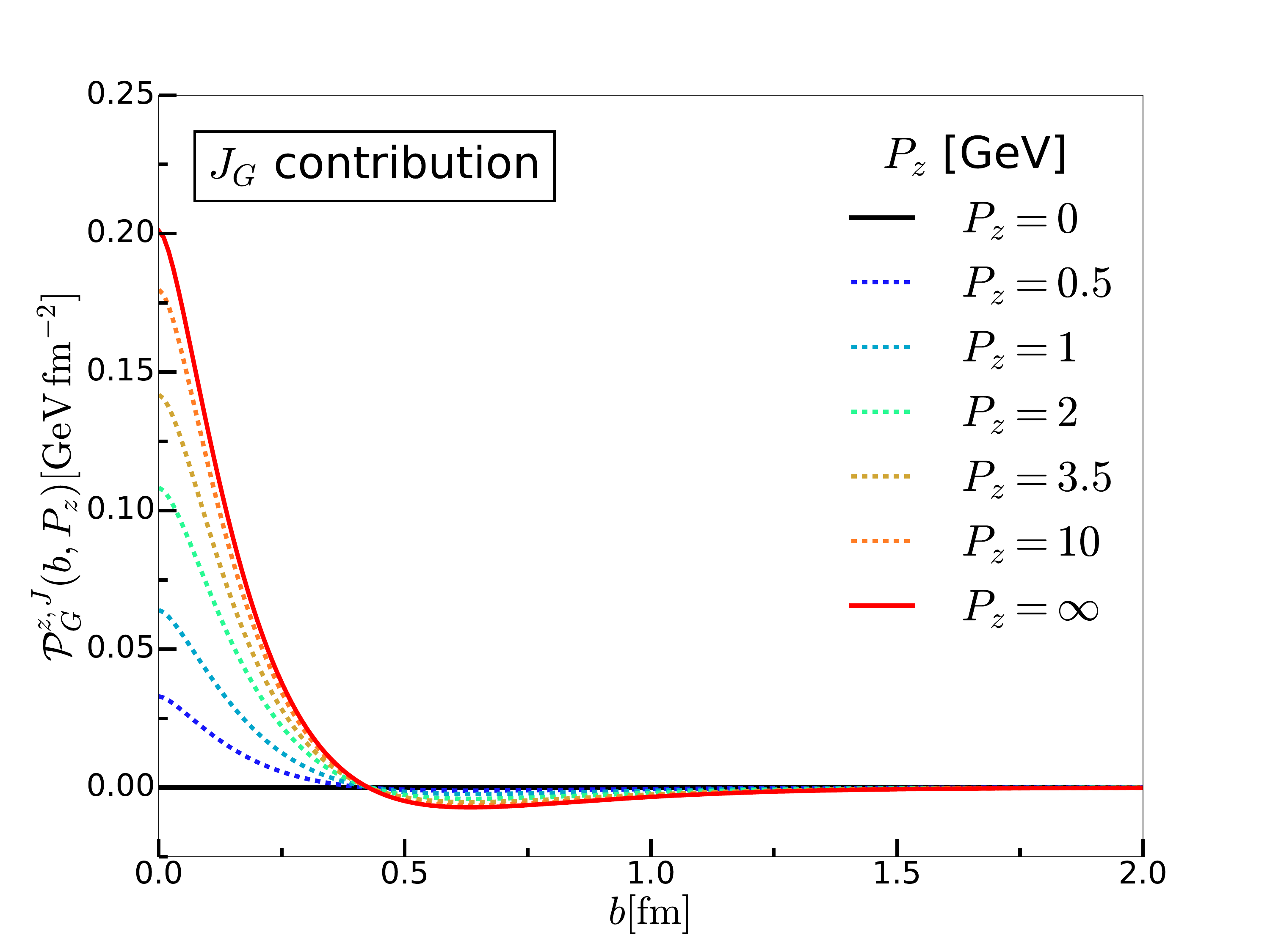}
  \end{minipage}
  
  \vspace{-0.35cm} 
  
  \begin{minipage}{0.35\textwidth}
    \centering
    \includegraphics[width=\textwidth]{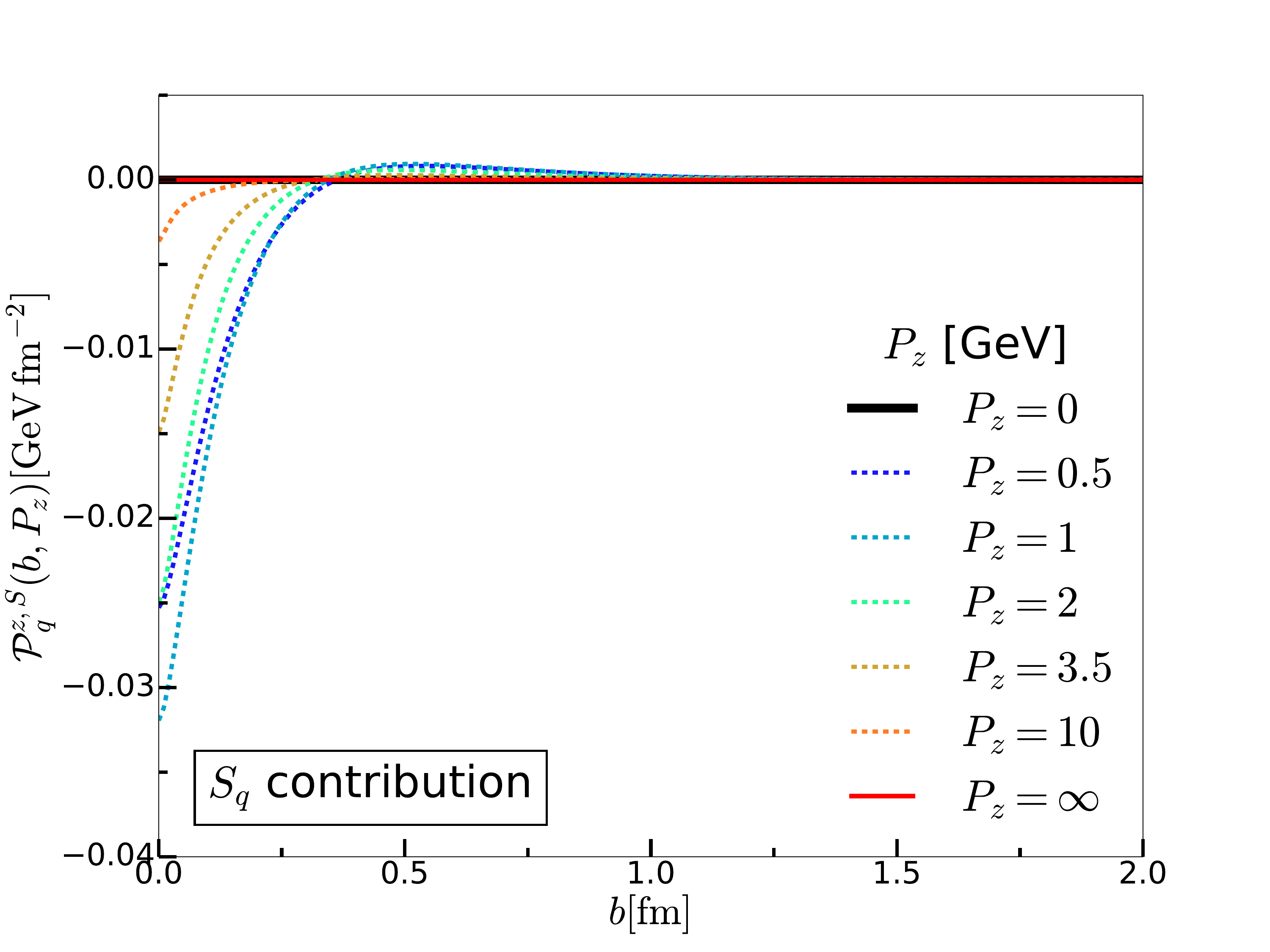}
  \end{minipage}
  \hspace{-0.6cm}
  \begin{minipage}{0.35\textwidth}
    \centering
    \includegraphics[width=\textwidth]{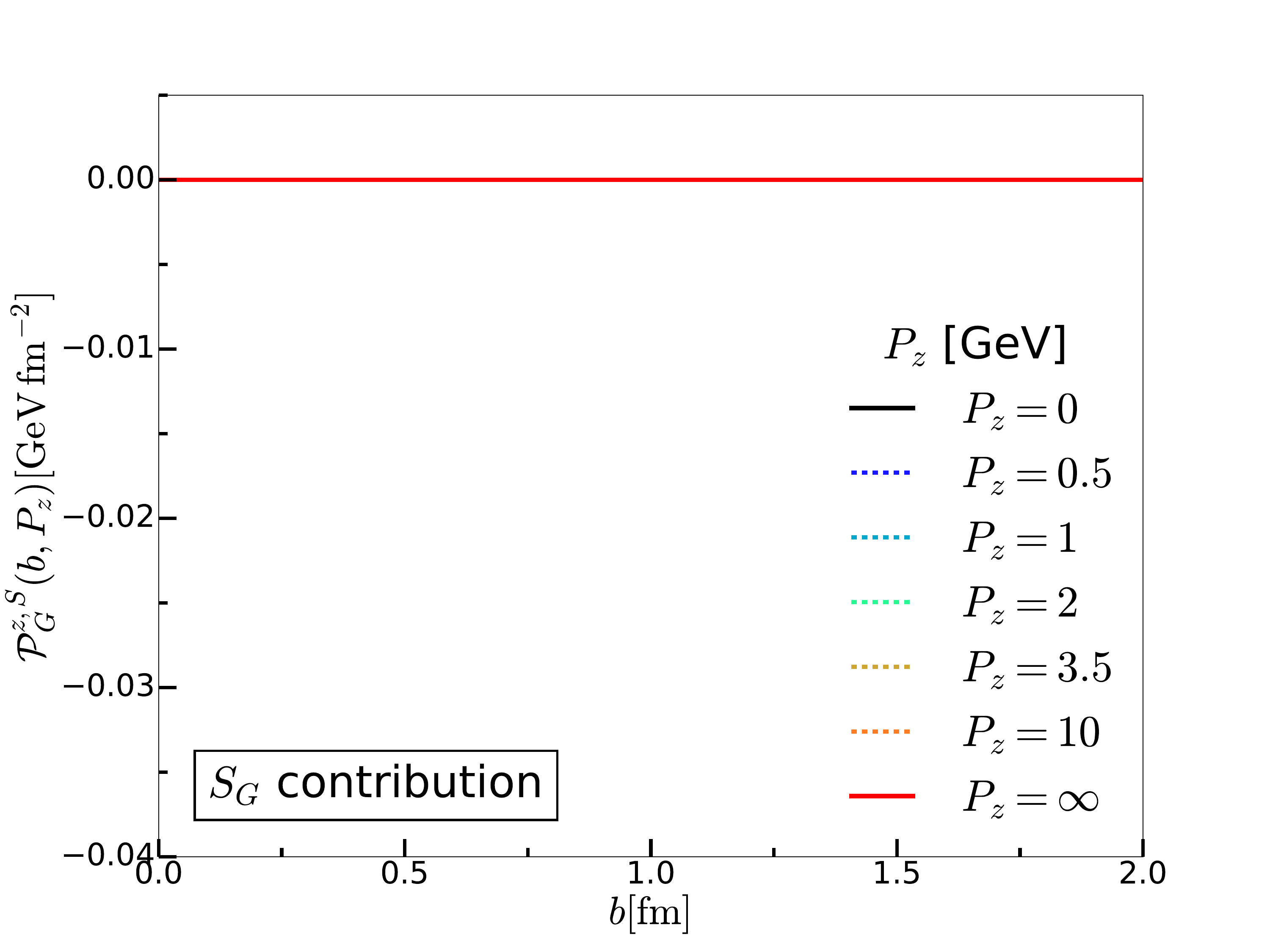}
  \end{minipage}

  \vspace{-0.35cm} 

  \begin{minipage}{0.35\textwidth}
    \centering
    \includegraphics[width=\textwidth]{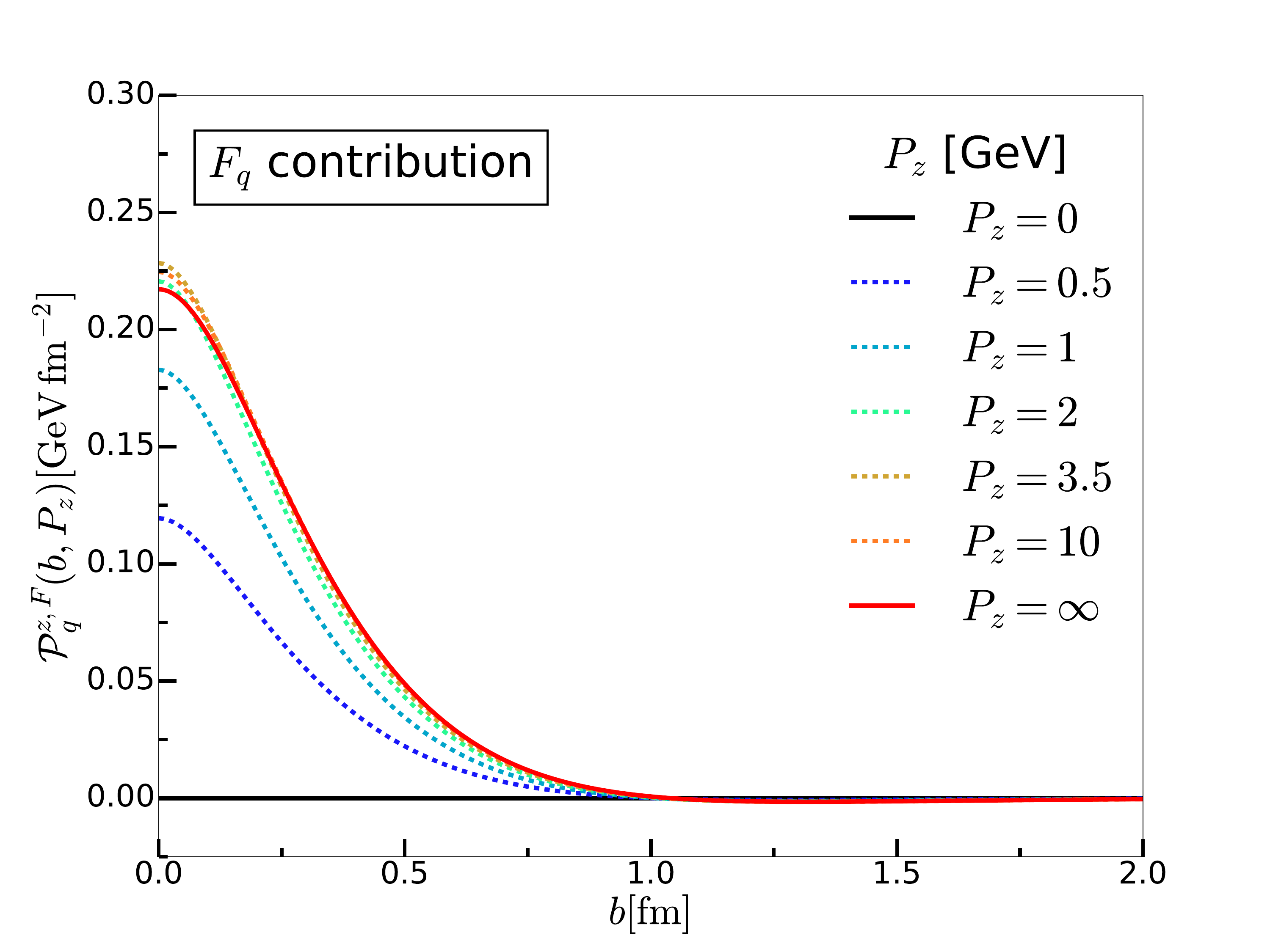}
  \end{minipage}
  \hspace{-0.6cm}
  \begin{minipage}{0.35\textwidth}
    \centering
    \includegraphics[width=\textwidth]{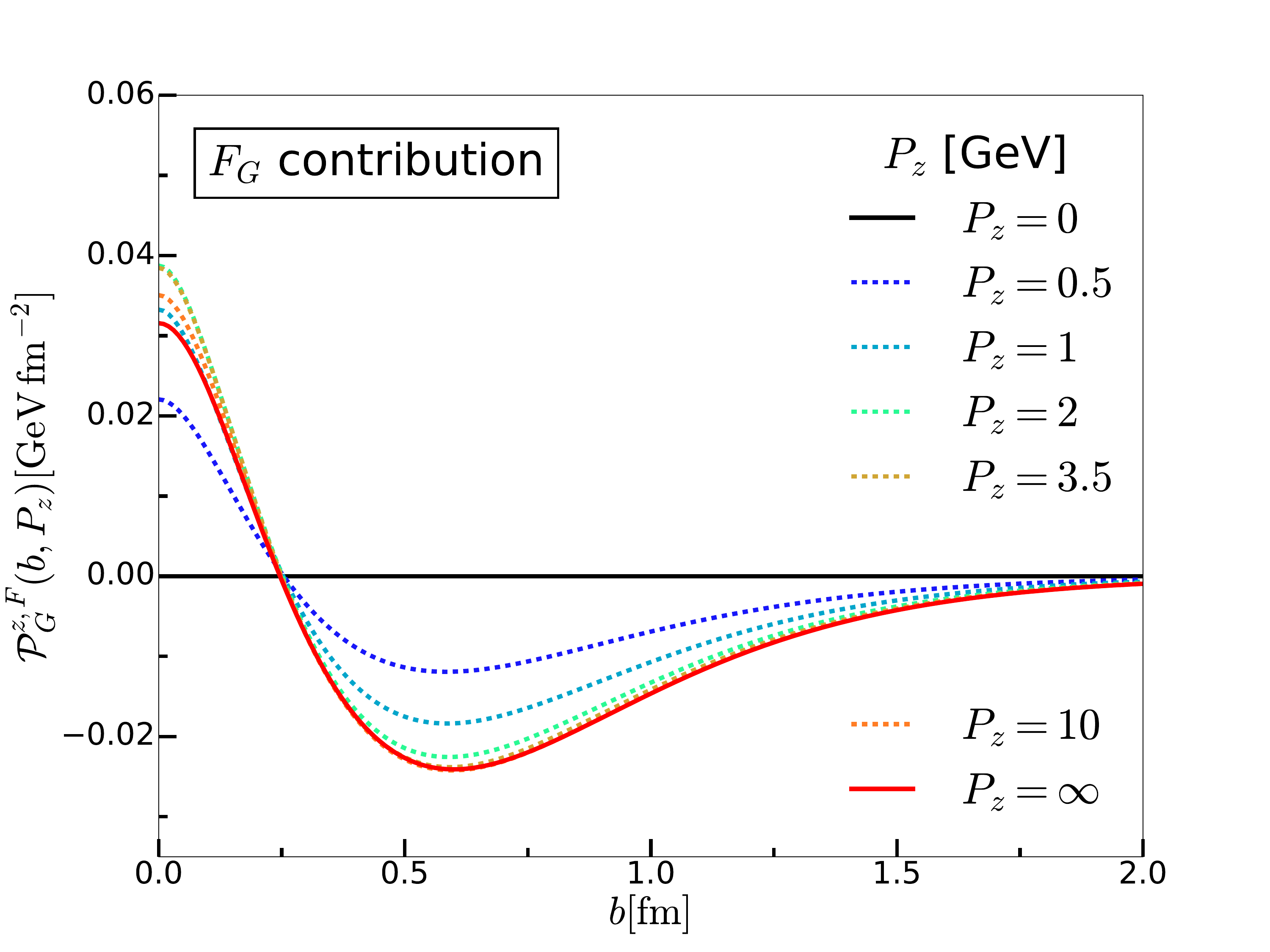}
  \end{minipage}
  
  \caption{Radial EF distributions of longitudinal momentum in an unpolarized nucleon for different values of the nucleon momentum. The total longitudinal momentum distribution is normalized to $M\beta_P$. The left (right) column corresponds to the quark (gluon) contribution. The first row depicts the sum of the four contributions shown in the other rows. Based on the simple multipole model of Ref.~\cite{Lorce:2018egm} for the EMT FFs.}

  \label{fig:4}
\end{figure*}

\begin{figure*}[htbp!]
  \centering
  \textbf{EF axial momentum flux distributions for the unpolarized nucleon}

  \vspace{0.2cm} 

  \begin{minipage}{0.42\textwidth}
    \centering
    Quark
  \end{minipage}
  \hspace{-0.05\textwidth}
  \begin{minipage}{0.42\textwidth}
    \centering
    Gluon
  \end{minipage}

  \begin{minipage}{0.40\textwidth}
    \centering
    \includegraphics[width=\textwidth]{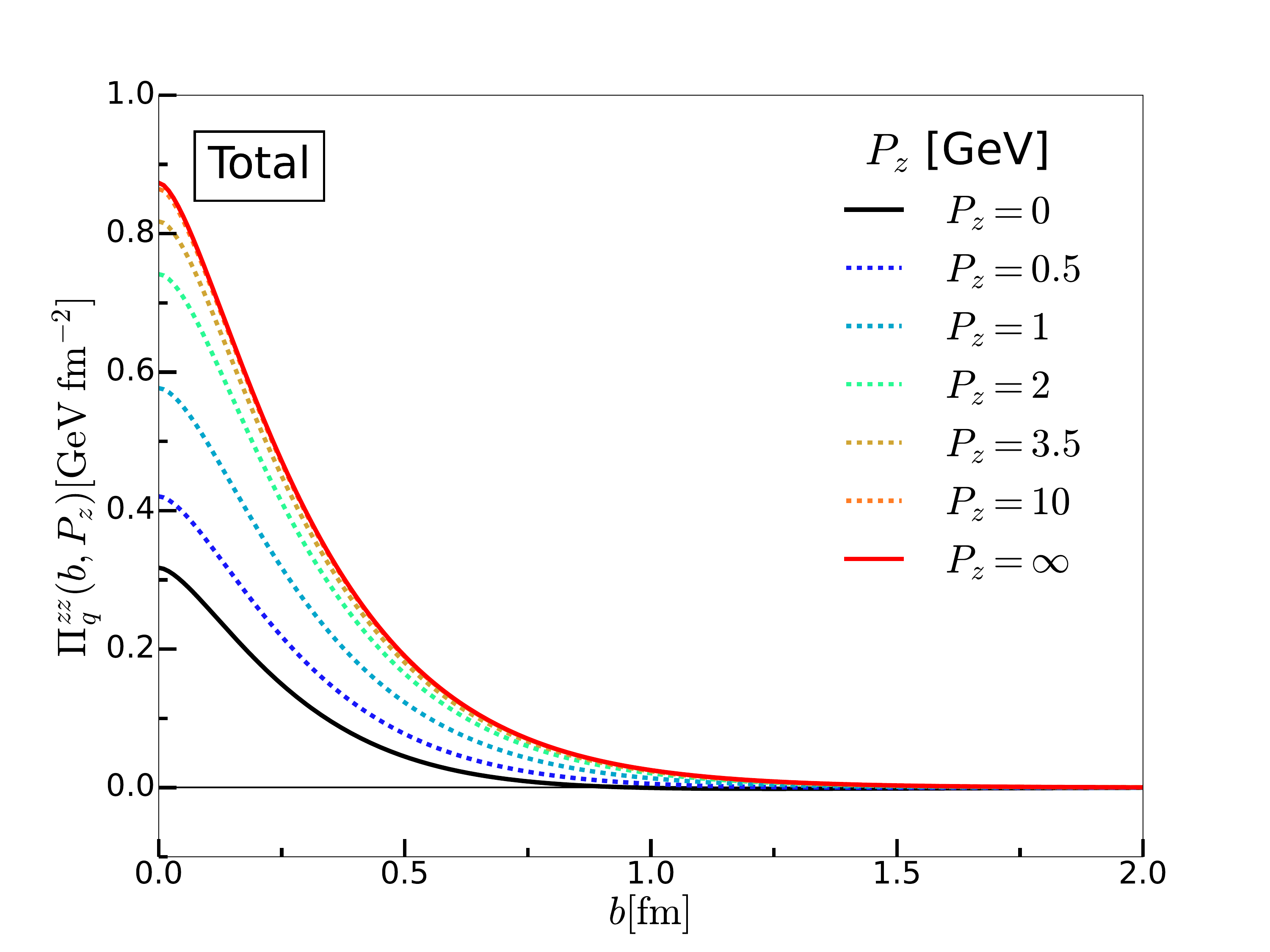}
  \end{minipage}
  \hspace{-0.6cm}
  \begin{minipage}{0.40\textwidth}
    \centering
    \includegraphics[width=\textwidth]{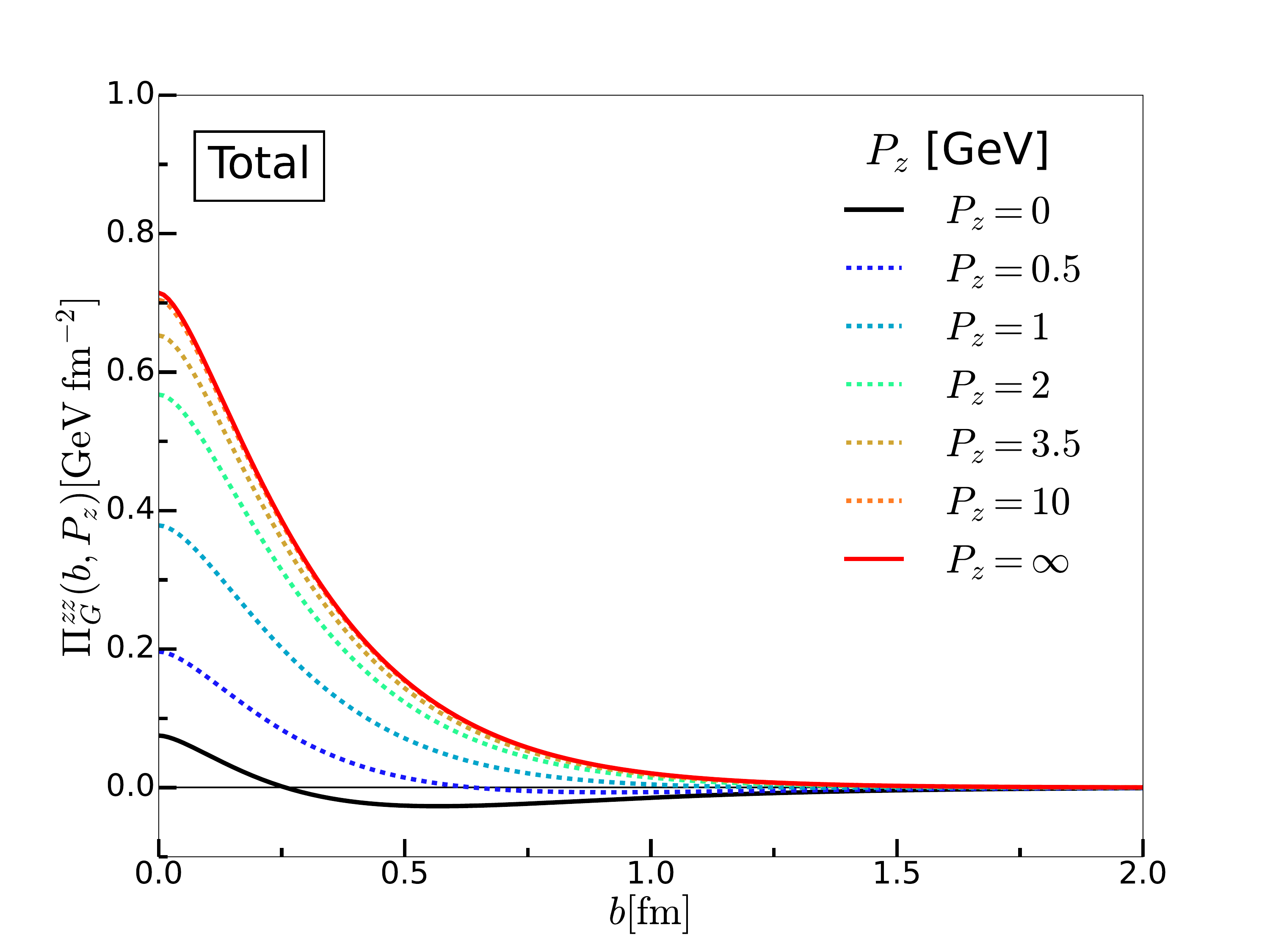}
  \end{minipage}
  
  \vspace{-0.4cm} 
  
  \begin{minipage}{0.40\textwidth}
    \centering
    \includegraphics[width=\textwidth]{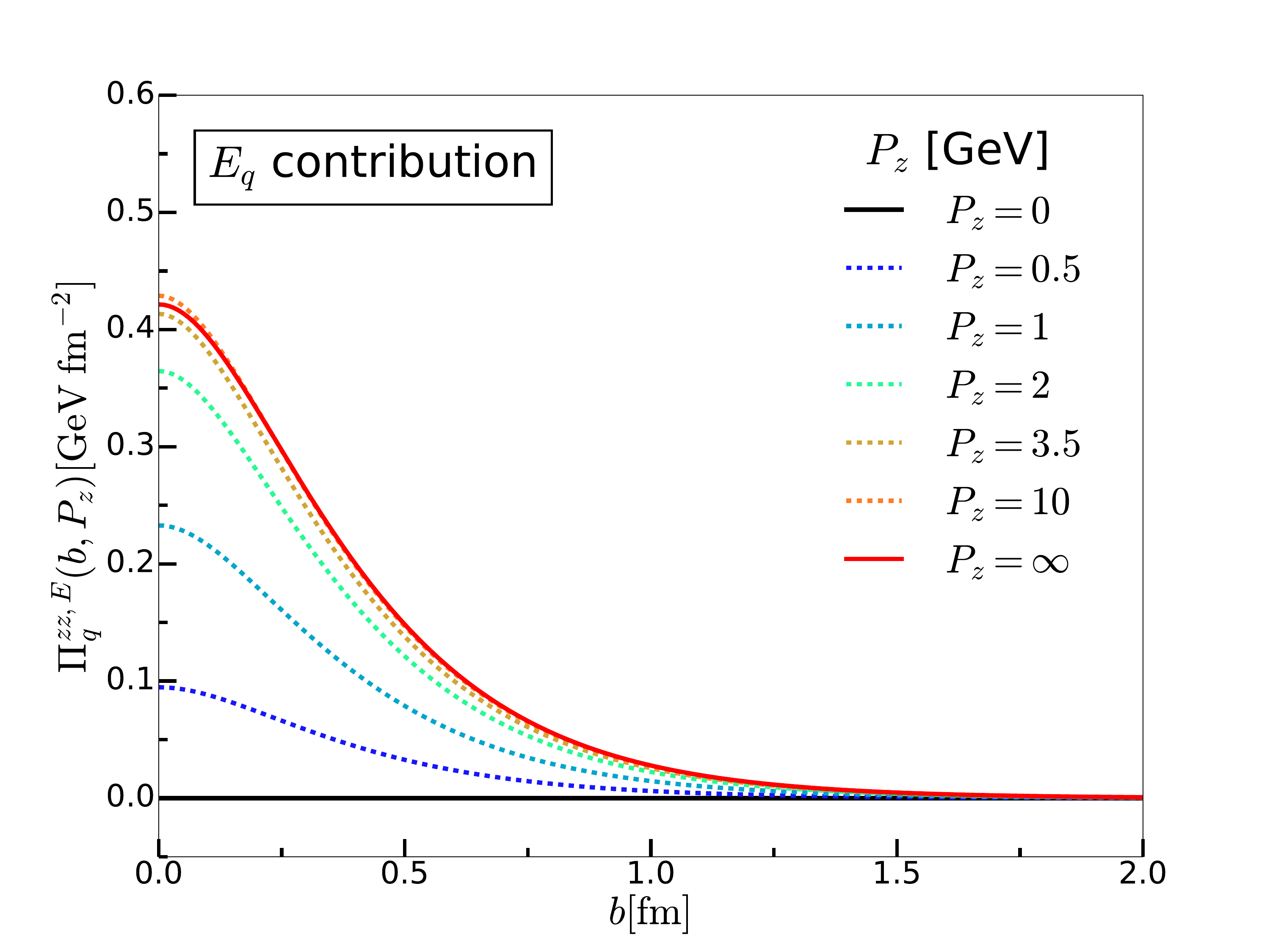}
  \end{minipage}
  \hspace{-0.6cm}
  \begin{minipage}{0.40\textwidth}
    \centering
    \includegraphics[width=\textwidth]{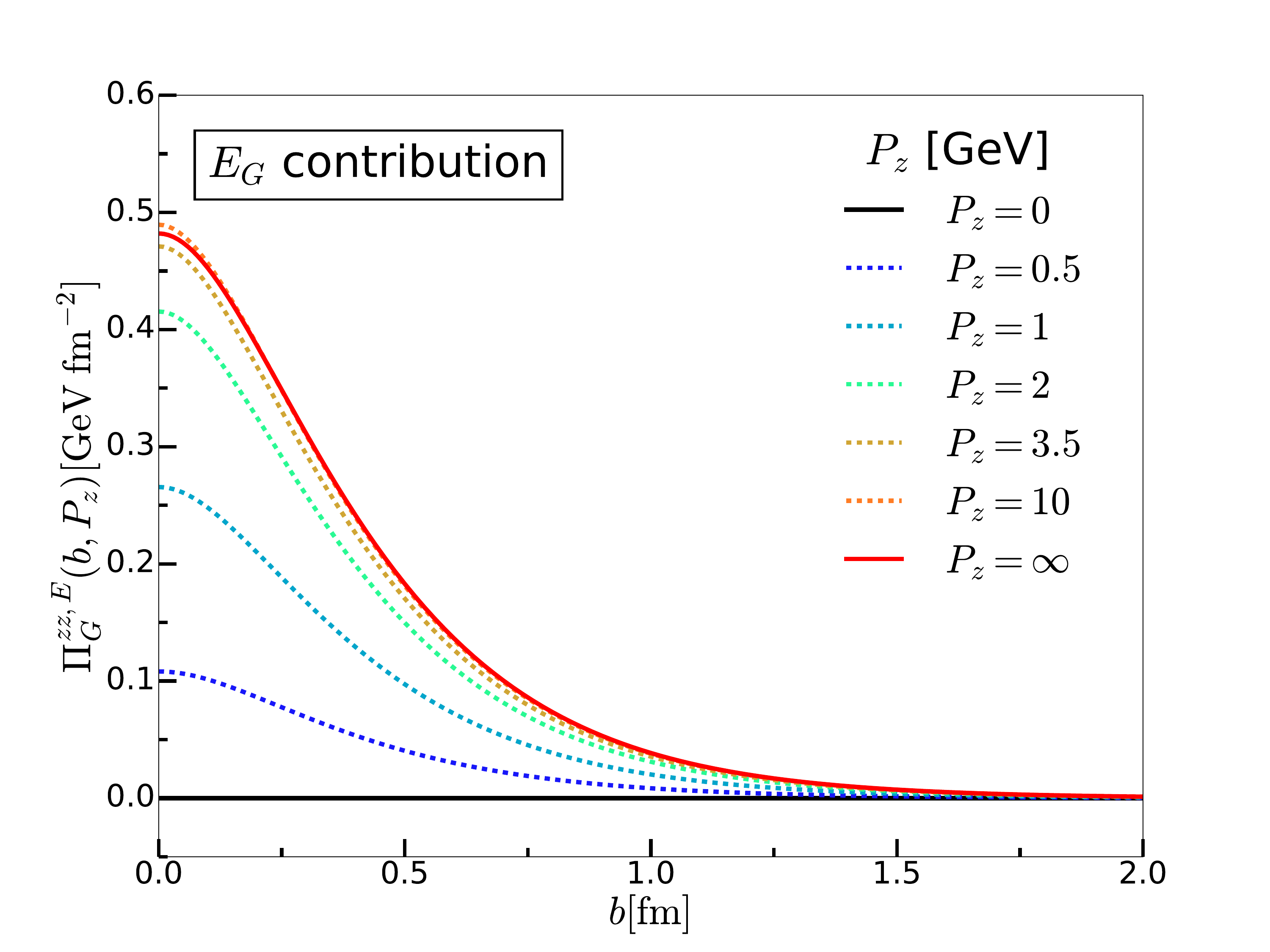}
  \end{minipage}
  
  \vspace{-0.4cm} 
  
  \begin{minipage}{0.40\textwidth}
    \centering
    \includegraphics[width=\textwidth]{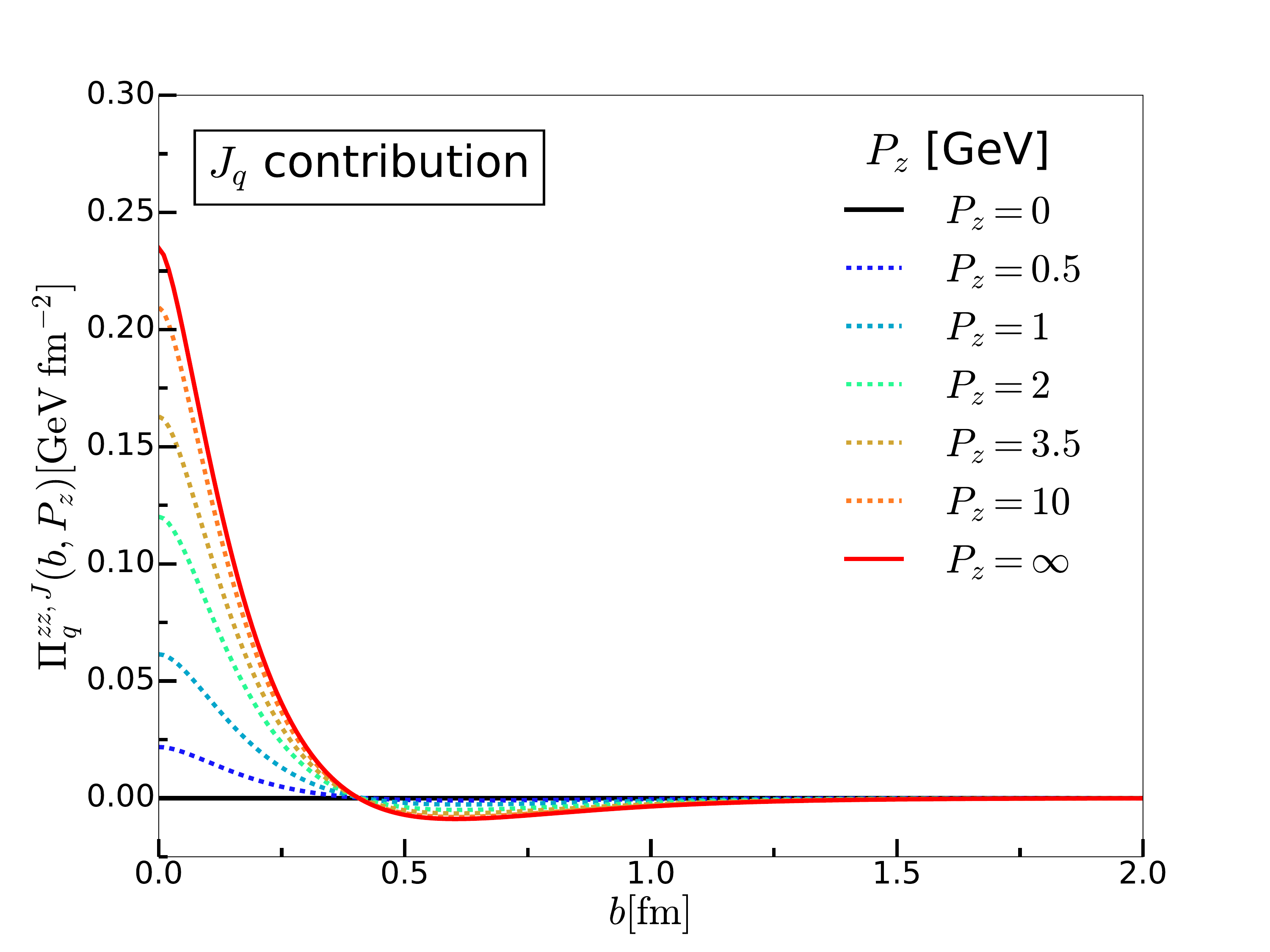}
  \end{minipage}
  \hspace{-0.6cm}
  \begin{minipage}{0.40\textwidth}
    \centering
    \includegraphics[width=\textwidth]{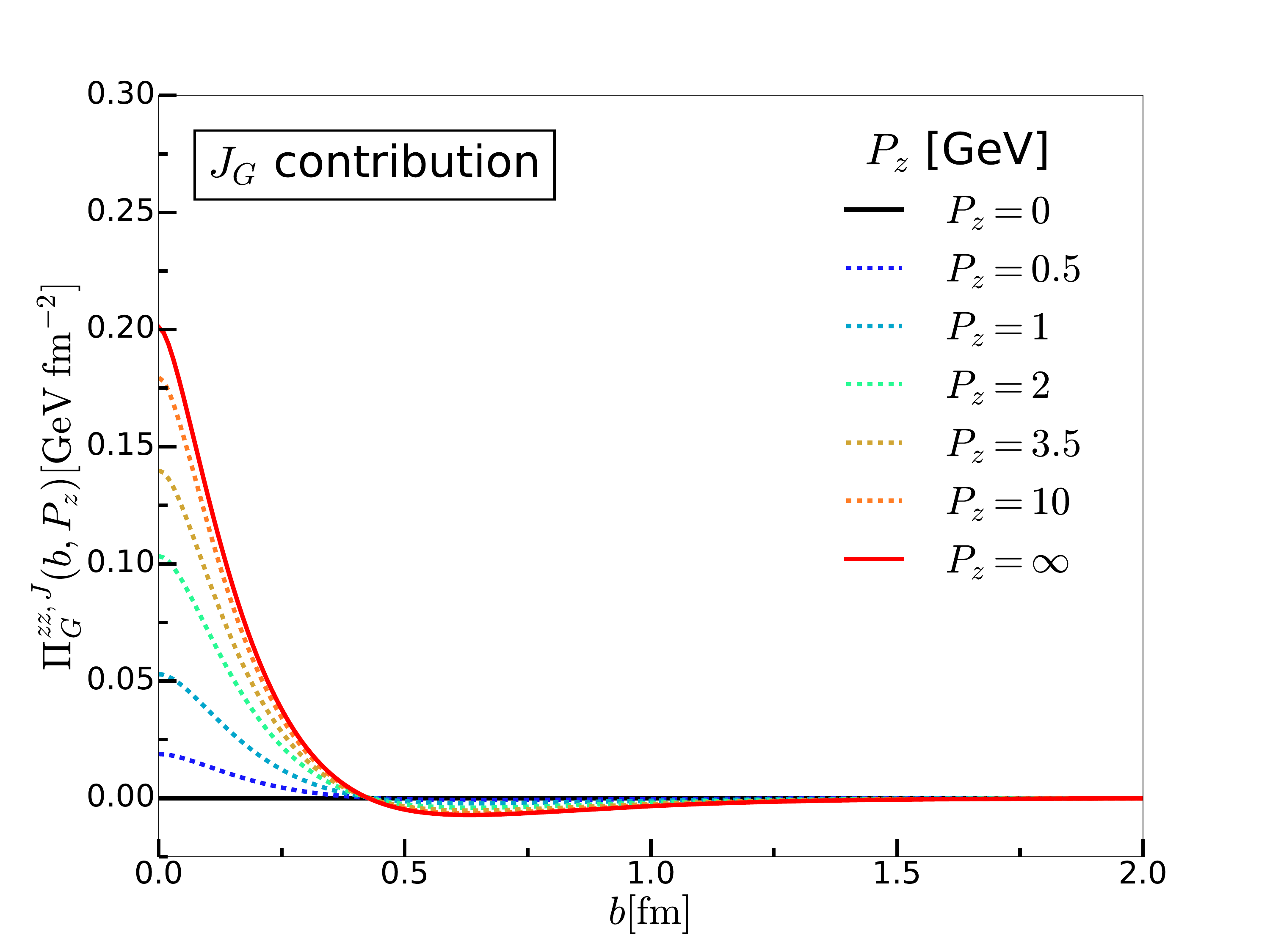}
  \end{minipage}
  
  \vspace{-0.4cm} 
  
  \begin{minipage}{0.40\textwidth}
    \centering
    \includegraphics[width=\textwidth]{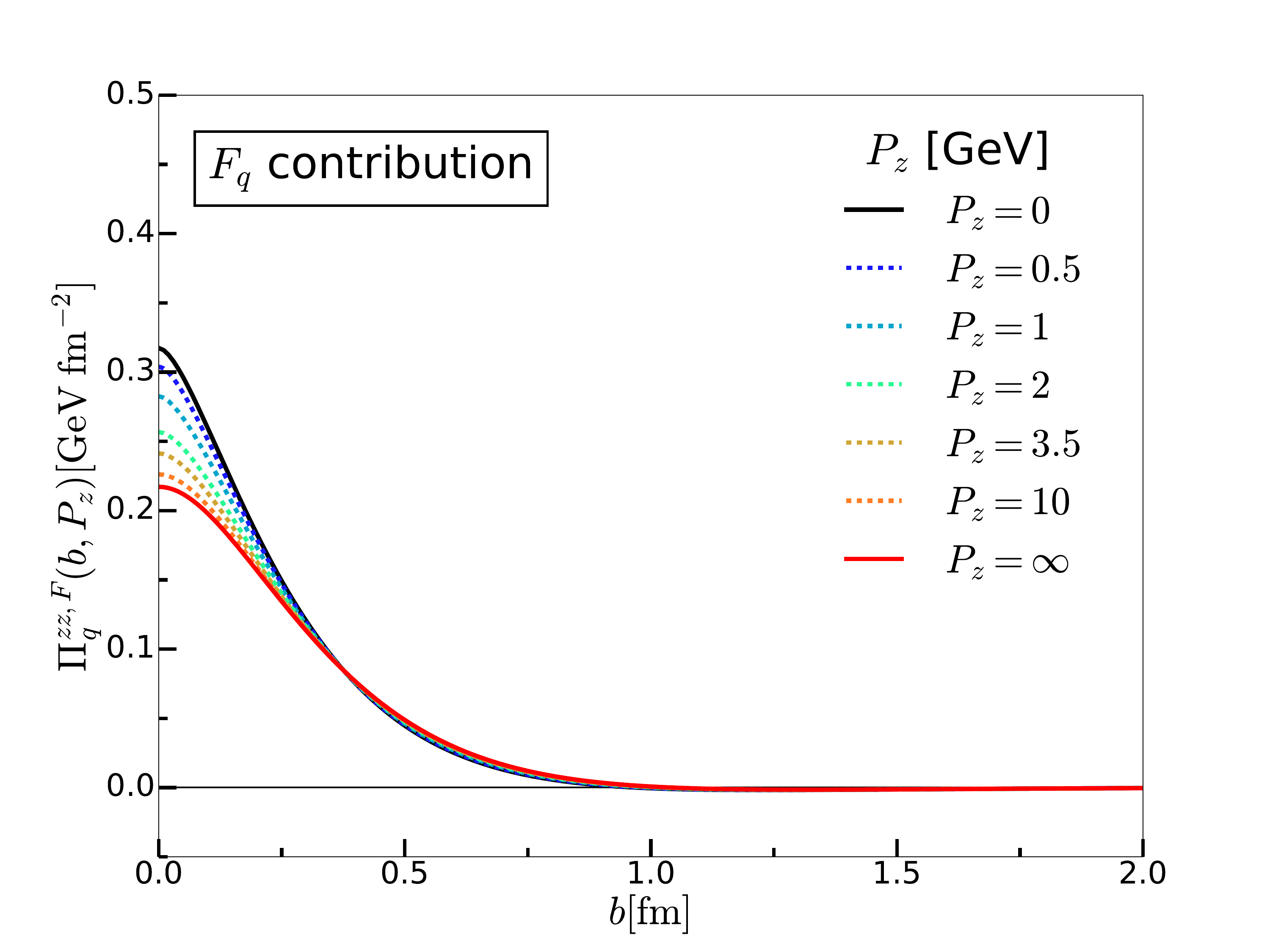}
  \end{minipage}
  \hspace{-0.6cm}
  \begin{minipage}{0.40\textwidth}
    \centering
    \includegraphics[width=\textwidth]{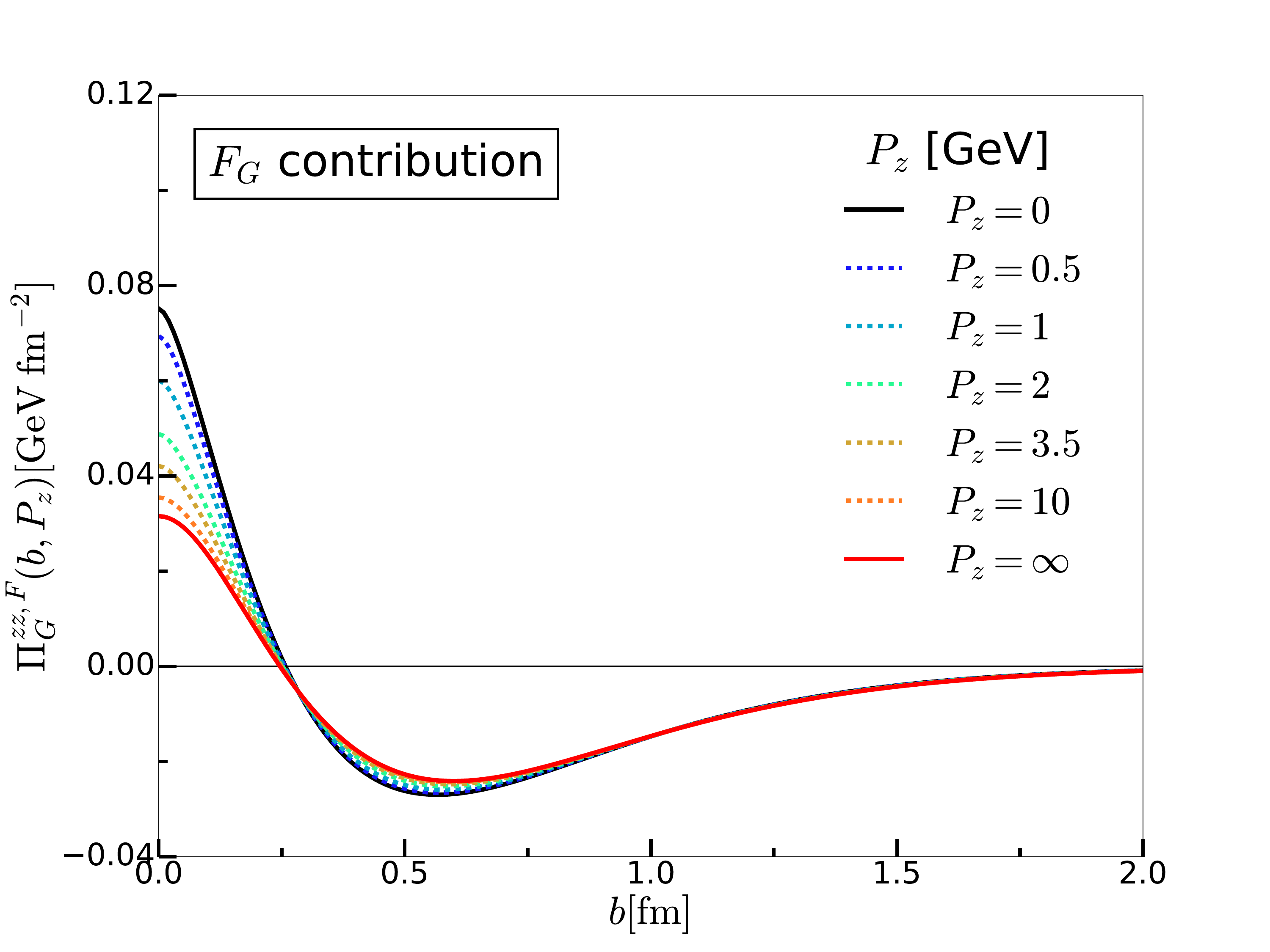}
  \end{minipage}
  
  \caption{Radial EF distributions of axial momentum flux in an unpolarized nucleon for different values of the nucleon momentum. The total axial momentum flux distribution is normalized to $M\beta^2_P$. The left (right) column corresponds to the quark (gluon) contribution. The first row depicts the sum of the three contributions shown in the other rows. Based on the simple multipole model of Ref.~\cite{Lorce:2018egm} for the EMT FFs.}

  \label{fig:7}
\end{figure*}

In Figs.~\ref{fig:4} and~\ref{fig:7}, we show the quark and gluon contributions to the EF distributions of, respectively, longitudinal momentum and axial momentum flux in an unpolarized nucleon at various values of $P_{z}$.
Since these distributions are axially symmetric, only the radial profiles are displayed. 
Note that the longitudinal momentum distributions are purely induced by the Lorentz boost. In the absence of Wigner rotation, only the contributions associated with $E_a$ and $F_a$ would play a role. At large $P_z$, the contribution associated with $S_a$ is the only one that is suppressed kinematically. The axial momentum flux distributions are similar to the energy distributions. The difference in the $P_z$-evolution between the contributions associated with $E_a$ and $F_a$ is essentially a reflection of the relative kinematical weight $\beta^2$ in Eqs.~\eqref{T00} and~\eqref{T33}.

\begin{figure*}[htbp]
  \centering
  \textbf{EF longitudinal momentum distributions for a transversely polarized nucleon}

  \vspace{0.2cm} 


  \begin{minipage}{0.37\textwidth}
    \centering
    Quark
  \end{minipage}
  \hspace{-0.05\textwidth}
  \begin{minipage}{0.37\textwidth}
    \centering
    Gluon
  \end{minipage}

  \begin{minipage}{0.35\textwidth}
    \centering
    \includegraphics[width=\textwidth]{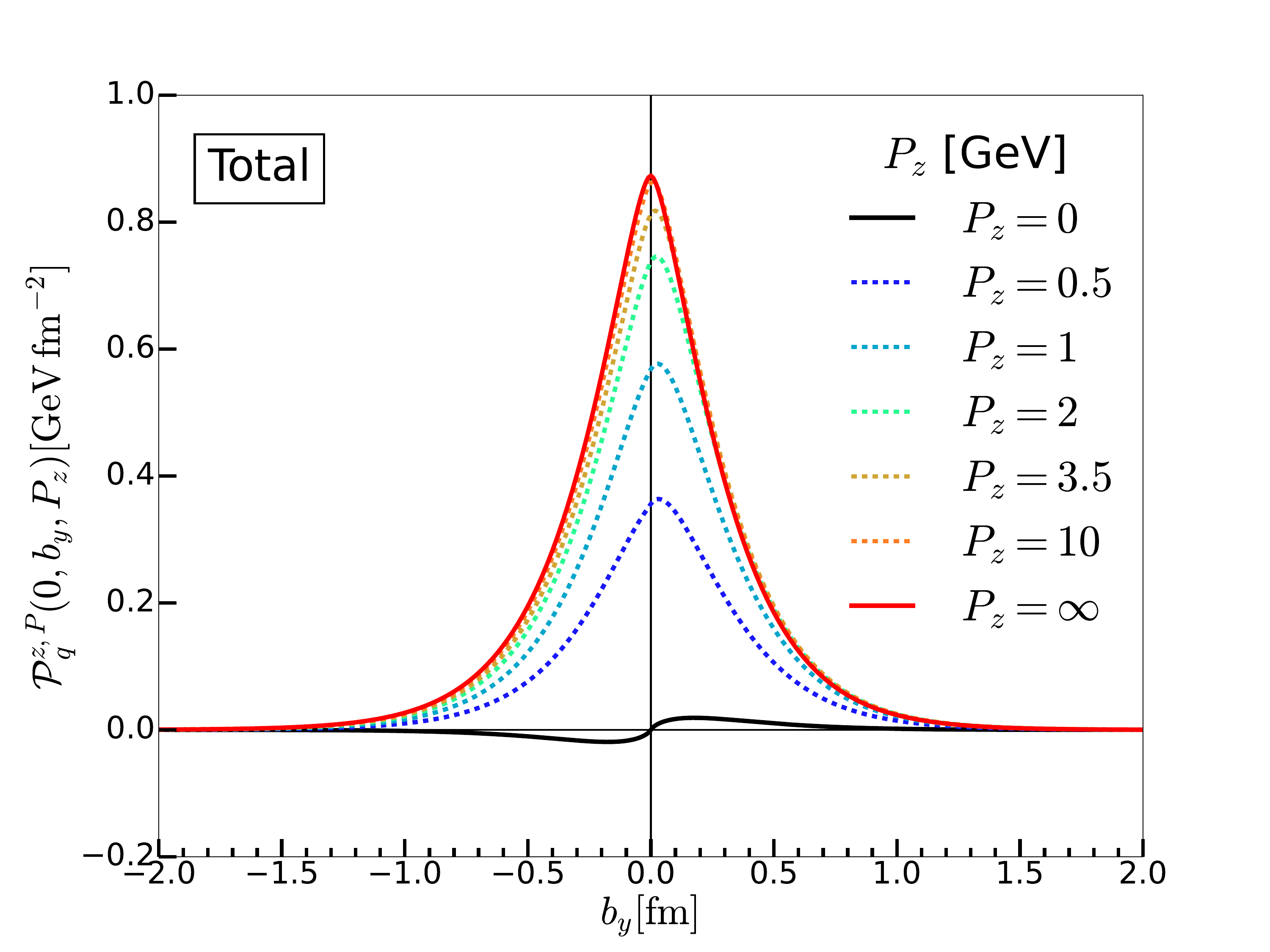}
  \end{minipage}
  \hspace{-0.6cm}
  \begin{minipage}{0.35\textwidth}
    \centering
    \includegraphics[width=\textwidth]{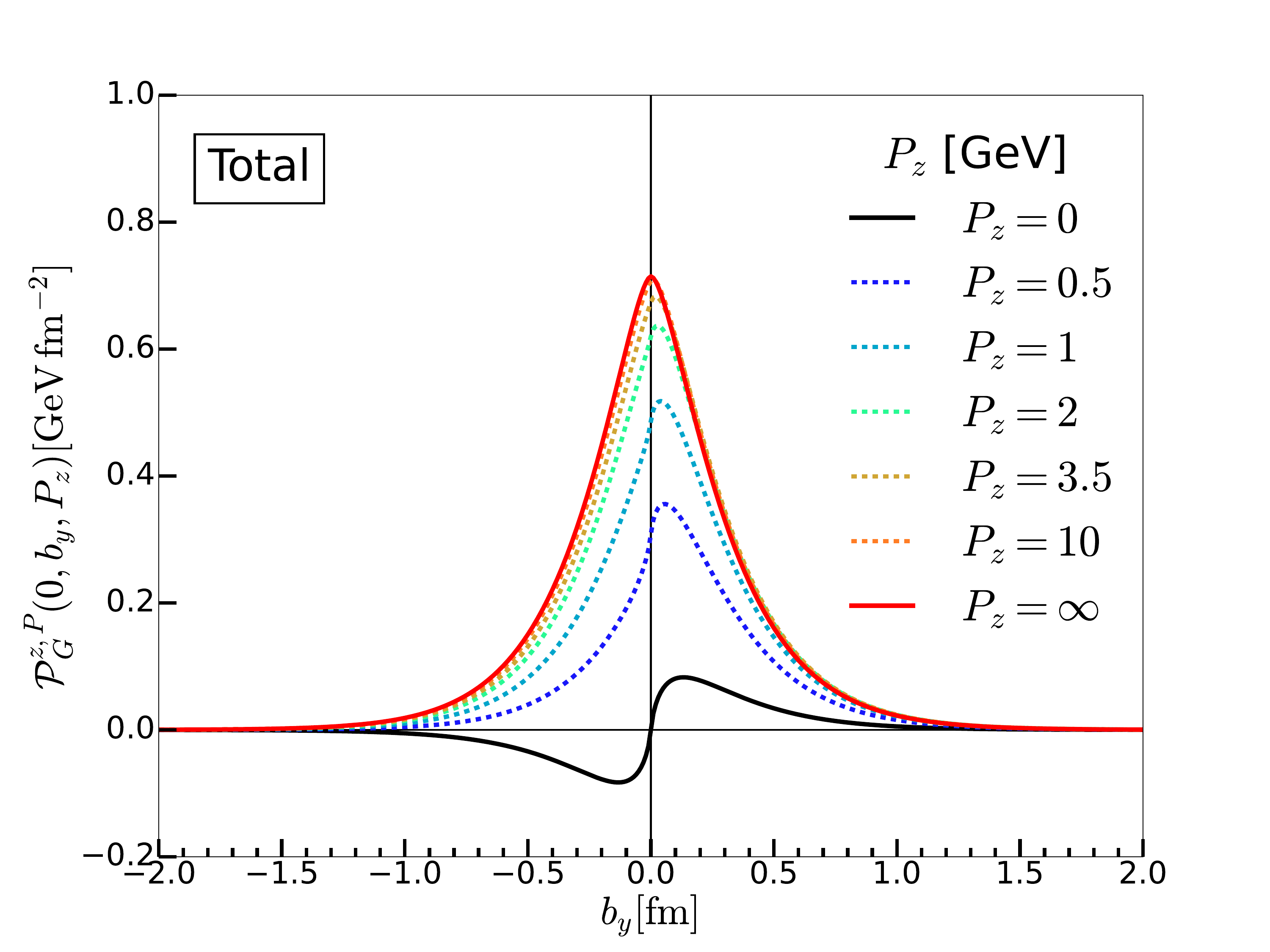}
  \end{minipage}
  
  \vspace{-0.35cm} 
  
  \begin{minipage}{0.35\textwidth}
    \centering
    \includegraphics[width=\textwidth]{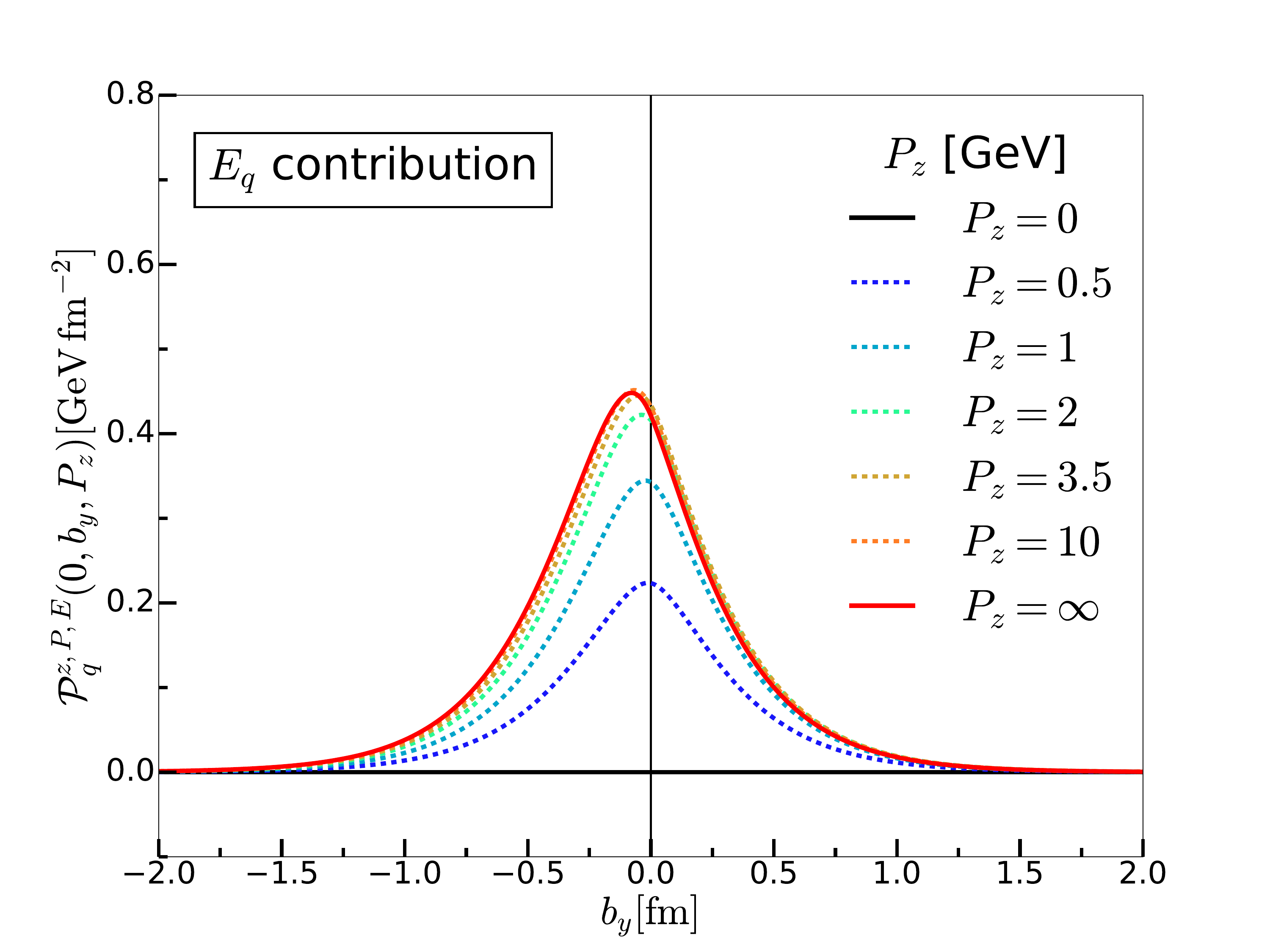}
  \end{minipage}
  \hspace{-0.6cm}
  \begin{minipage}{0.35\textwidth}
    \centering
    \includegraphics[width=\textwidth]{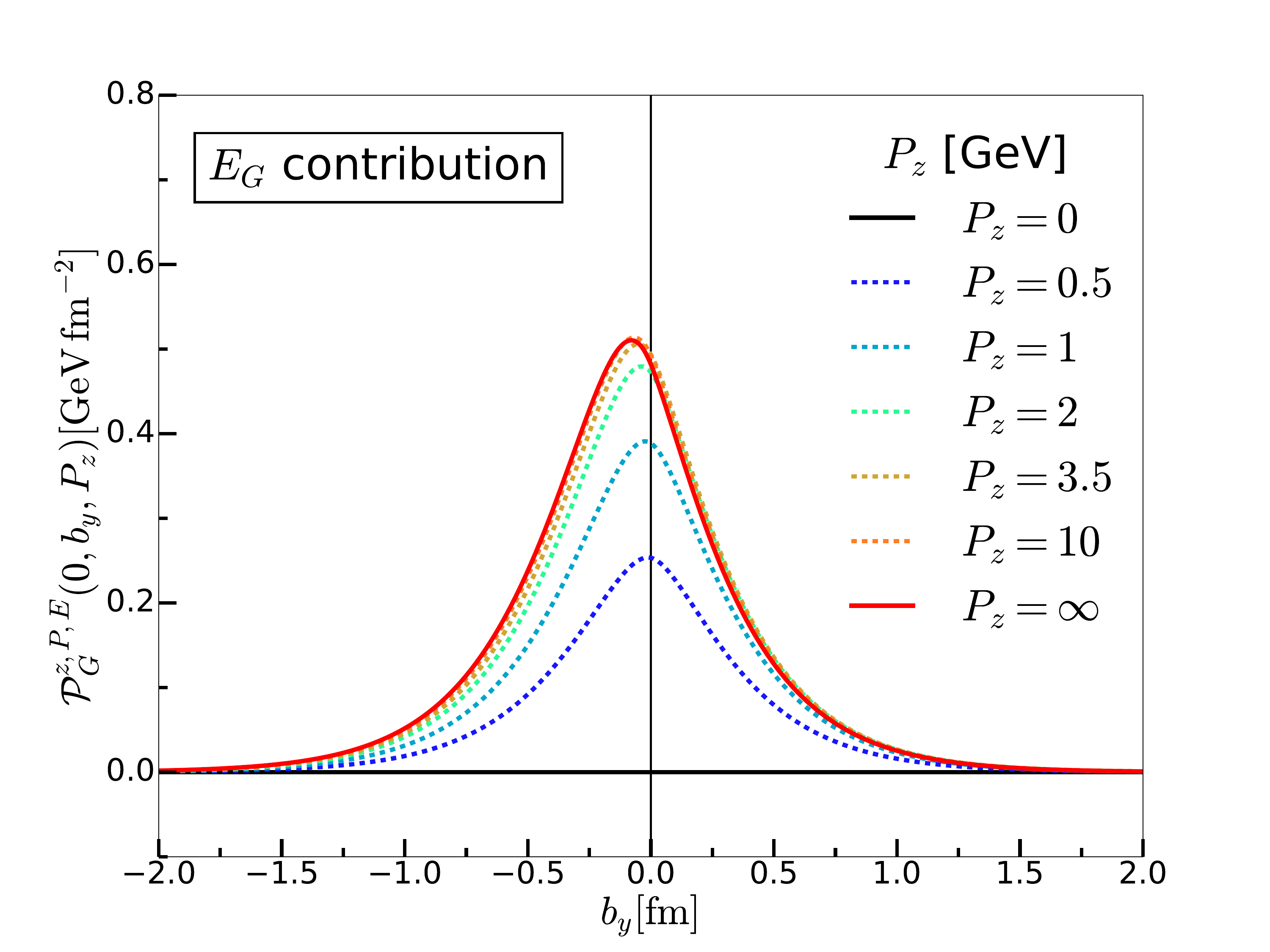}
  \end{minipage}
  
  \vspace{-0.35cm} 
  
  \begin{minipage}{0.35\textwidth}
    \centering
    \includegraphics[width=\textwidth]{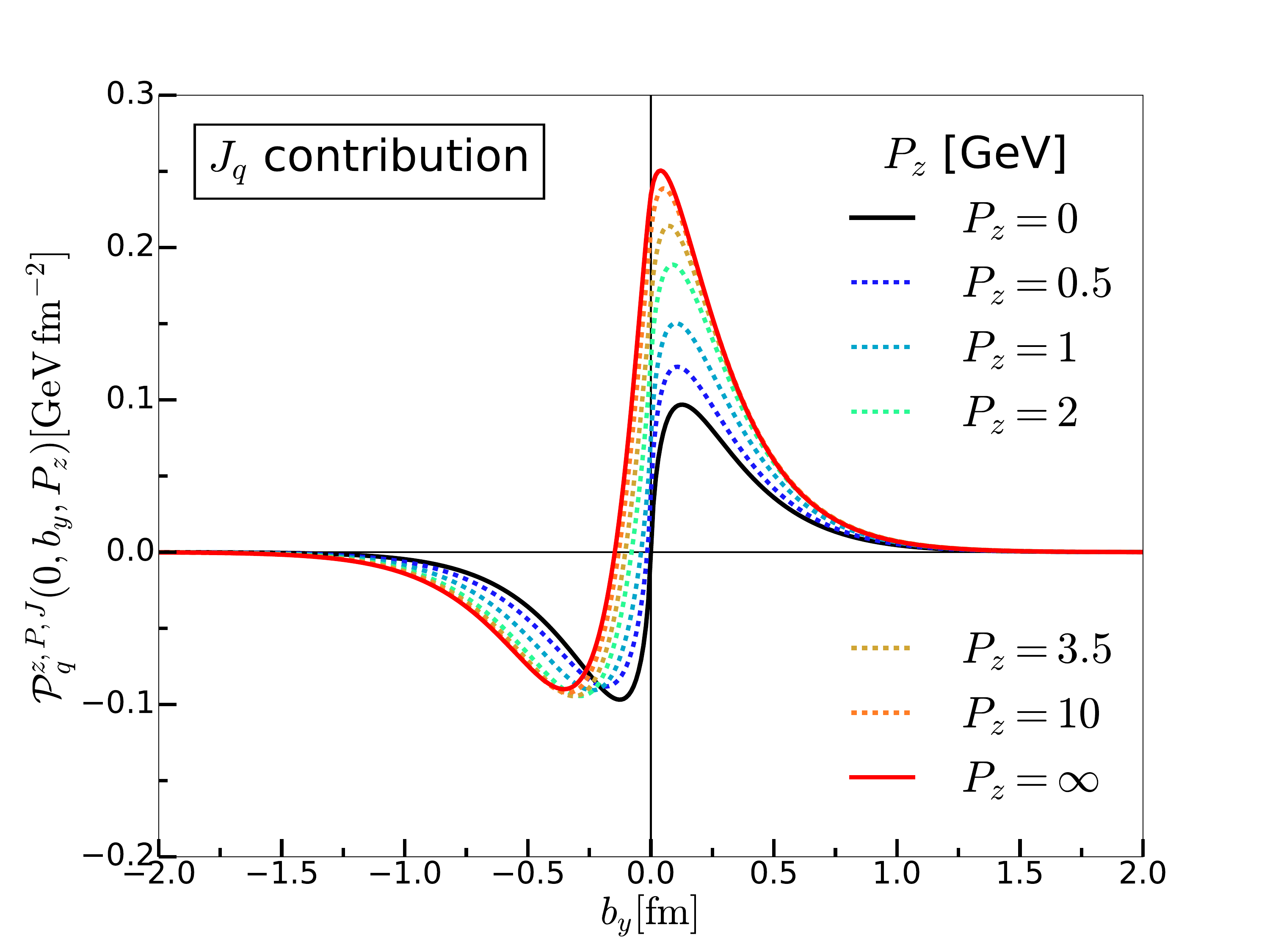}
  \end{minipage}
  \hspace{-0.6cm}
  \begin{minipage}{0.35\textwidth}
    \centering
    \includegraphics[width=\textwidth]{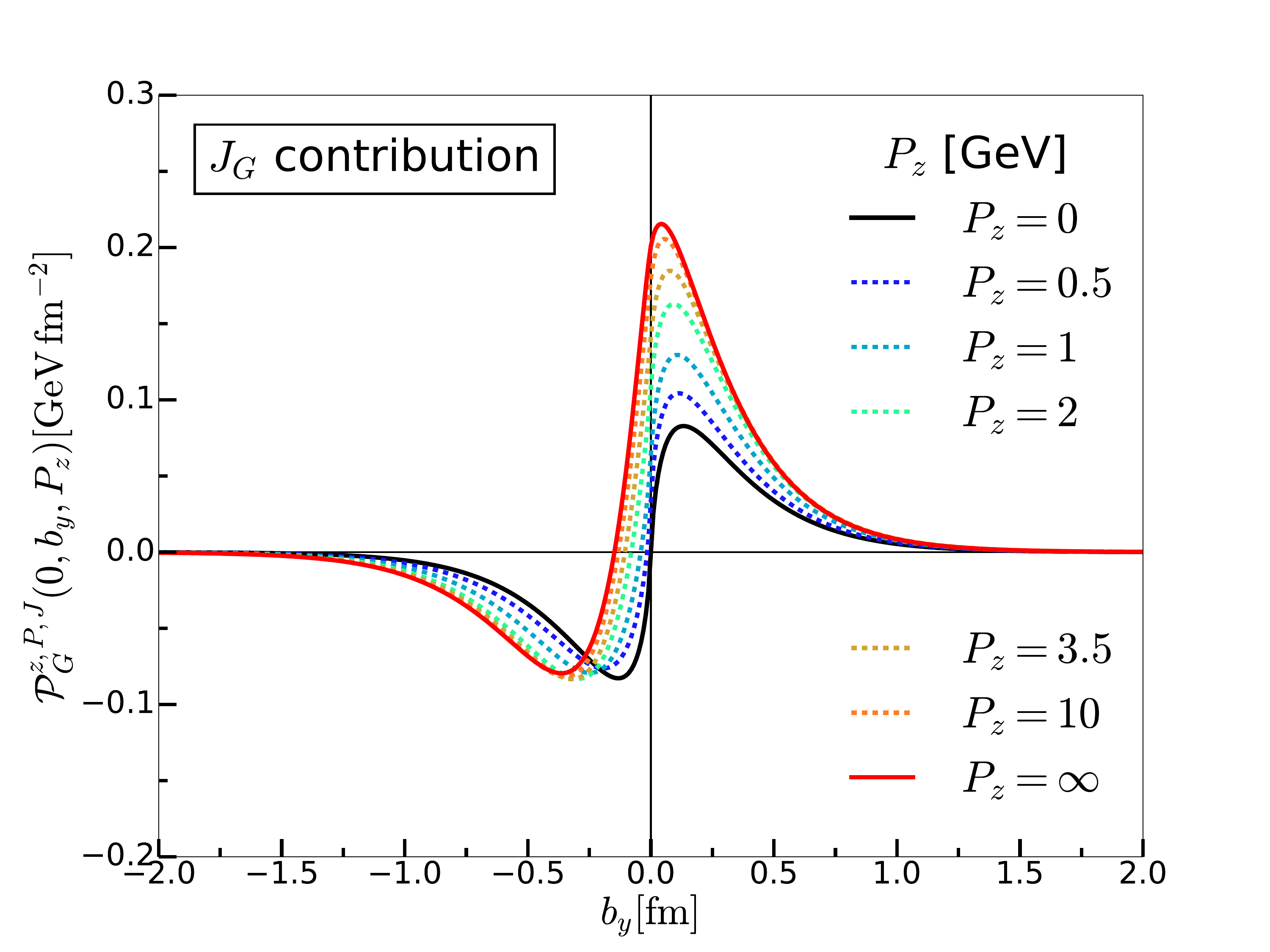}
  \end{minipage}
  
  \vspace{-0.35cm} 
  
  \begin{minipage}{0.35\textwidth}
    \centering
    \includegraphics[width=\textwidth]{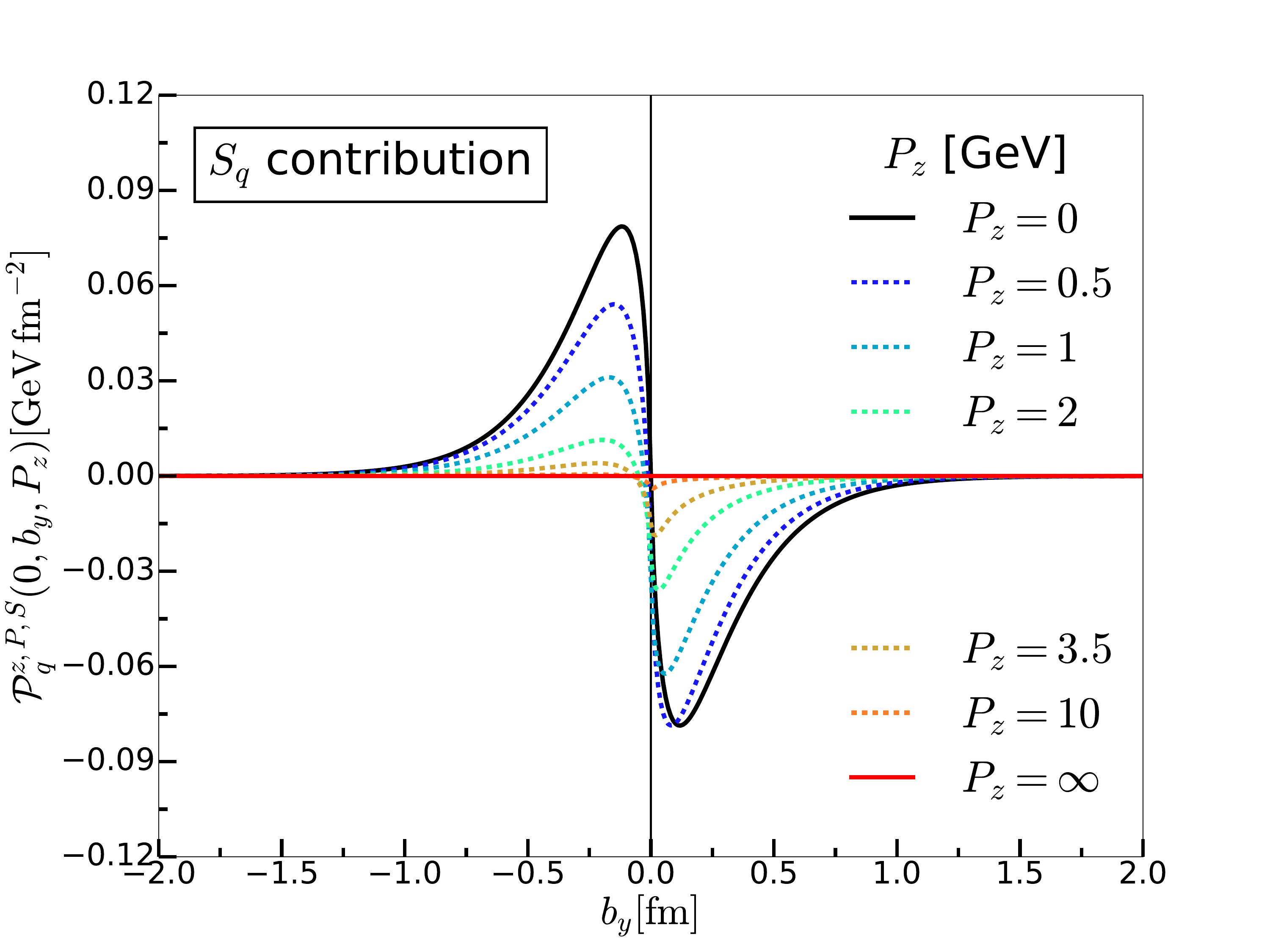}
  \end{minipage}
  \hspace{-0.6cm}
  \begin{minipage}{0.35\textwidth}
    \centering
    \includegraphics[width=\textwidth]{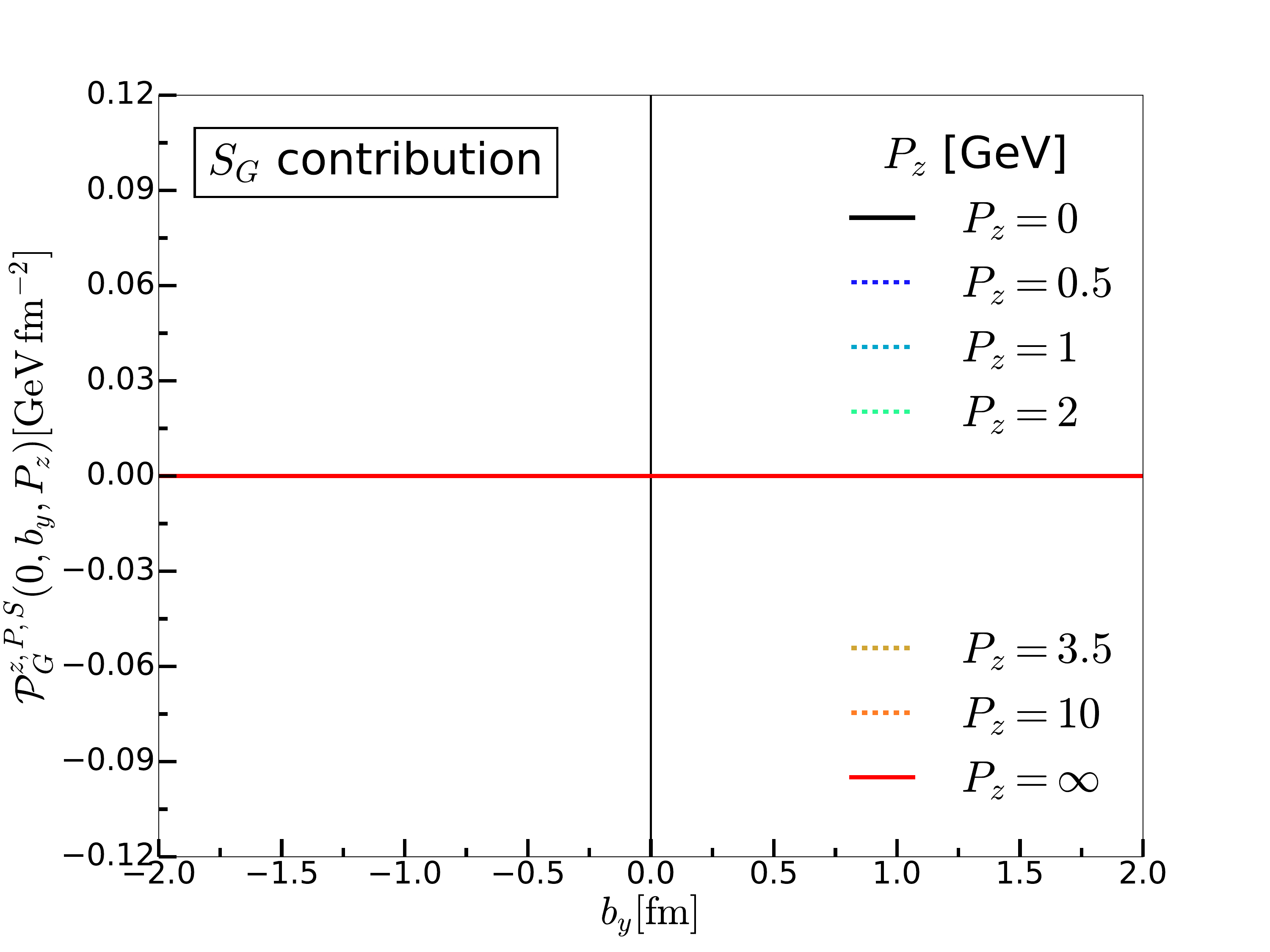}
  \end{minipage}

  \vspace{-0.35cm} 

  \begin{minipage}{0.35\textwidth}
    \centering
    \includegraphics[width=\textwidth]{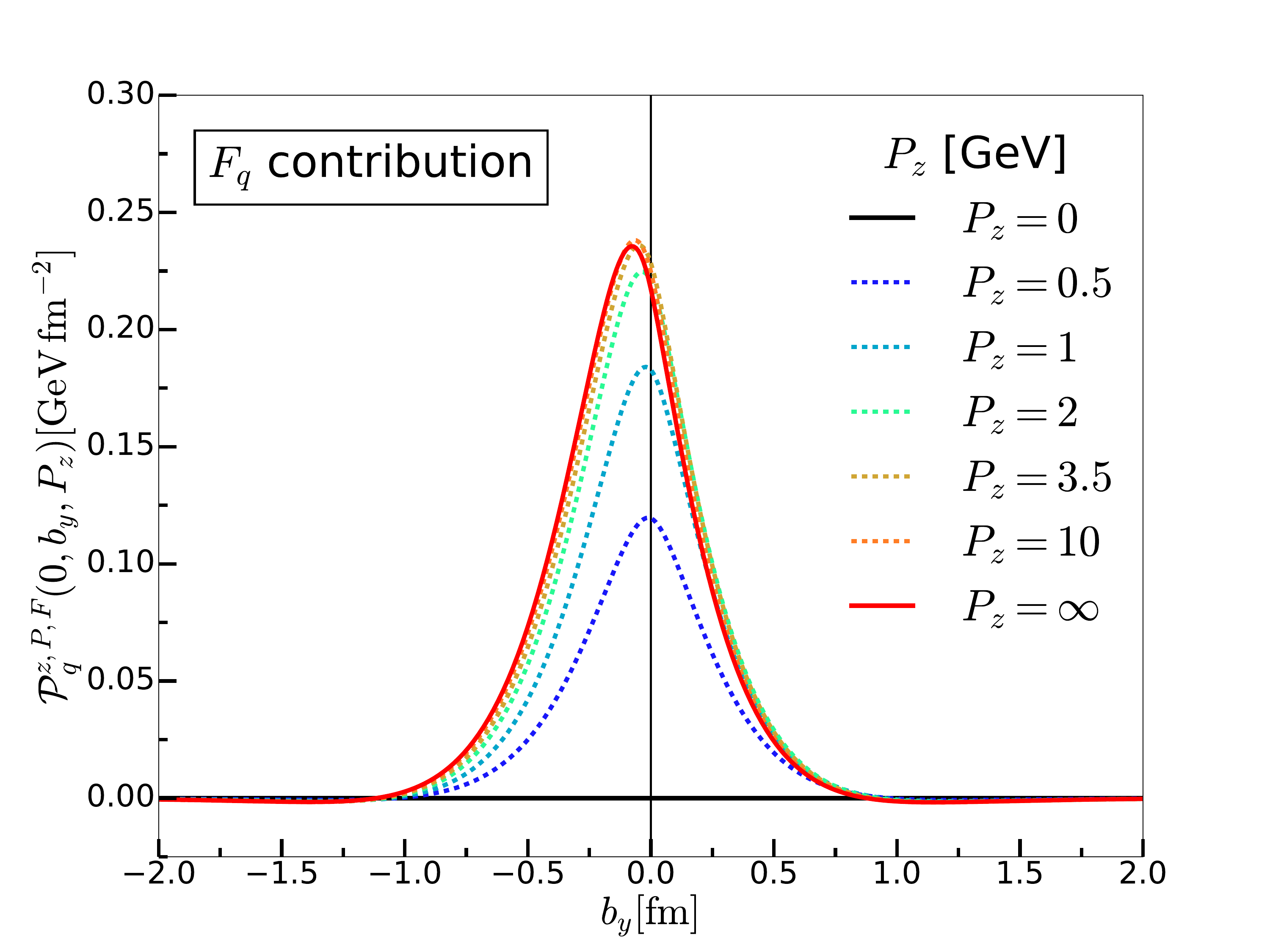}
  \end{minipage}
  \hspace{-0.6cm}
  \begin{minipage}{0.35\textwidth}
    \centering
    \includegraphics[width=\textwidth]{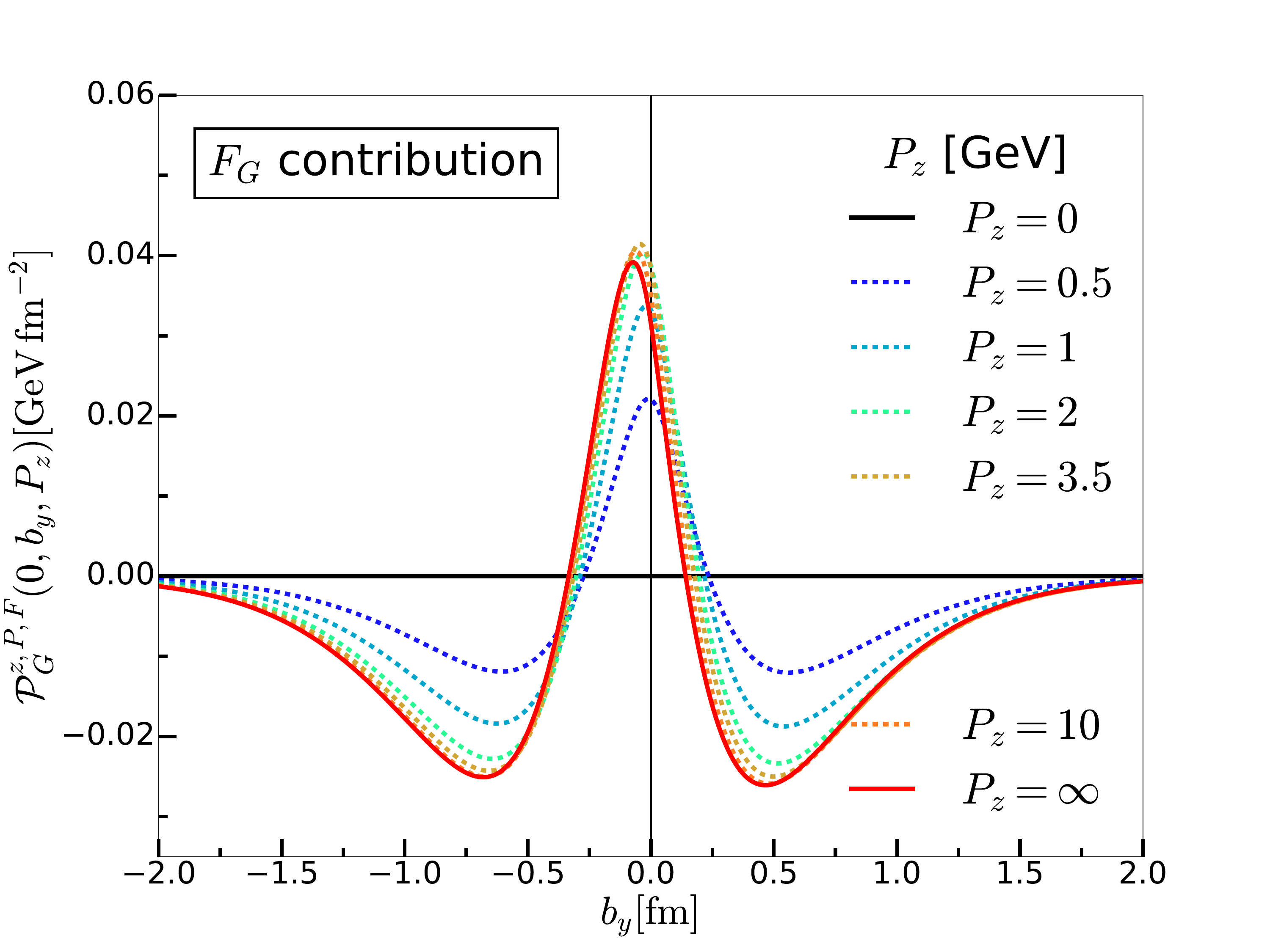}
  \end{minipage}

  \caption{EF distributions of longitudinal momentum at $b_x=0$ in a nucleon polarized along the $x$-axis for different values of the nucleon momentum. The total longitudinal momentum distribution is normalized to $M\beta_P$. The left (right) column corresponds to the quark (gluon) contribution. The first row depicts the sum of the four contributions shown in the other rows. Based on the simple multipole model of Ref.~\cite{Lorce:2018egm} for the EMT FFs.}

  \label{fig:5}
\end{figure*}

\begin{figure*}[htbp]
  \centering
  \textbf{EF axial momentum flux distributions for the transversely polarized nucleon}

  \vspace{0.2cm} 

  \begin{minipage}{0.42\textwidth}
    \centering
    Quark
  \end{minipage}
  \hspace{-0.05\textwidth}
  \begin{minipage}{0.42\textwidth}
    \centering
    Gluon
  \end{minipage}

  \begin{minipage}{0.40\textwidth}
    \centering
    \includegraphics[width=\textwidth]{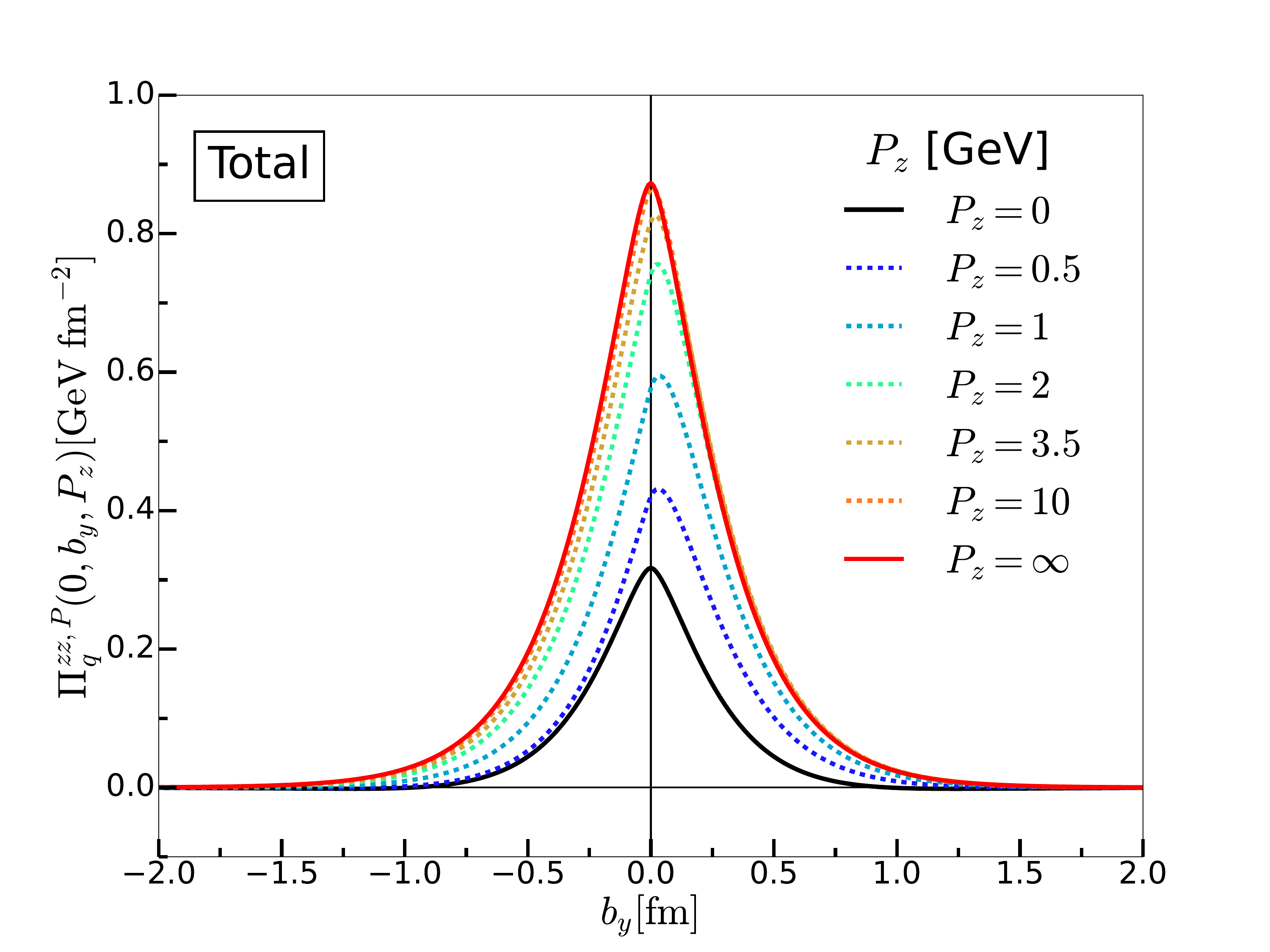}
  \end{minipage}
  \hspace{-0.6cm}
  \begin{minipage}{0.40\textwidth}
    \centering
    \includegraphics[width=\textwidth]{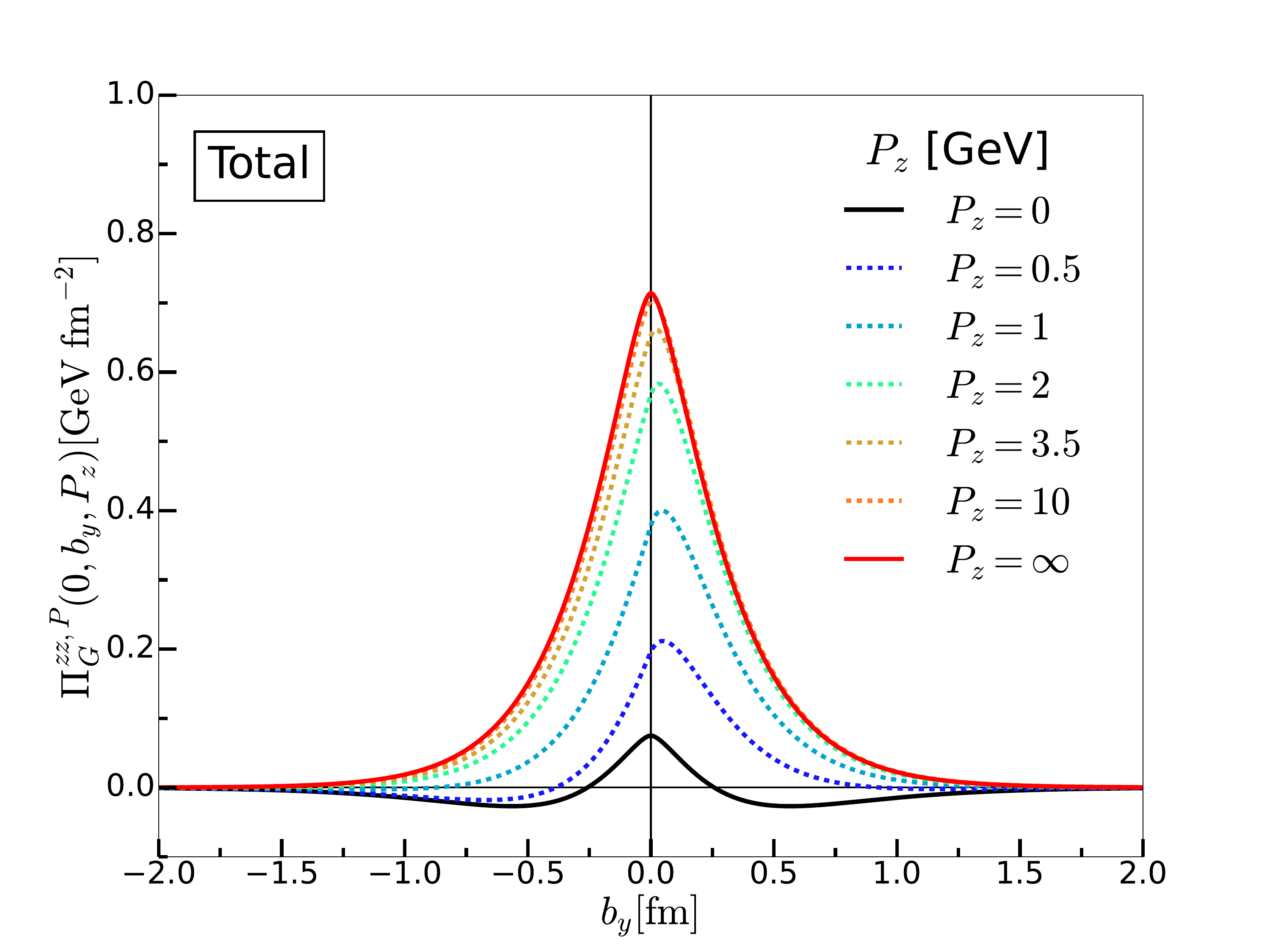}
  \end{minipage}
  
  \vspace{-0.4cm} 
  
  \begin{minipage}{0.40\textwidth}
    \centering
    \includegraphics[width=\textwidth]{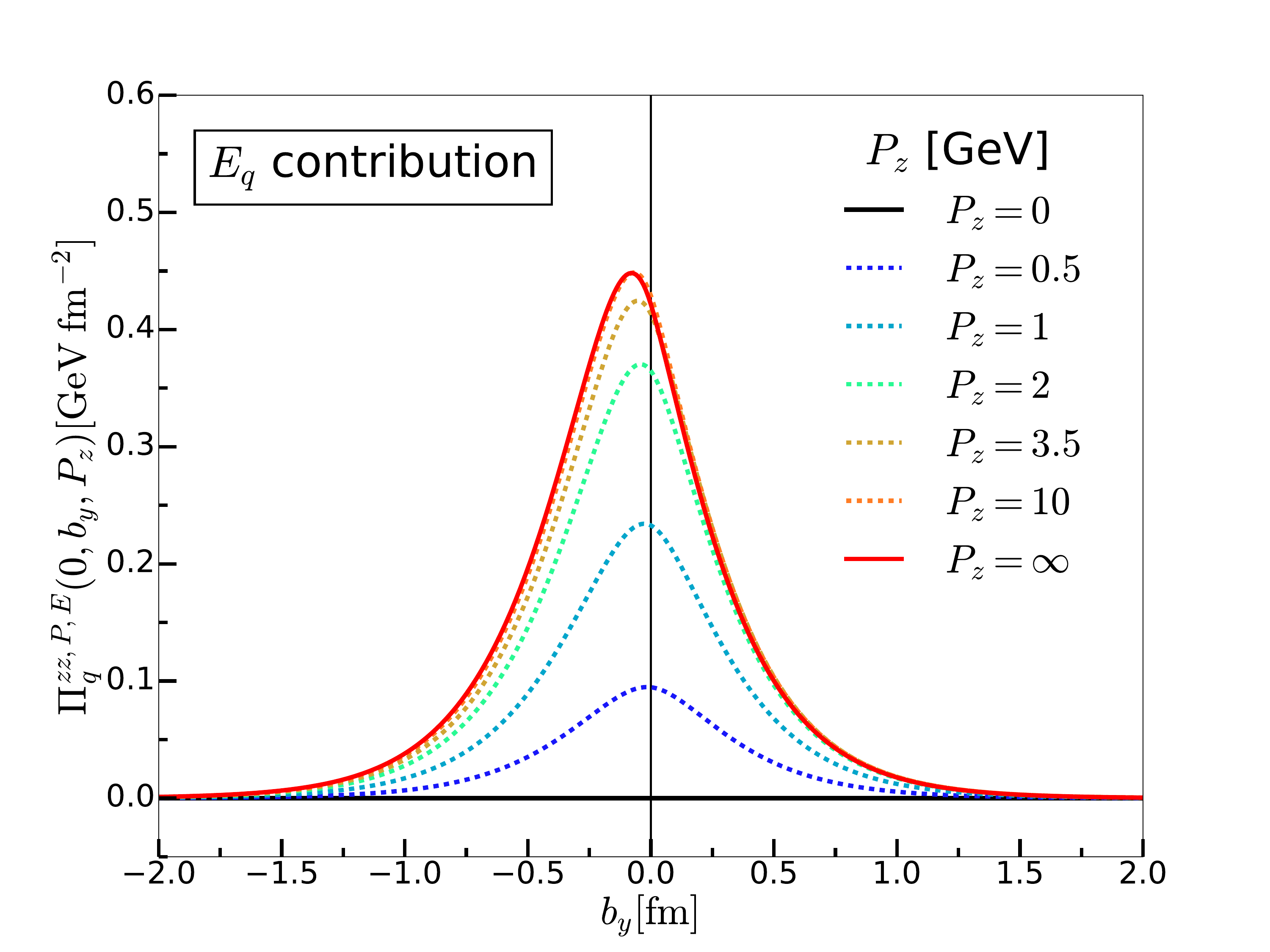}
  \end{minipage}
  \hspace{-0.6cm}
  \begin{minipage}{0.40\textwidth}
    \centering
    \includegraphics[width=\textwidth]{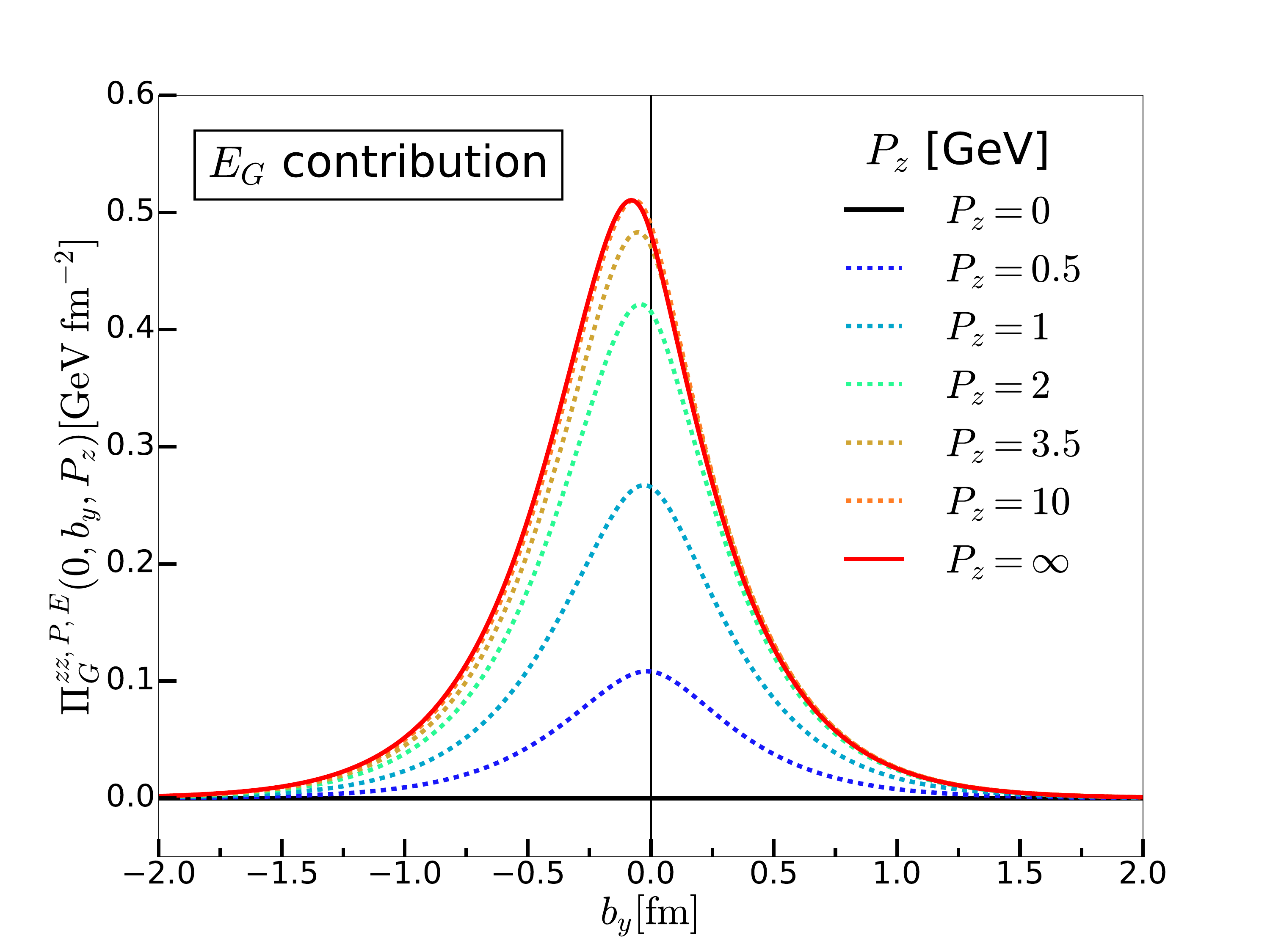}
  \end{minipage}
  
  \vspace{-0.4cm} 
  
  \begin{minipage}{0.40\textwidth}
    \centering
    \includegraphics[width=\textwidth]{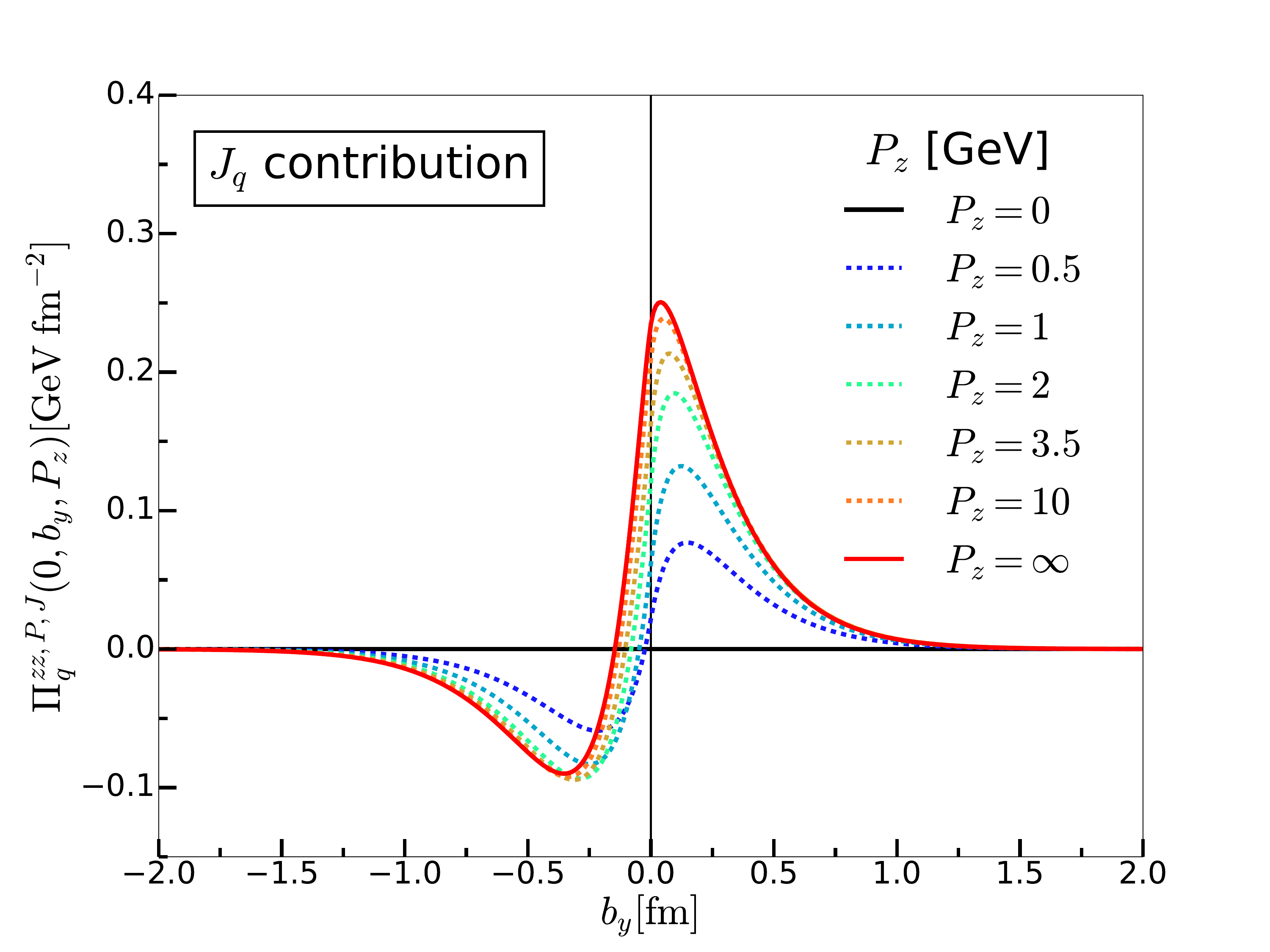}
  \end{minipage}
  \hspace{-0.6cm}
  \begin{minipage}{0.40\textwidth}
    \centering
    \includegraphics[width=\textwidth]{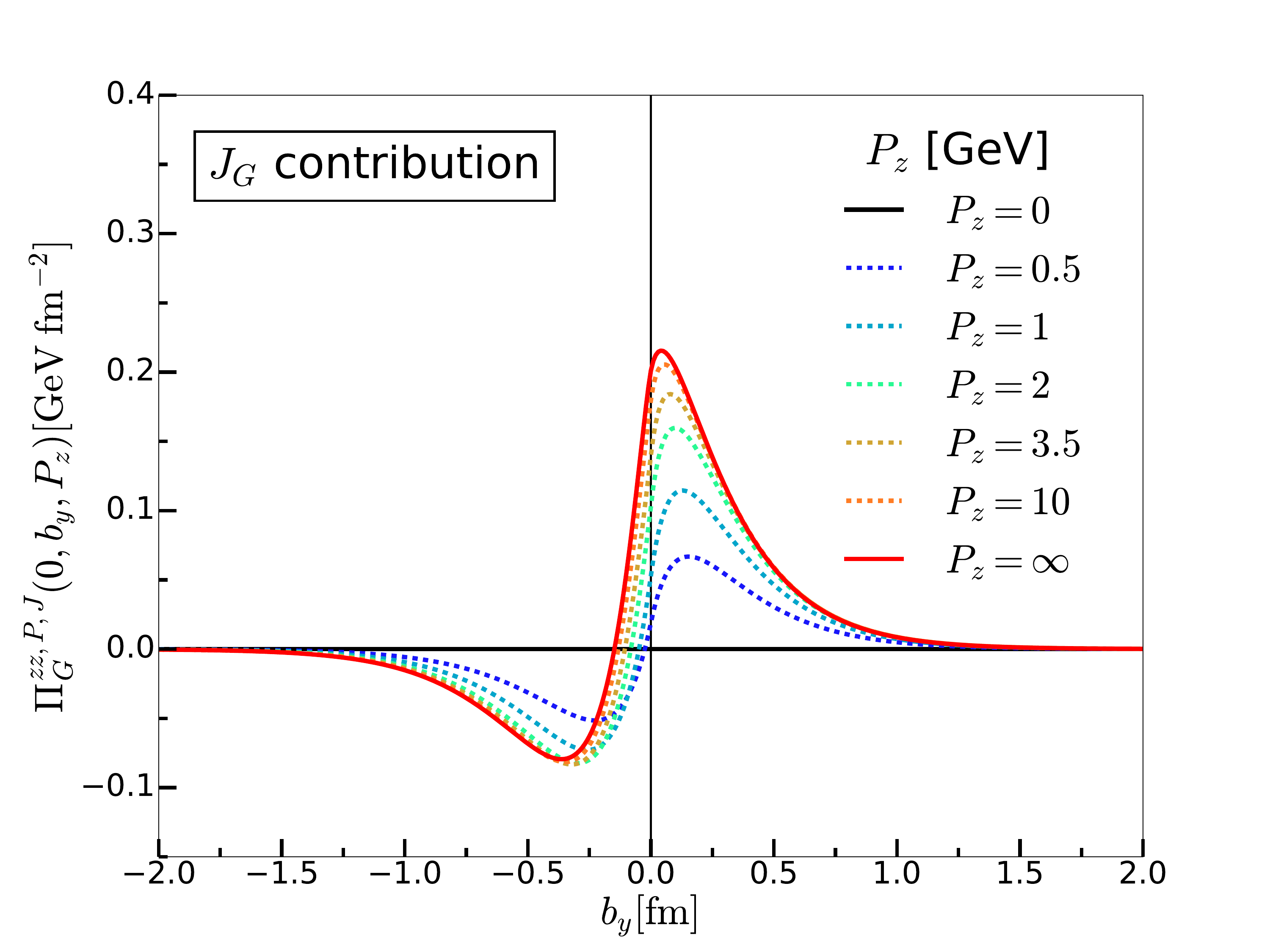}
  \end{minipage}
  
  \vspace{-0.4cm} 
  
  \begin{minipage}{0.40\textwidth}
    \centering
    \includegraphics[width=\textwidth]{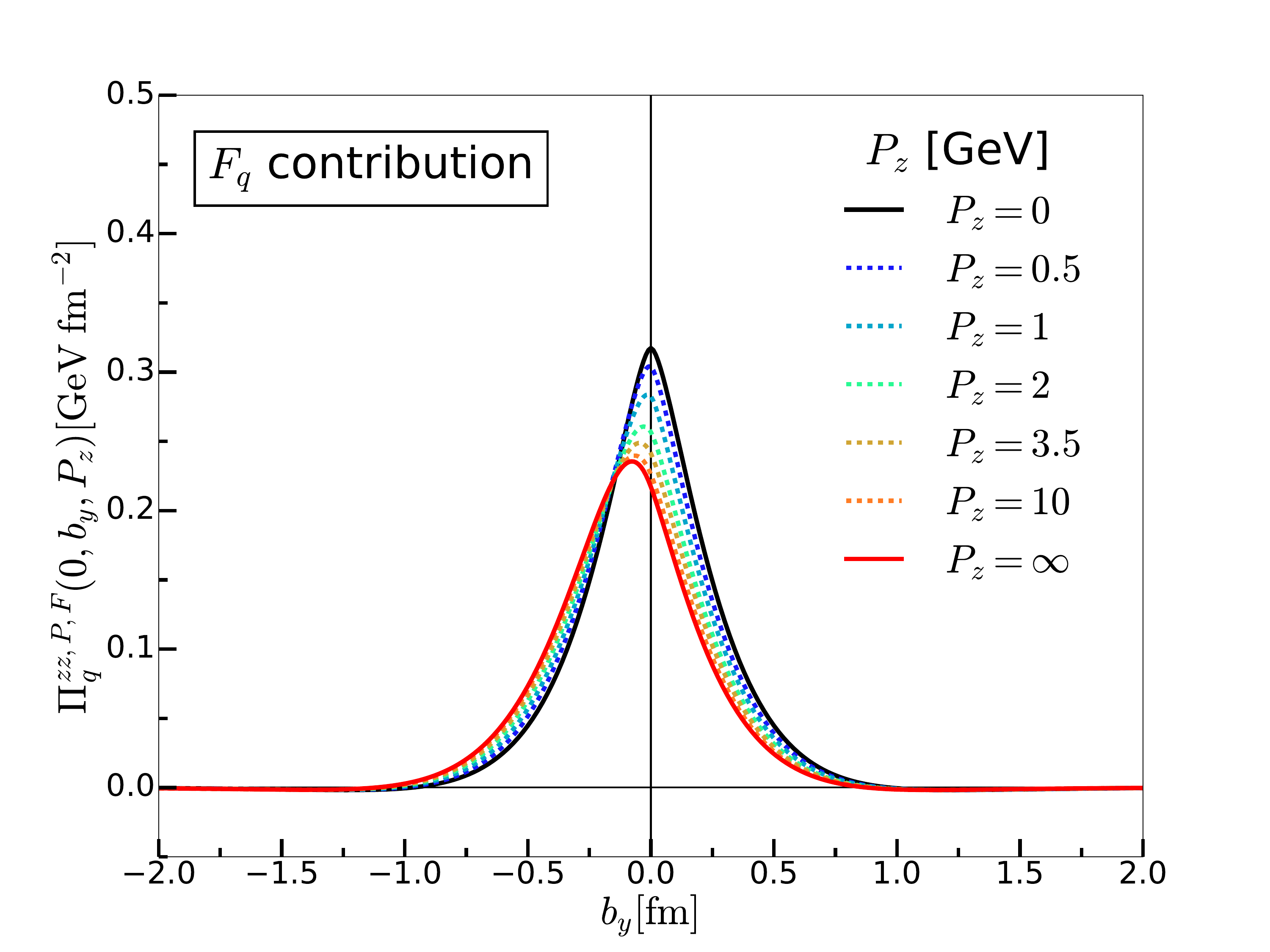}
  \end{minipage}
  \hspace{-0.6cm}
  \begin{minipage}{0.40\textwidth}
    \centering
    \includegraphics[width=\textwidth]{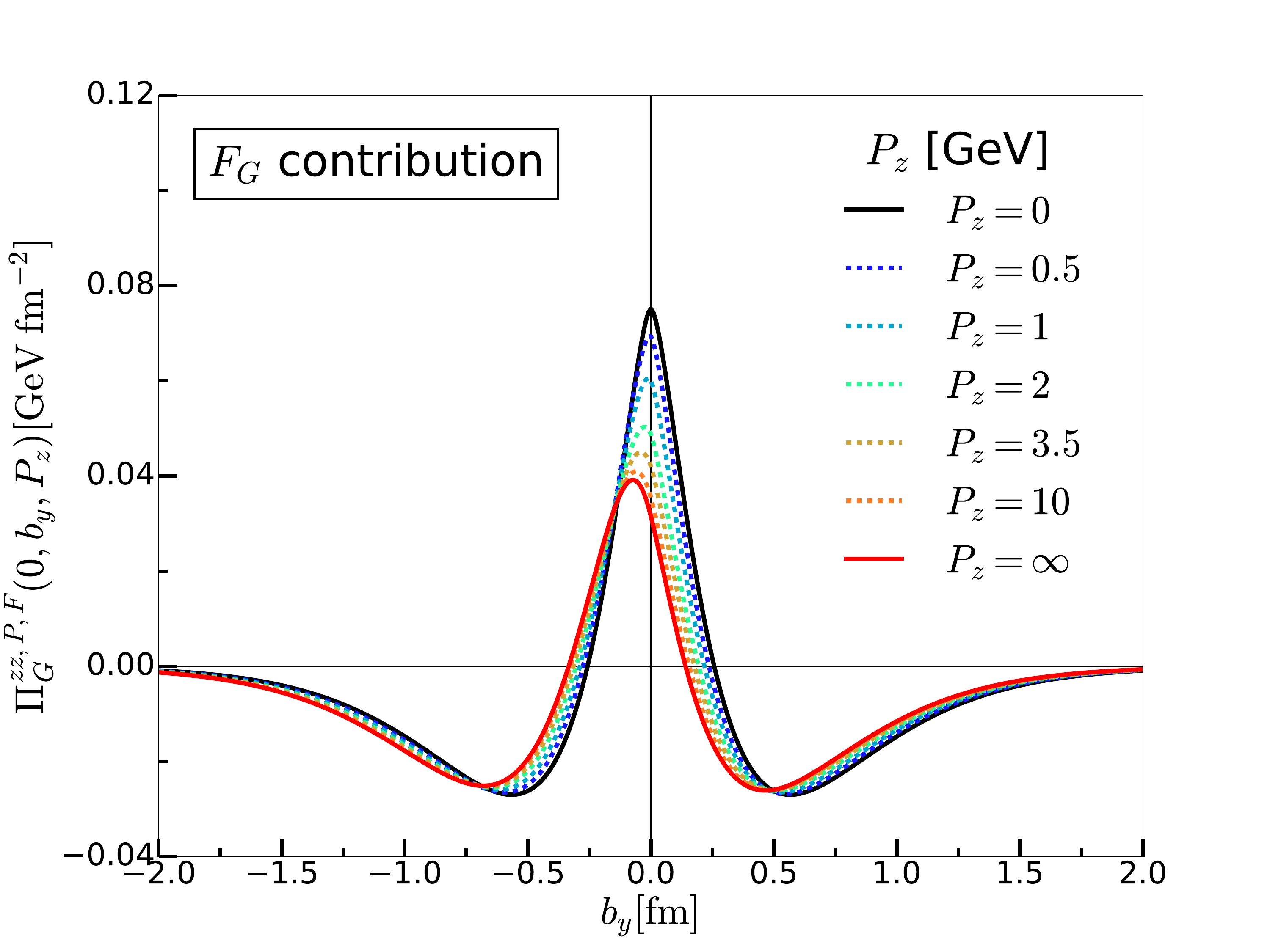}
  \end{minipage}

  \caption{EF distributions of axial momentum flux at $b_x=0$ in a nucleon polarized along the $x$-axis for different values of the nucleon momentum. The total axial momentum flux distribution is normalized to $M\beta^2_P$. The left (right) column corresponds to the quark (gluon) contribution. The first row depicts the sum of the three contributions shown in the other rows. Based on the simple multipole model of Ref.~\cite{Lorce:2018egm} for the EMT FFs.}

  \label{fig:8}
\end{figure*}

In Fig.~\ref{fig:5} and~\ref{fig:8},
we illustrate the same distributions for the nucleon polarized along the $x$-direction. Similar to the polarized EF energy distributions, the dipole shift induced by the contribution associated with $J_a$ goes 
in the opposite direction to those associated with $E_a$ and $F_a$. 

\bibliography{RED}
\bibliographystyle{apsrev}

\end{document}